\documentclass[twocolumn]{aastex61}
\usepackage{tcolorbox}
\usepackage{savesym}
\usepackage{natbib}
\usepackage{float}
\usepackage{comment}
\usepackage{subfigure}
\usepackage{longtable}
\usepackage{upgreek}
\usepackage{multirow}
\usepackage{graphicx}
\usepackage{chngcntr}

\savesymbol{tablenum}
\restoresymbol{SIX}{tablenum}
\definecolor{orcidlogocol}{HTML}{A6CE39}

\begin{document}
	\title{ARES IV\footnote{ARES: Ariel Retrieval of Exoplanets School}: Probing the atmospheres of the two warm small planets HD\,106315\,c and HD\,3167\,c with the HST/WFC3 camera} 
	
	\author{Gloria Guilluy}
	\affil{Dipartimento di Fisica, Università degli Studi di Torino, via Pietro Giuria 1, I-10125 Torino, Italy}
	\affil{INAF Osservatorio Astrofisico di Torino, Via Osservatorio 20, I-10025 Pino Torinese, Italy}
	
	\author{Am\'elie Gressier}
	\affil{LATMOS, CNRS, Sorbonne Universit\'e /UVSQ, 11 boulevard d’Alembert, F-78280 Guyancourt, France}
	\affil{Sorbonne Universit\'es, UPMC Universit\'e Paris 6 et CNRS, 
		UMR 7095, Institut d'Astrophysique de Paris, 98 bis bd Arago,
		75014 Paris, France}
	\affil{LESIA, Observatoire de Paris, Universit\'e PSL, CNRS, Sorbonne Universit\'e, Universit\'e de Paris, 5 place Jules Janssen, 92195 Meudon, France}

	\author{Sam Wright}
	\affil{Department of Physics and Astronomy, University College London, London, United Kingdom}
	
	\author{Alexandre Santerne}
	\affil{Aix Marseille Univ, CNRS, CNES, LAM, Marseille, France}
	
	\author{Adam Yassin Jaziri}  
	\affil{Laboratoire d'astrophysique de Bordeaux, Univ. Bordeaux, CNRS, B18N, all\'{e}e Geoffroy Saint-Hilaire, 33615 Pessac, France}
	
	\author{Billy Edwards}
	\affil{Department of Physics and Astronomy, University College London, London, United Kingdom}
	\author{Quentin Changeat}
	\affil{Department of Physics and Astronomy, University College London, London, United Kingdom}
	\author{Darius Modirrousta-Galian} 
	\affil{INAF – Osservatorio Astronomico di Palermo, Piazza del Parlamento 1, I-90134 Palermo, Italy}
	\affil{University of Palermo, Department of Physics and Chemistry, Via Archirafi 36, Palermo, Italy}
	
	\author{Nour Skaf} 
	\affil{LESIA, Observatoire de Paris, Universit\'e PSL, CNRS, Sorbonne Universit\'e, Universit\'e de Paris, 5 place Jules Janssen, 92195 Meudon, France}
	\affil{Department of Physics and Astronomy, University College London, London, United Kingdom}
	\author{Ahmed Al-Refaie} 
	\affil{Department of Physics and Astronomy, University College London, London, United Kingdom}
	
	\author{Robin Baeyens}
	\affil{Instituut voor Sterrenkunde, KU Leuven, Celestijnenlaan 200D bus 2401, 3001 Leuven, Belgium}
	
	\author{Michelle Fabienne Bieger}
	\affil{College of Engineering, Mathematics and Physical Sciences, Physics Building, University of Exeter
		North Park Road, Exeter, United Kingdom}
	
	\author{Doriann Blain}
	\affil{LESIA, Observatoire de Paris, Universit\'e PSL, CNRS, Sorbonne Universit\'e, Universit\'e de Paris, 5 place Jules Janssen, 92195 Meudon, France}
	
	\author{Flavien Kiefer}
	\affil{Sorbonne Universit\'es, UPMC Universit\'e Paris 6 et CNRS, 
		UMR 7095, Institut d'Astrophysique de Paris, 98 bis bd Arago,
		75014 Paris, France}
	
	\author{Mario Morvan}
	\affil{Department of Physics and Astronomy, University College London, London, United Kingdom}
	
	\author{Lorenzo V. Mugnai} 
	\affil{La Sapienza Universit\'a di Roma, Department of Physics, Piazzale Aldo Moro 2, 00185 Roma, Italy}
	
	\author{William Pluriel}  
	\affil{Laboratoire d'astrophysique de Bordeaux, Univ. Bordeaux, CNRS, B18N, all\'{e}e Geoffroy Saint-Hilaire, 33615 Pessac, France}
	
	\author{Mathilde Poveda}
	\affil{Laboratoire Interuniversitaire des Syst\`{e}mes Atmosph\'{e}riques (LISA), UMR CNRS 7583, Universit\'{e} Paris-Est-Cr\'eteil, Universit\'e de Paris, Institut Pierre Simon Laplace, Cr\'{e}teil, France}
	\affil{Maison de la Simulation, CEA, CNRS, Univ. Paris-Sud, UVSQ, Universit\'e Paris-Saclay, F-91191 Gif-sur-Yvette, France}

	\author{Tiziano Tsingales} 
	\affil{Laboratoire d'astrophysique de Bordeaux, Univ. Bordeaux, CNRS, B18N, all\'{e}e Geoffroy Saint-Hilaire, 33615 Pessac, France}
	
	\author{Niall Whiteford}
	\affil{Institute for Astronomy, University of Edinburgh, Blackford Hill, Edinburgh, EH9 3HJ, UK}
	\affil{Centre for Exoplanet Science, University of Edinburgh, Edinburgh, EH9 3FD, UK}

	\author{Kai Hou Yip}
	\affil{Department of Physics and Astronomy, University College London, London, United Kingdom}
	
	\author{Benjamin Charnay}
	\affil{LESIA, Observatoire de Paris, Universit\'e PSL, CNRS, Sorbonne Universit\'e, Universit\'e de Paris, 5 place Jules Janssen, 92195 Meudon, France}
	
	\author{J\'{e}r\'{e}my Leconte}  
	\affil{Laboratoire d'astrophysique de Bordeaux, Univ. Bordeaux, CNRS, B18N, all\'{e}e Geoffroy Saint-Hilaire, 33615 Pessac, France}
	
	\author{Pierre Drossart}  
	\affil{Sorbonne Universit\'es, UPMC Universit\'e Paris 6 et CNRS, 
		UMR 7095, Institut d'Astrophysique de Paris, 98 bis bd Arago,
		75014 Paris, France}
	\affil{LESIA, Observatoire de Paris, Universit\'e PSL, CNRS, Sorbonne Universit\'e, Universit\'e de Paris, 5 place Jules Janssen, 92195 Meudon, France}

	\author{Alessandro Sozzetti}		
	\affil{INAF Osservatorio Astrofisico di Torino, Via Osservatorio 20, I-10025 Pino Torinese, Italy}
	
	\author{Emmanuel Marcq}
	\affil{LATMOS, CNRS, Sorbonne Universit\'e /UVSQ, 11 boulevard d’Alembert, F-78280 Guyancourt, France}
	
	\author{Angelos Tsiaras}
	\affil{Department of Physics and Astronomy, University College London, London, United Kingdom}
	
	\author{Olivia Venot}  
	\affil{Laboratoire Interuniversitaire des Syst\`{e}mes Atmosph\'{e}riques (LISA), UMR CNRS 7583, Universit\'{e} Paris-Est-Cr\'eteil, Universit\'e de Paris, Institut Pierre Simon Laplace, Cr\'{e}teil, France}

	\author{Ingo Waldmann}
	\affil{Department of Physics and Astronomy, University College London, London, United Kingdom}
	
	\author{Jean-Philippe Beaulieu}
	\affil{School of Physical Sciences, University of Tasmania,
		Private Bag 37 Hobart, Tasmania 7001 Australia}
	\affil{Sorbonne Universit\'es, UPMC Universit\'e Paris 6 et CNRS, 
		UMR 7095, Institut d'Astrophysique de Paris, 98 bis bd Arago,
		75014 Paris, France}
	
	%%ABSTRACT
	
	\begin{abstract}
		We present an atmospheric characterization study of two medium sized planets bracketing the radius of Neptune: HD\,106315\,c (R$_{\rm{P}}$=4.98 $\pm$ 0.23 R$_{\oplus}$) and HD\,3167\,c (R$_{\rm{P}}$=2.740$_{-0.100}^{+0.106}$ R$_{\oplus}$). We analyse spatially scanned spectroscopic observations obtained with the G141 grism (1.125 - 1.650 $\mu$m) of the Wide Field Camera 3 (WFC3) onboard the Hubble Space Telescope. We use the publicly available \verb+Iraclis+ pipeline and TauREx3 atmospheric retrieval code and we detect water vapor in the atmosphere of both planets with an abundance of $\log_{10}[\mathrm{H_2O}]=-2.1^{+0.7}_{-1.3}$ ($\sim$5.68$\sigma$) and $\log_{10}[\mathrm{H_2O}]=-4.1^{+0.9}_{-0.9}$  ($\sim$3.17$\sigma$) for HD\,106315\,c and HD\,3167\,c, respectively.  The transmission spectrum of HD\,106315\,c shows also a possible evidence of ammonia absorption ($\log_{10}[\mathrm {NH_3}]=-4.3^{+0.7}_{-2.0}$, $\sim$1.97$\sigma$ -even if it is not significant-), whilst carbon dioxide absorption features may be present in the atmosphere of HD\,3167\,c in the $\sim$1.1-1.6~$\mu$m wavelength range ($\log_{10}[\mathrm{CO_{2}}]= -2.4^{+0.7}_{-1.0}$, $\sim$3.28$\sigma$). However the CO$_2$ detection appears significant, it must be considered carefully and put into perspective. Indeed, CO$_2$ presence is not explained by 1D equilibrium chemistry models, and it could be due to possible systematics.
		The additional contribution of clouds, CO and CH$_4$ are discussed. HD\,106315\,c and HD\,3167\,c will be interesting targets for upcoming telescopes such as the James Webb Space Telescope (JWST) and the Atmospheric Remote-Sensing Infrared Exoplanet Large-Survey (Ariel).

	\end{abstract}
	
	%%KEYWORDS
	
	%%INTRODUCTION
	
	\section{Introduction} 
	\label{sec_intro}
	%\begin{comment}
	%Since the first discovery of an extra-solar planet, 51 Pegasi b \citep{Mayor1995} in 1995,  the number of detected exoplanets has increased enormously.  Most of the exoplanets discovered early on, named `hot Jupiters', had masses and radii comparable to those of Jupiter and very short orbital periods (less than 10 days). 
	
	High precision photometry with the NASA's Kepler space mission revealed the existence of a large population of transiting planets with radii between those of the Earth and Neptune, and with period shorter than 100 days \citep[e.g.][]{Borucki2011,Batalha2013,Howard2012,Fressin2013,Dressing2013,Petigura2013}. 
	%Considering K2 data and accounting for planet detection efficiency, \citet[][]{Fulton2017} and \citet{FultonePetigura2018} found that
	%the radius distribution of small planets is bimodal with a paucity of planets with radii in the range of 1.5-2~R$_\oplus$. 
	Thanks to more precise measurements of the stellar radii of the Kepler field, first via spectroscopy \citep{Petigura2017, Fulton2017} and then via Gaia Data Release 2 data \citep{FultonePetigura2018}, it was then discovered that
		the radius distribution of small planets is bimodal with a paucity of planets with radii in the range of 1.5-2~R$_\oplus$.
	The right peak of this bimodal distribution (2-5~R$_\oplus$) is made up of sub-Neptune (2-4~R$_\oplus$), and Neptunes planets (R$\geq$4~R$_\oplus$). For this population of planets, a broad range of scenarios are possible, including water-worlds, rocky super-Earths and planets with H- and He- dominated atmospheres \citep[e.g.][]{Leger2004, Valencia2006,Rogers2010A,Rogers2010B,Rogers2011,Rogers2015, Zeng2019_waterworlds}. Atmospheric measurements are needed to understand their composition. To date, very few atmospheric studies concerning this class of planets have been conducted (see Table~\ref{table0}), but a larger number of observations will be necessary to put constraints on the planetary formation and migration theories and link the larger gas giants
	to the smaller terrestrial planets. 
	
	\begin{table*}[]
		\caption{Planets with size between 2-5~R$_\oplus$ with published atmospheric characterization studies. }
		\centering
		\begin{tabular}{c|c|l}
			\hline\hline
			Planet &  Chemical Species & Reference \\
			\hline
			%55\,Cnc\,e & HCN (perhaps) & \citet{Tsiaras2016b}\\
			%\hline
			GJ\,3470\,b & H$_2$O & \citet{Fisher2018,Benneke2019} \\
			& H & \citet{ Bourrier2018} \\
			\hline
			GJ\,436\,b & Flat spectrum (clouds or hazes) & \citet{knutsonGJ} \\
			& H &\citet{Bourrier2016}\\
			\hline
			GJ\,1214\,b & Flat spectrum (clouds or hazes) & \citet{kreidbergGJ} \\
			\hline
			HD\,97658\,b & Flat spectrum (clouds or hazes) & \citet{knutsonHD} \\
			\hline
			HAT-P-11\,b & He & \citet{Allart2018, Mansfield2018} \\
			& H$_2$O & \citet{Fraine2014,Fisher2018, Chachan_2019} \\
			& CH$_4$ (maybe) & \citet{Chachan_2019}\\
			\hline
			%HAT-P-26 b & H$_2$O & \citet{MacDonald_2019,Wakeford2017} \\
			%& metal hydrides & \citet{MacDonald_2019} \\
			%\hline
			K2-18\,b & H$_2$O & \citet{TsiarasK218b,benneke2019water}\\
			\hline
			
			%Wasp-69\,b & Na %&\citet{Casasayas-Barris2017}\\
			%& He & \citet{Nortmann2018}\\
			%& H$_2$O &\citet{Tsiaras2018}\\
			
			%\hline
			
		\end{tabular}
		\label{table0}
	\end{table*}
	\par
	An element of comparison, which allows us to better understand the atmospheric physics of sub-Neptune and Neptune-type exoplanets, can be found in our Solar System, more precisely in Uranus and Neptune. These ice giants  can be used as (cold) template for listing the physical phenomena present in this class of planets, and a good understanding of them would give access to more accurate extrapolations for different temperatures of the planets.
	One element to emphasis is the large differences in atmospheric composition between Uranus and Neptune, reviewed in \cite{Moses2020}. The observability of chemical compounds is defined by equilibrium chemistry in the hot interior, modified in the upper atmosphere by transport-induced quenching as well as photochemistry. The dynamic activity of the planet (modelized by an eddy diffusion coefficient for simplified mixing calculations) can therefore have a direct effect on the observable composition. Such effects could have to be considered for this class of planets, especially for warm sub-Neptune and Neptune-type planets.\\
	In this paper we analyse the transmission spectra of the Neptune-type HD\,106315\,c and of the sub-Neptune HD\,3167\,c, using publicly available observations from the Hubble Space Telescope (HST) Wide Field Camera 3 (WFC3) operating in its spatial scanning mode.
	\begin{table}[ht]
		\caption{Stellar and planetary parameters used in our analysis.}             
		\label{table1}     
		\centering 
		\footnotesize
		%\resizebox{0.45\textwidth}{!}{%
		\begin{tabular}{l | c c }          
			\hline\hline                       
			Parameters & HD\,106315\,c &  HD\,3167\,c \\    
			\hline
			\multicolumn{3}{l}{\textbf{Stellar parameters}}\\
			\hline
			Stellar type & F5V\footnote{\citet{Houk1999}.} & K0V \\
			$[$Fe/H$]_{\star}$ & $-0.276 \pm0.083$ & $0.03\pm0.03$ \\
			T$_{\rm eff}$ [K] & $6256\pm 51$  & $5286\pm40$   \\     
			log$_{10}$ $g_{\star}$ [cgs] & $4.235\pm0.030$  & $4.53 \pm0.03 $ \\
			R$_{\star}$ [R$_{\odot}$] & $1.31\pm 0.04$ & $0.835\pm 0.026$\\
			M$_{\star}$[M$_{\odot}$] & 1.079$\pm$0.037 & 0.877$\pm$0.024\\
			\hline
			\multicolumn{3}{l}{\textbf{Planetary and transit parameters}} \\
			\hline
			M${\rm_P}$ [M$_{\oplus}$] & $14.6\pm4.7$ & $8.33^{+1.79}_{-1.85}$ \\
			R$_{\rm P}$/R$_\star [\%] $ &  $3.481\pm0.099$ & $3.006_{-0.055}^{+0.065}$  \\
			R$_{\rm P}$[R$_{\oplus}$] & $4.98\pm0.23$ & 2.740$_{-0.100}^{+0.106}$\\
			P [days] & $21.05731\pm0.00046$ & $29.84622_{-0.00091}^{+0.00098}$ \\
			i [deg] & $88.17\pm0.11$  & $89.6\pm0.2$ \\
			a/R$_\star$  & $25.10\pm0.79$  & $46.5\pm1.5$  \\
			T$_{\rm 0}$ [BJD$_{\rm TDB}]$ & $2457569.0211\pm0.0053$& $2457394.97831\pm0.00085$\\
			e & 0.052 $\pm$ 0.052 & 0.05$_{-0.04}^{+0.07}$\\
			$\omega$ & 157 $\pm$ 140& 178$
			_{-136}^{+134}$\\
			\hline  
			Reference &  This work, \S~\ref{Sect2} & \citet{Gandolfi2017} \\ 
			\hline
		\end{tabular}%}
		%\tablefoot{
		%\tablefoottext{a}{}
		%  }
	\end{table}	
	
	The first small warm planet we studied in this paper is HD\,106315\,c. With a mass of 14.6$\pm$4.7~M$_{\oplus}$, a radius of 4.98$\pm$0.23~R$_{\oplus}$, and a density of 0.65$\pm$0.23~g~cm$^{-3}$, it orbits its F5V host star with a period of 21.05731$\pm$0.00046~day (this work, Table~\ref{table1}). Its equilibrium temperature, computed by assuming an albedo of 0.2 \citep{CrossfieldTrends}, is 835$\pm$20~K. The planet has a inner-smaller companion HD\,106315\,b (R${\rm_P}$=2.18$\pm$0.33~R$_{\oplus}$, this work). The discovery of this multi-planetary system was simultaneously announced by \citet{Crossfield2017} and \citet{Rodriguez2017} using data from the K2 mission.
	Due to the paucity of radial velocities measurements, %Since HD\,106315 is a fast rotator ($v \, \sin i= 13.2 \pm 1.0$~km s$^{-1}$, \citealt{Crossfield2017}) 
	both teams were not able to derive a precise measurement of the planetary mass, and only the High Accuracy Radial velocity Planet Searcher (HARPS) radial velocity observations by \citet{Barros2017} allowed a mass estimation. 
	More recently, \citet{Zhou2018} reported also an obliquity measurement ($\lambda=-10\buildrel{\circ}\over{.} {9}_{-3.8}^{+3.6}$) for HD\,106315\,c from Doppler tomographic observations gathered with the Magellan Inamori Kyocera Echelle (MIKE), HARPS, and Tillinghast Reflector Echelle Spectrograph (TRES). Given the brightness of the host star (V=8.951$\pm$0.018~mag, \citealt{Crossfield2017}), the atmospheric scale height (H$\sim$518 $\pm$ 174~km, calculated by assuming a primary mean molecular weight of 2.3~amu), and the contribution to the transit depth of 1 scale height (40 $\pm$ 14~ppm, calculated by using the relationship that the change in transit depth due to a molecular feature scales as $2\,H\,R_{\rm{p}}/R_\star^2$, \citealt{Brown2001}), HD\,106315\,c represents a golden target on which to perform transmission spectroscopy, and thus, to provide constraints on not only the planetary interior, but also the formation and evolution history. 
	%At the time of submission of this work, \citet{Kreidberg_2020}, presented a transmission spectrum of HD\,106315\,c from optical to infrared wavelength by combining different transits events of HD\,106315\,c gathered with HST/WFC3, K2 and Spitzer. Their analysis is based on the same HST/WFC3 data, analyzed independently. They interpreted the small amplitude feature (30~ppm) detected in their spectrum as due to a possible presence of water vapor in the planetary atmosphere. Based on the Bayes factor values obtained (of 1.7, 2.6 depending on prior assumptions), they referred to this detection as `tentative'.  %As a first step, due to inconsistencies among the planetary parameters published in the above-mentioned papers, we preferred to refine HD\,106315\,c's ephemeris (see \S~\ref{Sect2} and Table~\ref{table1}) before analysing the HST data.

	The other small-size planet we analyzed in this work is HD\,3167\,c. It was discovered orbiting its host star, together with an inner planet HD\,3167\,b (R${\rm_P}$=1.574 $\pm$ 0.054~R$_{\oplus}$), by \citet{Vanderburg2016}.  \citet{Gandolfi2017} and \citet{Christiansen2017} then revised the system parameters and determined radii and masses for the two exoplanets.
	HD\,3167\,c has a mass of M${\rm_P}$=$8.33^{+1.79}_{-1.85}$~M$_{\oplus}$, a radius of R${\rm_P}$=2.740$_{-0.100}^{+0.106}$~R$_{\oplus}$ \citep{Gandolfi2017}, and a temperature of T$_{\rm{eq}}$=518$\pm$12~K (assuming an albedo of 0.2). It orbits its K0V host star with a period of $29.84622_{-0.00091}^{+0.00098}$~days. Given a mean density of $\rho$=2.21$_{-0.53} ^{+0.56}$~g\,cm$^{-3}$, \citet{Gandolfi2017} quoted that HD\,3167\,c should have had a solid core surrounded by a thick atmosphere. The brightness of the host star (V=8.94$\pm$ 0.02~mag, \citet{Vanderburg2016}) combined with the atmospheric scale height (171 $\pm$ 40~km, calculated by assuming a primary mean molecular weight of 2.3~amu), and with the contribution to the transit depth of one scale height (18 $\pm$ 4~ppm, this work) make the planet a suitable target for atmospheric characterization.

	We used the publicy avaiable Python package \verb+Iraclis+ \citep{Iraclis} to analyze the raw HST/WFC3 images of the two warm small planets.
	%We used the publicly available Python package \verb+Iraclis+ \citep{Iraclis} to analyze the raw HST/WFC3 images of the two Neptune-like planets. 
	In \S~\ref{Sect2} we present the different steps we performed to obtain our 1D transmission spectra from the raw images. We then explain (\S~\ref{Sect3}) the modeling of the extracted spectra carried out by using the publicly available spectral retrieval algorithm TauREx3 \citep{Waldmann_2015,Waldmann_2015_2,al-refaie_taurex3}. In \S~\ref{discussion}, we discuss our findings underling possible limitations of our data-analysis and due to WFC3's narrow spectral coverage. We draw also some interpretations of the interior compositions of the two exoplanets, and we put our results in comparison with other low spectral resolution studies \citep[e.g. those arising from ARES, i.e.][]{Edwards2020, Skaf2020, Pluriel2020_ares}. We then simulate possible future studies with the upcoming space-borne instruments, such as the James Webb Space Telescope (JWST), and Ariel. Finally, we conclude (\S~\ref{Conclusion}) by highlighting the importance of future atmospheric characterisation both from the ground and from the space. %which leverages highly accurate line lists, ExoMol \citep{Tennyson2016}, HITEMP \citep{Rothman2010}, and HITRAN \citep{Rothman2013}, along with Bayesian analysis. \GG{to be completed..} 

	\section{Data analysis} \label{Sect2}
	%\subsection{HD\,106315\,c Orbital Ephemeris and transit parameters refinement}
	%\label{Santerne}
	From the comparison of the above-mentioned papers \citep{Crossfield2017,Rodriguez2017,Barros2017, Zhou2018} a discrepancy emerges in the light-curve parameters of HD\,106315\,c, and in particular in the value of the planetary radius (R$_{\rm{P}}$). On one hand, the photometric studies by \citet{Crossfield2017,Rodriguez2017}, and \citet{Barros2017} seem to converge toward a lower planetary radius ($\sim$4~R$_{\oplus}$), but with big error bars (this is probably a consequence of having a light curve with a high impact parameter). More precisely, \citet{Crossfield2017} measured a planetary radius of 3.95$^{+0.42}_{-0.39}$~R$_{\oplus}$, \citet{Rodriguez2017} of 4.40$^{+0.25}_{-0.27}$~R$_{\oplus}$, and \citet{Barros2017} of 4.35$\pm$0.23~R$_{\oplus}$. On the other hand, the independent spectroscopic analysis by \citet{Zhou2018} resulted in a higher R$_{\rm{P}}$ value with smaller uncertainties (i.e. R$_{\rm{P}}$=4.786$\pm$0.090~R$_{\oplus}$).
	To overcome these inconsistencies, before looking at the HD\,106315\,c's HST/WFC3 data, we decided to perform a combined analysis, using both spectroscopic and photometric observations. 
	%The absence of mass measurements in both \citet{Crossfield2017} and \citet{Rodriguez2017}, is problematic for the interpretation of the transmission spectrum of HD\,106315\,c, and so for the TauREx analysis. For this reason, as a first step, we decided to use the planetary and stellar parameters from \citet{Barros2017}. However, we were not able to obtain a good fit of our HST data (see Figure~\ref{Barros_fig}), and this was probably due to some differences between the findings of \citet{Barros2017}, and those of the aforementioned previous papers (e.g. the eccentricity value found by \citealt{Barros2017} is greater than that found by \citealt{Rodriguez2017}). For this reason, we decide to refine HD\,106315\,c's ephemeris and transit parameters.
	More precisely, we included in our analysis ESO/HARPS radial velocities \citep{Barros2017}, space-based K2 data and three ground-based transits, namely one observation gathered with the Las Cumbres Observatory (LCO) telescopes \citep{Barros2017} and two with the EULER telescope \citep{Lendl2017}. We modeled these data by employing the Markov
	chain Monte Carlo Bayesian Planet Analysis and Small Transit Investigation Software (PASTIS) code \citep{Pastis2014} as done in \citet{Barros2017}. 
	The improved system's parameters are listed in Table~\ref{table1}. In particular, if we compare our results to the previous papers, trying to break the above-mentioned inconsistency on the R$_{\rm{P}}$ value, we note that our planetary radius is in agreement with that found by the spectroscopic analysis of \citet{Zhou2018}.
	\vspace{1cm}\\
	Our analysis is based on four %three\footnote{Due to the lack of a sufficient number of points in the initial part of the transit, and to the presence of huge systematics that could not be fitted with the model employed for the other visits, we were unable to obtain a usable transmission spectrum for the last HST visit of HD\,106315\,c taken on November 23$^{\rm rd}$ 2019. For this reason, in the rest of the paper we consider only three transits of HD\,106315\,c. \label{footnote_1}}
	and five transit observations of HD\,106315\,c and HD\,3167\,c, respectively (Table~\ref{table2}). Both were obtained with the G141 infrared grism (1.125 - 1.650 $\mu$m) of the HST/WFC3. 
	The observations were part of the HST proposal GO 15333 (PI: Ian Crossfield) and were downloaded from the public Mikulski Archive for Space Telescopes (MAST) archive. An independent analysis of the same dataset for HD\,106315\,c and HD\,3167\,c, with different pipelines, is presented by \citet{Kreidberg_2020} and \citet{Evans2020}, respectively.
	We analyzed and extracted white and spectral light-curves from the raw HST/WFC3 images using \verb+Iraclis+ \citep{Iraclis}.
	This tool includes multiple different steps: 
	\begin{itemize} \itemsep0em 
		\item [$\ast$] Data reduction and calibration (\S~\ref{Sect2.1})
		\item [$\ast$] Light-curves extraction (\S~\ref{Sect2.2})
		\item [$\ast$] Limb-darkening coefficients calculation (\S~\ref{Sect2.3})
		\item [$\ast$] White light-curves fitting (\S~\ref{Sect2.4})
		\item [$\ast$] Spectral light-curves fitting (\S~\ref{Sect2.5})
	\end{itemize}
	Each transit was observed over six and seven HST orbits for HD 106315 c and HD 3167 c, respectively. We used both forward (increasing row number) and reverse (decreasing row number) scanning.

	\begin{center}
		\begin{table}
			\caption{Proposal information for the data used in our analysis}
			\label{table2}
			\footnotesize
			\centering
			%\resizebox{0.45\textwidth}{!}{%
			%\resizebox{\columnwidth}{!}{
			\begin{tabular} {c | c c | c c}
				\hline
				\hline
				Planet & Proposal ID & Proposal PI &  \multicolumn1{p{10mm}}{\centering{Transits used}} & \multicolumn1{p{14mm}}{\centering{HST orbit used}} \\ 
				\hline
				HD\,106315\,c & 15333 & Crossfield I. &4 &20 \\%3\footnote{\mbox{See note \ref{footnote_1}}} & 15 \\
				HD\,3167\,c & 15333 & Crossfield I. &5 &  28 \\
				\hline
			\end{tabular}%}
		\end{table}     
	\end{center}
	
	\subsection{Data reduction and calibration}\label{Sect2.1}
	
	The first step of the \verb+Iraclis+ pipeline is the reduction and calibration of the HST/WFC3 raw images. This part of the analysis consists of several operations: zero-read subtraction, reference pixels correction, non-linearity correction, dark current subtraction, gain conversion, sky background subtraction, flat-field correction, bad
	pixels/cosmic rays correction and wavelength calibration \citep{Tsiaras2016,Tsiaras2016b,Tsiaras2018}.
	
	\subsection{Light-curve extraction}
	\label{Sect2.2}
	After the reduction and calibration of the raw images, we extracted the wavelength-dependent light-curves. In performing this operation the geometric distortions caused by the tilted detector of the WFC3/IR channel are taken into account, as explained in \citet{Tsiaras2016}. \\
	Two kinds of light-curve were extracted:
	\begin{itemize}
		\item [$\ast$] a $white$ $light$-$curve$: calculated from a broad wavelength band ($1.088$ -- $1.68\;\mu$m) covering the whole wavelength range of WFC3/G141,
		\item [$\ast$] a set of $spectral$ $light$-$curves$: extracted using a narrow band with a resolving power at $1.4\;\upmu$m of 70. The bins were selected such that the signal to noise is approximately uniform across the planetary spectrum. We ended up with 25 bands, with bin-widths in the range 188.0-283.0~nm. 
	\end{itemize}

	\subsection{Limb darkening coefficients}
	\label{Sect2.3}
	The stellar limb darkening effect is modelled using the non-linear formula with four terms from \citet{2000A&A...363.1081C}. The coefficients are calculated by fitting the stellar profile from an ATLAS model \citep{1970SAOSR.309.....K,2011MNRAS.413.1515H} and by using the stellar parameters presented in Table~\ref{table1}. Table~\ref{table3} shows the limb-darkening coefficients calculated for the white light-curve (between 1.125 - 1.650 $\mu$m).

	\begin{table*} [htbp]
		\caption{White light-curve fitting results for HD\,106315\,c and HD\,106315\,c.}
		\footnotesize
		\begin{tabular}{c| c c c c c c c c c}
			\hline \hline
			Planet & Visit & T$_0$ (HJD$\_$UTC) & (R$_{\rm P}$/R$_\star)^2$ ($\%$) & \multicolumn{4}{c}{Limb darkening coefficient } & $n^{\mathrm{for}}_W$ & $n^{\mathrm{rev}}_W$\\
			& & &  & a$_1$ & a$_2$ &  a$_3$ &  a$_4$\\
			\hline
			HD\,106315\,c & 1 & ${2458453.3973_{-0.0002}^{+0.0003}}$ & $0.113_{-0.002}^{+0.002}$& \multirow{4}{*}{0.8} &  \multirow{4}{*}{-0.8} & \multirow{4}{*}{0.9} & \multirow{4}{*}{-0.4} &1341046587$_{-20470}^{+30705}$ & 1340876749$_{-24006}^{+ 27435}$\\
			& 2 & $2458474.4537 _{-0.0003}^{+0.0003}$  & $0.104 _{-0.002}^{+0.003}$ & & & &  &1340747906$_{-21939}^{+32909}$ & 1340594761$_{-21530}^{+32295}$ \\
			& 3 & ${2458516.5668_{-0.0003}^{+0.0003}}$  & $0.108_{-0.003}^{+0.003}$ & & & & & 1341132987$_{-35040}^{+52560}$ & 1341001153$_{-35449}^{+53174}$ \\
			& 4 & $2458811.3661_{-0.0022}^{+0.0007}$ &0.105$^{+0.003}_{-0.003}$ & & & & & 1340514245$^{+38572}_{-44082}$ & 1340404691$^{+38600}_{-44115}$\\
			\hline
			HD\,3167\,c &	1 & $2458260.52574^{+0.00016}_{-0.00014} $ &  $0.092^{+0.002}_{-0.002}$& \multirow{4}{*}{0.9} &  \multirow{4}{*}{-0.8} & \multirow{4}{*}{0.9} & \multirow{4}{*}{-0.4} & 1204464323$^{+31822}_{-27844}$
			& 1204408711$^{+31047}_{-27166}$\\
			&2 & $2458320.2132^{+0.0018}_{-0.0016}$  &  $0.094^{+0.002}_{-0.002}$ & & & &  & 1204727344$^{+37561}_{-32866}$
			& 1204655124$^{+37385}_{-32711}$\\
			&3 & $2458648.52966^{+0.00017}_{-0.00019}$  &  $0.085^{+0.003}_{-0.002}$ & & & &  & 1204169898$^{+46776}_{-40929}$
			&  1204128316$^{+41679}_{-41679}$\\
			&4 & $2458708.220^{+0.005}_{-0.003}$  &  $0.095^{+0.003}_{-0.003}$ & & & &  & 1204871733$^{+60247}_{-40165}$
			& 1204812454$^{+53852}_{-47121}$\\
			& 5 & 2459036.5327$^{+0.0019}_{-0.0022}$ & 0.095$^{+0.001}_{-0.001}$ & & & & &  1204456150$^{+21950}_{-19206}$
			& 1204407529$^{+21612}_{-18910}$ \\
			\hline
		\end{tabular}
		\label{table3}
	\end{table*}
	
	\begin{figure*}[htpb]
		\centering
		\resizebox{\textwidth}{!}{
			\begin{subfigure}[]{}	
				\includegraphics[width=7cm]{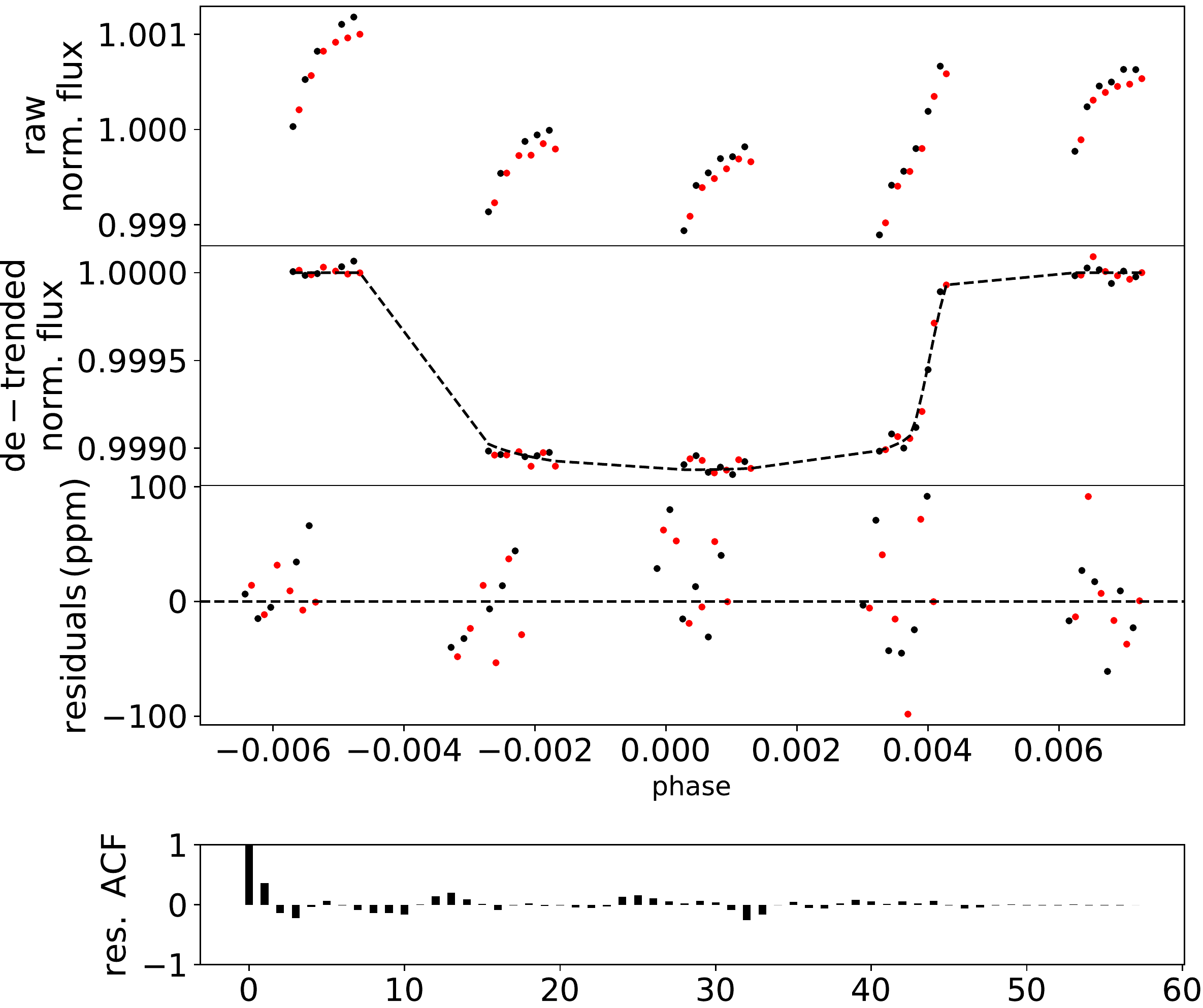}
			\end{subfigure}%
			~
			\begin{subfigure}[]{}	
				\includegraphics[width =7cm]{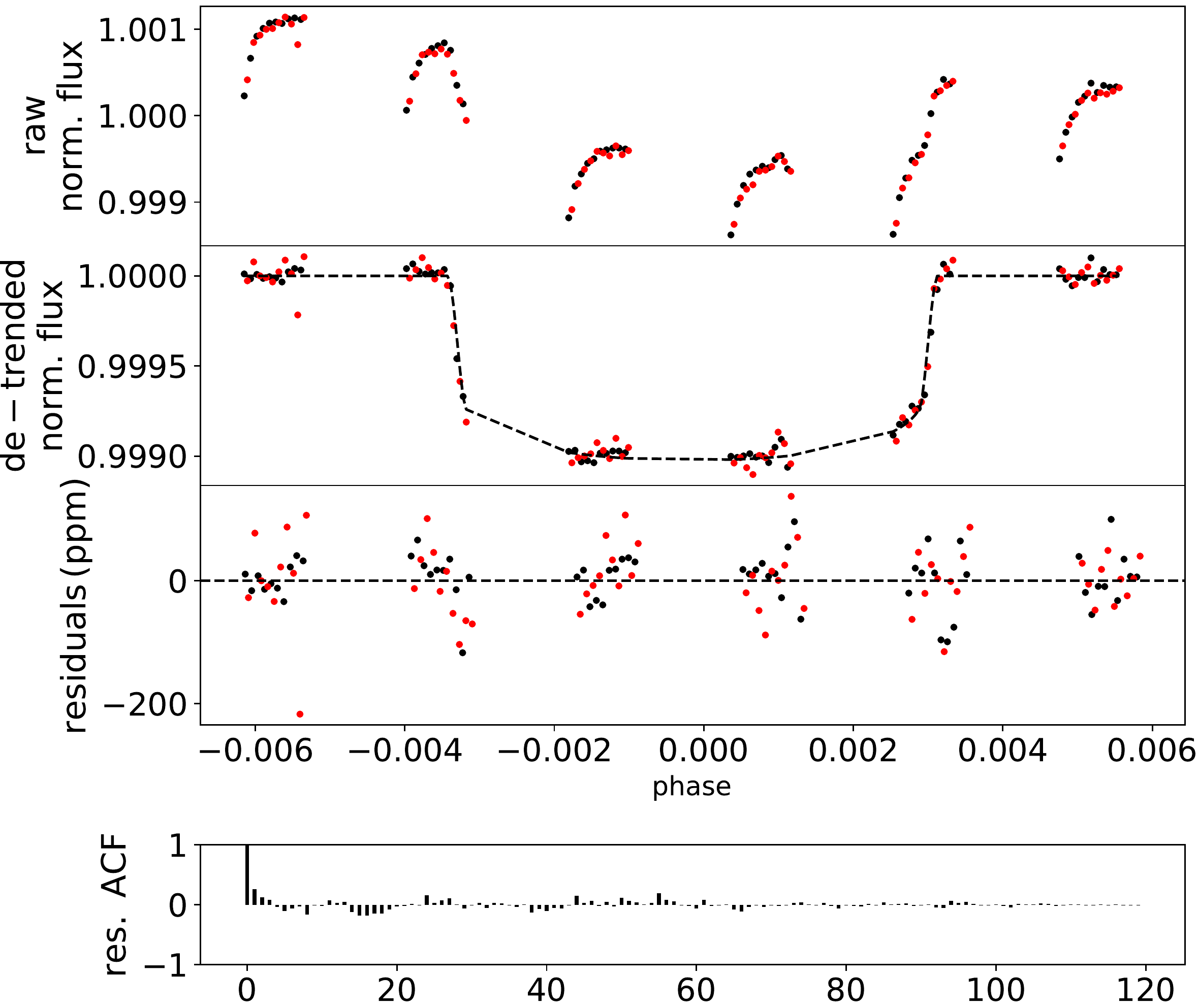}
		\end{subfigure}}
		
		\caption{Results of the white light-curve analysis for the first considered transit of HD\,106315\,c (a) and the first considered transit of HD\,3167\,c (b). Top panel: normalized raw light-curve for the forward (black) and reverse (red) scans. Second panel: light-curves divided by the best-fit systematic effects model. Third panel: Fitting residuals. Bottom panel: auto-correlation function of residuals. } 
		\label{fig2}
	\end{figure*}
	
	\subsection{White light-curves fitting}
	\label{Sect2.4}
	The products of the previous steps are the white and spectral light-curves. To continue our characterization of the two exoplanet atmospheres we then created transmission spectra which were obtained by fitting the light-curves with a transit model. 
	However, before fitting the extracted white and spectral light-curves, we had to consider the time-dependent systematics introduced by HST: one long-term `ramp' (which affects all the visits) with a linear (and, in some cases, a quadratic) trend and one short-term `ramp' (which affects every HST orbit) with an exponential trend. %Figure~\ref{fig1} shows these systematics in the raw white light-curve but they are also present in all the spectral light-curves.
	
	In order to remove all these systematics, we fitted the white light-curves using the transit python package $PyLightcurve$,  i.e. we used a transit model multiplied by a model for the systematics \citep{Tsiaras2016, Tsiaras2018}:
	\begin{equation}
	\resizebox{.9\linewidth}{!}{$\rm{n}^{\mathrm{scan}}_W \left[ 1- \rm{r}_{a1} \rm{(t-  T_0)+r_{a2}(t-  T_0)^2}\right]\cdot \left[ 1-r_{b_1} e^{-r_{b_2}(t-t_0)} \right]$}
	\label{eq1}
	\end{equation}
	where $t$ is time, $T_0$ is the mid-transit time, $t_0$ is the starting time of each HST orbit, $r_{a1}$ and $r_{a2}$ are the linear and quadratic systematic trend's slope, $r_{b_1}$ and $r_{b_2}$ are the exponential systematic trend's coefficients, and $n^{\mathrm{scan}}_W$ is a normalisation factor that changes for forward scanning ($n^{\mathrm{for}}_W$), and for reverse scanning ($n^{\mathrm{rev}}_W$).
	%For HD\,106315\,c's visits we fitted the white light-curves for both visit-long linear and orbit-long exponential ramps. 
	Second order (quadratic) visit-long ramps were also fitted for HD\,3167\,c visits because they were more affected by systematics. 
	%\textcolor{red}{G. In two Angelos paper there is written a sentence like this: 'After an initial fit, we scaled-up the uncertainties on the individual data points, in order for their median to match the standard deviation of the residuals, and fitted again', but are we doing this re-fitting operation? Is an automatic operation?}\\
	The parameter space was sampled via emcee \citep{emcee}. We used 300000 emcee iterations, 200 walkers, and 100000 burned iterations. We employed this set up for all the visits for both the planets. The only exception  is represented by the fourth visit of HD\,106315\,c were we had to use 200000 iterations to obtain a good fit to our data.\\
	Figure~\ref{fig2} shows the light-curves for the first transits of both exoplanets divided by the best-fit systematic model. (The same plots for the other transits are shown in appendix in Figure~\ref{figA1}). In the fit we took T$_0$ and R$_{\rm {P}}$/R$_\star$ as free parameters, and we used fixed values for $P$, $\omega$, $i$, $a/R_{\star}$, and $e$ parameters, as
	reported in Table~\ref{table1}. We made this choice because we miss ingress/egress observations in some visits.
	%In Figure~\ref{fig2}, only 5 and 6 orbits are plotted for HD\,106315\,c and HD\,3167\,c, respectively, instead of 6 and 7 (Figure~\ref{fig1}).
	For both the planets we decided to eliminate data gathered during the first HST orbit and the first two points of each orbit because of the stronger systematics that affect them. %These effects originate from different sources than those in successive orbits and therefore cannot be corrected with this method. 
	An incorrect fitting of the behavior of the instrument at this stage would have introduced additional uncertainties in the final values of the transit parameters. 
	Processing visits~3 and~4 for HD\,3167\,c required additional steps; this was on account of poor initial fitting due to HD\,3167\,b also transiting the stellar disk during these observations. Strong auto-correlation in the fit residuals for visits~3 and~4 led to an investigation of the orbits for both the transiting planets in the HD 3167 system: b and c. Theoretical transit light curves were plotted for all four HD 3167 observation windows, again using $PyLightcurve$ and taking parameters for both planets from \citet{Gandolfi2017}. The theoretical light curves showed no overlap between transits for the first two visits but contamination of the third and fourth visits by concurrent transits of HD 3167 b. In both cases this effect was limited to a single HST orbit in each affected visit. These two orbits were then disregarded, leaving six orbits for each of visits~1,~2, and~5 five orbits apiece for vists~3 and~4. These affected orbits can be seen in the appendix in figure \ref{figA2}.  
	The final fitting results and their uncertainties can be found in Table~\ref{table3}. 
	\begin{figure*}[htpb]
		\centering
		\resizebox{0.9\textwidth}{!}{
			\begin{subfigure}[]{}	
				\includegraphics[width =8cm]{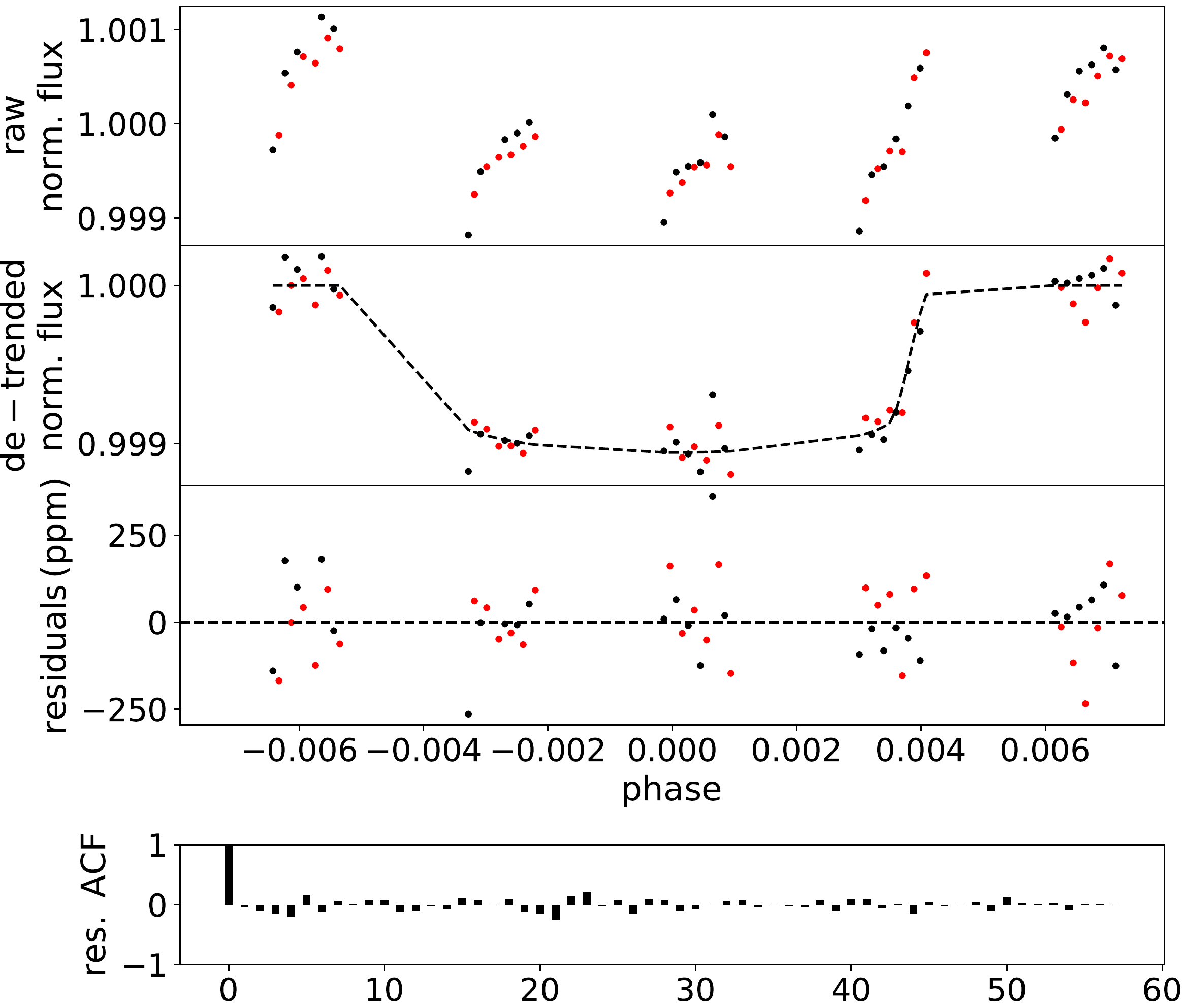}
				
			\end{subfigure} %
			\begin{subfigure}[]{}
				\includegraphics[width =8cm]{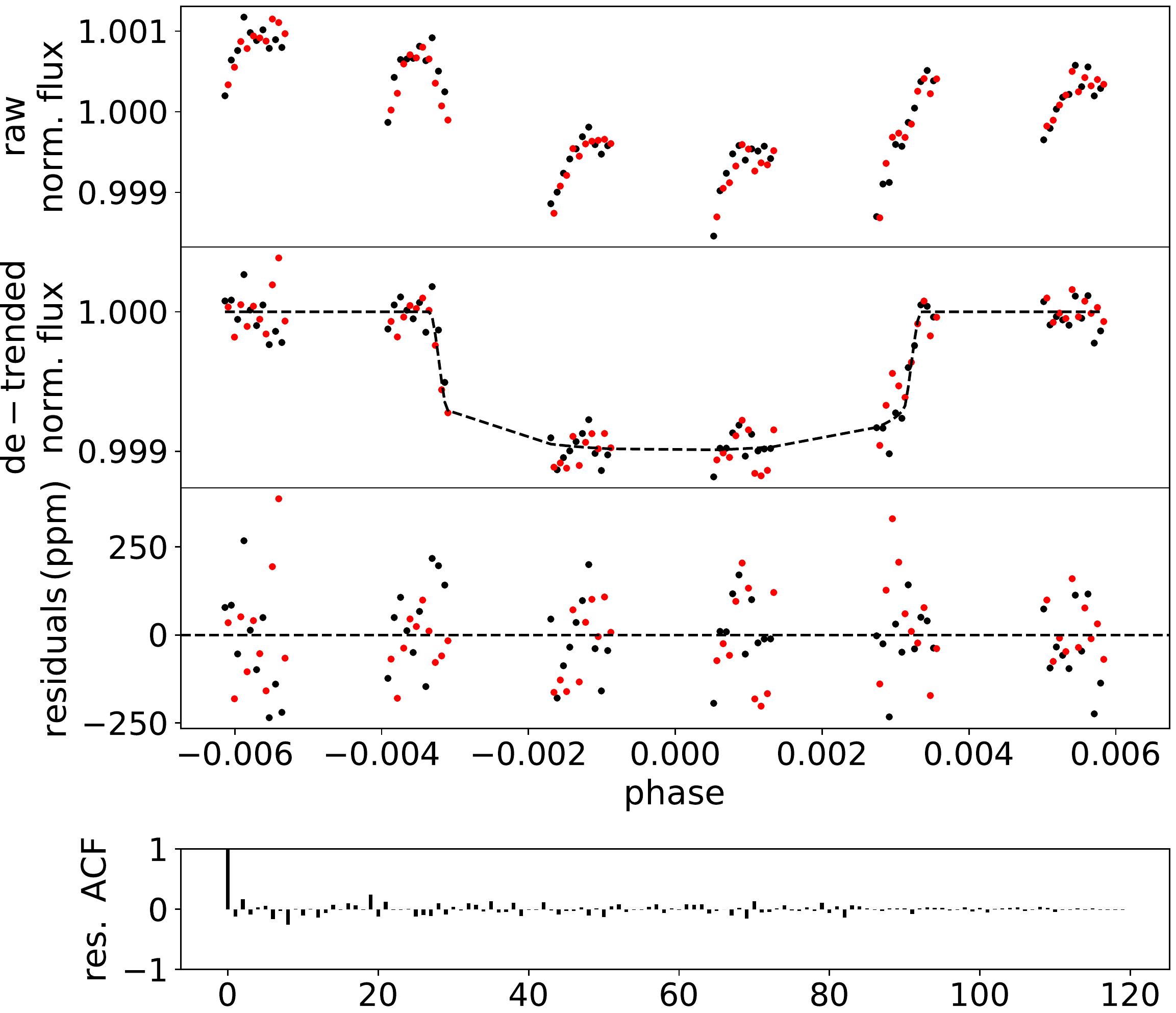}
		\end{subfigure}} 
		\caption{10th bin ($\lambda \sim 1.3\;\upmu$m) spectral light-curve for the first transit of both HD\,106315\,c (a) and of HD\,3167\,c (b). } 
		\label{fig3}
	\end{figure*}
	\subsection{Spectral light-curves fitting}
	\label{Sect2.5}
\begin{figure*}
		\centering
		\begin{subfigure}[]{}
			\includegraphics[width=0.65\textwidth] {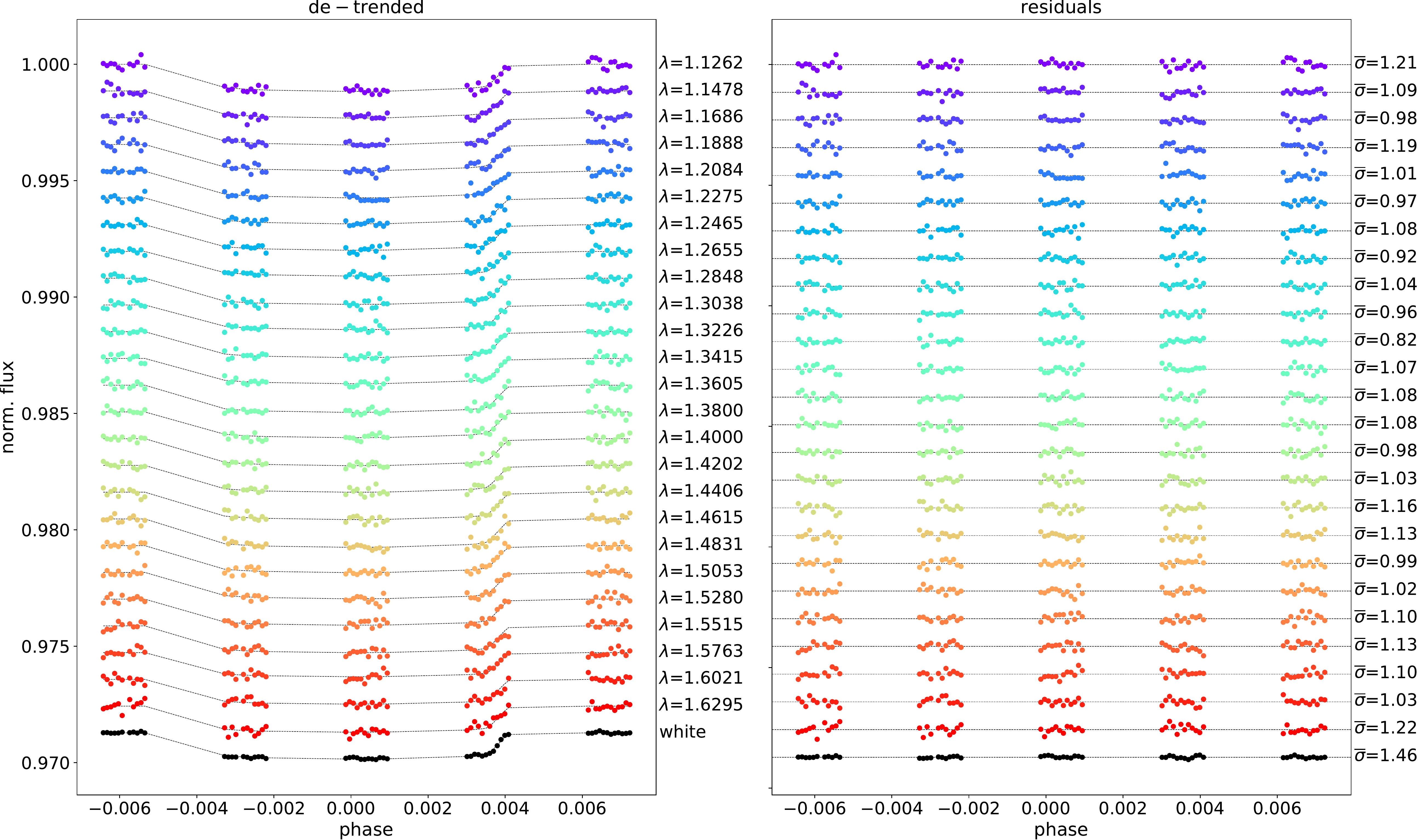} 
			
		\end{subfigure}
		\begin{subfigure}[]{}
			\includegraphics[width=0.65\textwidth] {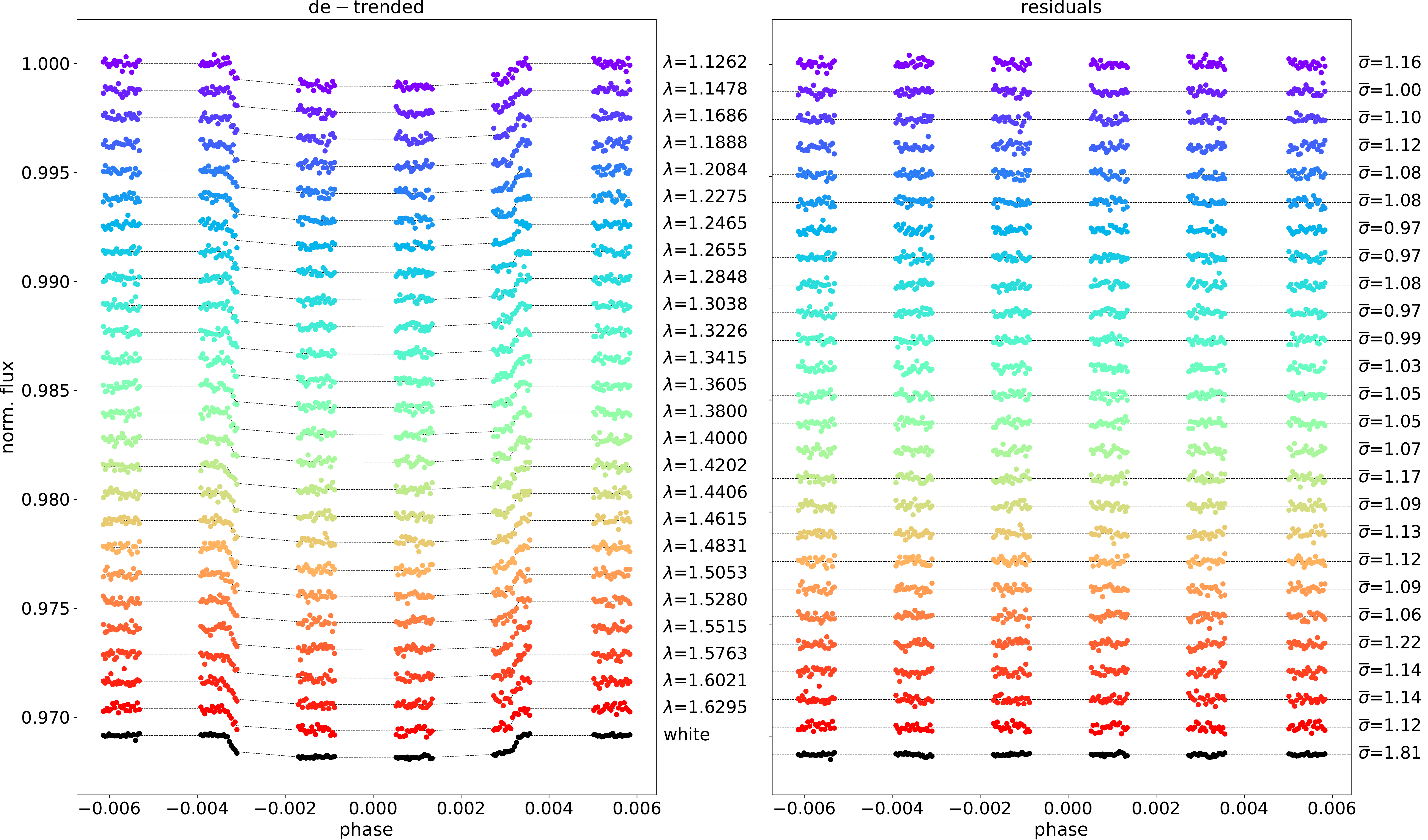} 
		\end{subfigure}
		\caption{Analysis of the HD\,106315\,c (a) and HD\,3167\,c (b) white and spectral light-curves (left panels), for the first transit, plotted with an offset for clarity. Left panels: over-plotted white (black points) and spectral (coloured points) light-curves. Right panels: over-plotted residuals, $\bar{\sigma}$ indicates the ratio between the standard deviation of the residuals and the photon noise. The reason for some $\bar{\sigma}$ fall below 1 because of the small number of datapoints. Hence the measured standard deviation is not always representative of standard deviation of the underlaying distribution. These values are displayed as an indication for the goodness of fit, highlighting the differences between the different wavelengths and most importantly differences between the white light curve and spectral light curves.}
		\label{fig4} % it is fig 3, but label are consistents
	\end{figure*}

\begin{figure}
		\begin{subfigure}[]{}
			\includegraphics[width=\linewidth]{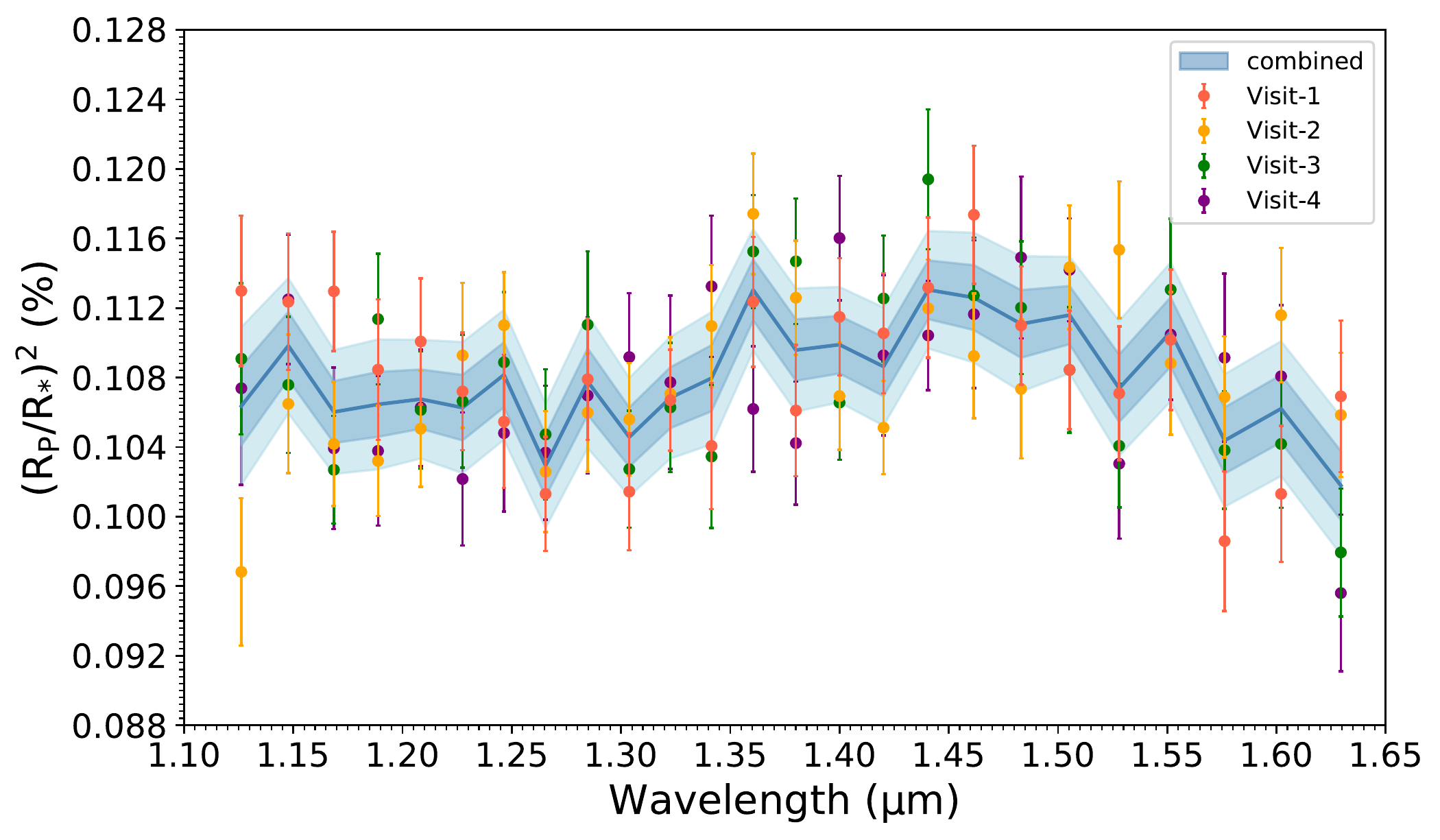}
		\end{subfigure}
		\begin{subfigure}[]{}
			\includegraphics[width=\linewidth]{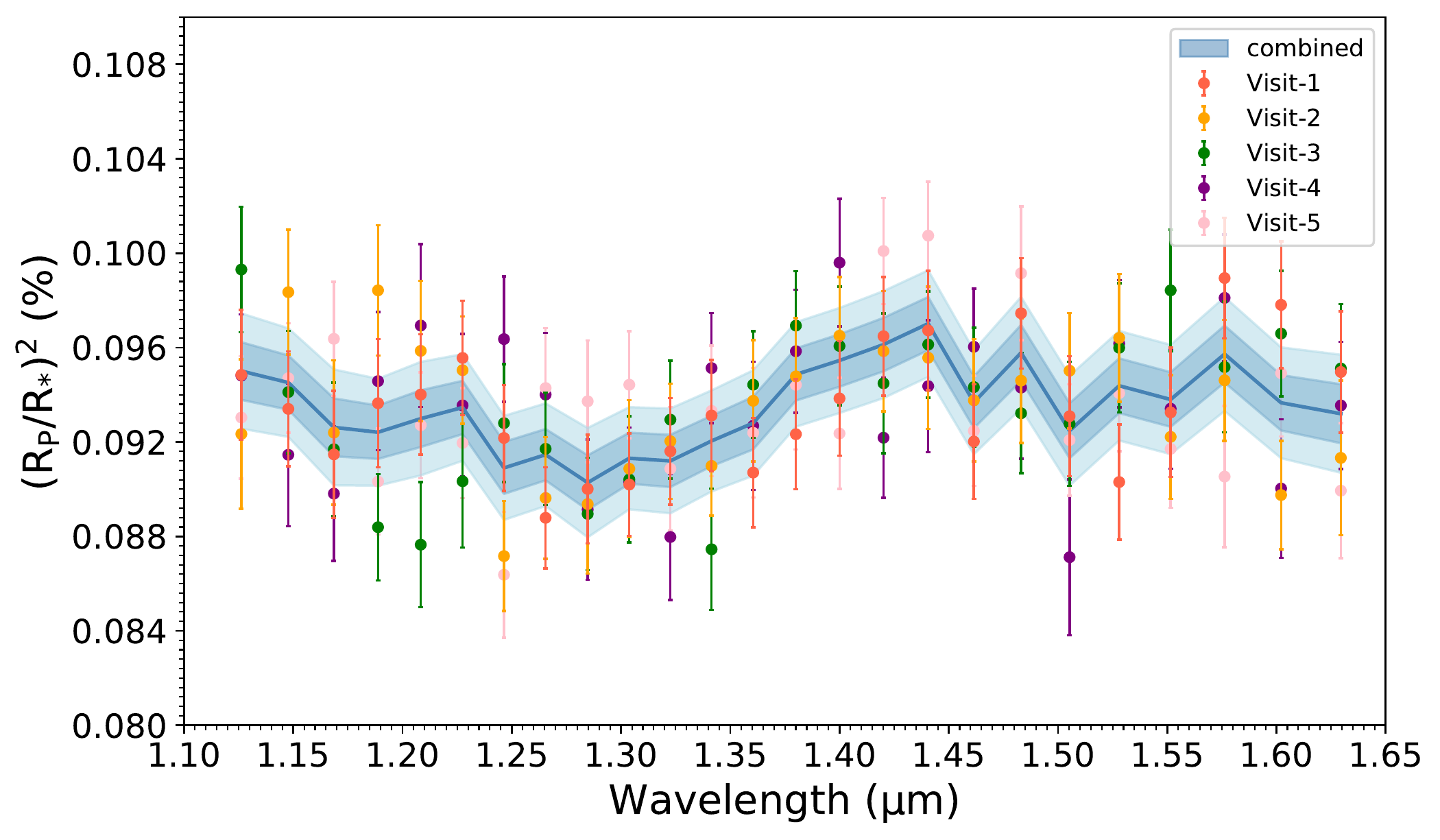}
		\end{subfigure}
		\caption{Spectra per visit and final weighted average with 1$\sigma$ and 2$\sigma$ uncertainty ranges for HD\,106315\,c (a) and HD\,3167\,c (b). %First, we subtracted each spectrum by the value of the white light-curve depth and then we  added the weighted average of all white light-curve values.
		}
		\label{fig5} %actually it is fig 4...
	\end{figure}

	In order to correct for the systematics present in the spectral light-curves, we used the $divide$ $white$ $method$ introduced by \citet{KreidbergB2014b}, i.e. each spectral light-curve was fitted with a model that includes the white light curve and its best-fit model:
	\begin{equation}
	\rm{n^{\mathrm{scan}}_\lambda \left[ 1-r_a (t-T_0) \right] \frac{\mathrm{LC}_W}{\mathrm{M}_W}}
	\end{equation}
	
	where  $r_a$ is the coefficient of a wavelength-dependent linear slope along each HST visit, ${LC}_W$ is the white light-curve, ${M}_W$ is  best fitting model to the white light-curve, $n^{\mathrm{scan}}_\lambda$ is the normalisation factor we used (it changes to $n^{\mathrm{for}}_\lambda$, when the scanning direction is upwards, and to $n^{\mathrm{rev}}_\lambda$ when it is downwards). As Table~\ref{table3} shows we obtained big numbers for these normalization factors, these is due because the light curve are in units of electrons, thus the large values are reasonable. In the spectral light-curve fitting, the only free parameter is R$_{\rm {P}}$/R$_\star$, while the other parameters are the same as we used for the white light-curve fitting.
	Using the white light-curve as a comparison has the advantage that the residuals from fitting one of the spectral light-curves (see Figure~\ref{fig3}) do not show trends similar to those in the white light-curve (see Figure~\ref{fig2}). All the spectral and white light-curves we obtained, for the first HST visit of each planet, are plotted in Figure~\ref{fig4}.
	As for the white light-curves fitting, the parameters space was sampled by using the emcee method. In this case we used 50000 emcee iterations, 100 walkers and  20000 burned iterations.\\
	Starting from the spectral light-curves, the final spectra were extracted and combined from the spectral light-curves by computing the average of the transit spectra weighted by their respective uncertainties. First we subtracted each spectrum by the corresponding white light-curve depth, and then we computed the weighted average of all the transit observations. Finally, we added the weighted average of all white light-curves values to the averaged spectrum. The white light transit depths were consistent between transits, except for visit 3 for HD\,3167\,c (0.0291$\pm$0.0005 compared to the weighted mean 0.03058$\pm$0.00015). This is probably due to remaining systematics or to stellar activity. We obtained a final spectrum with an increased S/N ratio (Table~\ref{table4}, and Figure~\ref{fig5}) which we then used for atmospheric retrieval.

	\begin{table}[htb]%\[!htb\]
		\caption{Transit depth (R$_{\rm P}$/R$_\star)^2$ for the different wavelength channels, where R$_{\rm P}$ is the
			planetary radius, R$_\star$ is the stellar radius, and $\lambda$, is the center value of each wavelength channel.} 
		\label{table4}  
		\begin{subtable}
			\raggedright
			\footnotesize
			\begin{tabular}[t]{c|c}
				\hline\hline                       
				\multicolumn{2}{c}{HD\,106315\,c} \\
				\hline
				$\lambda$ & (R$_{\rm P}$/R$_\star)^2$ \\
				$\mu$m  &   \%  \\
				\hline
				1.1263&  0.1064$\pm$0.0027\\
				1.1478&  0.1098$\pm$0.0019\\
				1.1686&  0.1060$\pm$0.0018\\
				1.1888&  0.1065$\pm$0.0019\\
				1.2084&  0.1068$\pm$0.0017\\
				1.2275&  0.1063$\pm$0.0019\\
				1.2465&  0.1082$\pm$0.0019\\
				1.2655&  0.1029$\pm$0.0018\\
				1.2848&  0.1078$\pm$0.0019\\
				1.3038&  0.1046$\pm$0.0017\\
				1.3226&  0.1068$\pm$0.0018\\
				1.3415&  0.1080$\pm$0.0019\\
				1.3605&  0.1130$\pm$0.0018\\
				1.3801&  0.1096$\pm$0.0018\\
				1.4000&  0.1099$\pm$0.0017\\
				1.4202&  0.1086$\pm$0.0017\\
				1.4406&  0.1130$\pm$0.0017\\
				1.4615&  0.1126$\pm$0.0019\\
				1.4831&  0.1111$\pm$0.0019\\
				1.5053&  0.1116$\pm$0.0017\\
				1.5280&  0.1074$\pm$0.0019\\
				1.5516&  0.1106$\pm$0.0020\\
				1.5762&  0.1044$\pm$0.0019\\
				1.6021&  0.1062$\pm$0.0019\\
				1.6295&  0.1018$\pm$0.0020\\
				\hline
			\end{tabular}
		\end{subtable}%
		\begin{subtable}
			\raggedleft
			\footnotesize
			\begin{tabular}[t]{c | c}
				\hline\hline                       
				\multicolumn{2}{c}{HD\,3167\,c} \\
				\hline
				$\lambda$ & (R$_{\rm P}$/R$_\star)^2$ \\
				$\mu$m  &   \%  \\
				\hline
				1.1263&  0.0950$\pm$0.0012\\
				1.1478&  0.0945$\pm$0.0012\\
				1.1686&  0.0926$\pm$0.0012\\
				1.1888&  0.0924$\pm$0.0011\\
				1.2084&  0.0930$\pm$0.0012\\
				1.2275&  0.0935$\pm$0.0011\\
				1.2465&  0.0909$\pm$0.0011\\
				1.2655&  0.0915$\pm$0.0011\\
				1.2848&  0.0903$\pm$0.0012\\
				1.3038&  0.0913$\pm$0.0011\\
				1.3226&  0.0912$\pm$0.0011\\
				1.3415&  0.0920$\pm$0.0011\\
				1.3605&  0.0928$\pm$0.0011\\
				1.3801&  0.0949$\pm$0.0011\\
				1.4000&  0.0955$\pm$0.0011\\
				1.4202&  0.0961$\pm$0.0011\\
				1.4406&  0.0970$\pm$0.0011\\
				1.4615&  0.0937$\pm$0.0011\\
				1.4831&  0.0958$\pm$0.0012\\
				1.5053&  0.0925$\pm$0.0012\\
				1.5280&  0.0944$\pm$0.0012\\
				1.5516&  0.0938$\pm$0.0012\\
				1.5762&  0.0957$\pm$0.0012\\
				1.6021&  0.0937$\pm$0.0012\\
				1.6295&  0.0932$\pm$0.0013\\
				\hline
			\end{tabular}
		\end{subtable}
	\end{table}

	\section{Atmospheric characterisation} \label{Sect3}
	\begin{table}
		\caption{Fit evaluation criteria and maximum a-posteriors retrieval results.}

		%\resizebox{0.45\textwidth}{!}{%
		\noindent\makebox[0.45\textwidth]{
			\footnotesize
			\begin{tabular}{l c c c}	\hline\hline \multicolumn1{p{20mm}}{\centering{Retrieved parameters}}&  bounds & \textbf{HD\,106315\,c}&\textbf{HD\,3167\,c}\\
				\hline 
				T$_{\rm P}$ (K) & $\pm$60$\%$ T$_{\rm eq}$ & 630$^{+ 326}_{- 115}$ & 440$^{+ 119 } _{ -79 }$\\
				R$_{\rm P}$ (R$_{\rm J}$)  & $\pm$50$\%$ R$_{\rm P}$   & 0.395$^{+ 0.009}_{-0.021}$ & 0.246$^{+ 0.002 } _{- 0.002 }$ \\ 
				log$_{10}$[H$_2$O]  & [-12 ; -1]& $-2.1^{+ 0.7}_{-1.3}$ & $-4.1^{+ 0.9 } _{ -0.9 }$\\
				log$_{10}$[NH$_{3}$]  & [-12 ; -1]& $-4.3^{+ 0.7}_{- 2.0}$ & $<-5$ \\
				log$_{10}$[CO$_{2}$]  & [-12 ; -1]& unconstrained& $-2.4^{+ 0.7 } _{ -1.0}$ \\
				log$_{10}$[CO]  &       [-12 ; -1] & unconstrained & unconstrained  \\
				log$_{10}$[CH$_{4}$]  & [-12 ; -1] &$<-5$  &  $<-5$ \\
				log$_{10}$[P$_{\rm clouds}$/1Pa]  & [-2 ; 6]&3.7$^{+ 1.4}_{- 1.3}$ &5.3$^{+ 0.5 } _{ -0.5 }$\\
				
				$\mu$ (derived)& & 2.38$^{+ 0.52}_{-0.07}$& 2.44$^{+ 0.66 } _{-0.13}$ \\
				\hline
				ADI & - & 15.97 & 9.58\\
				$\Delta_{\rm E2}$& - & 6.07 & 6.65  \\ 
				$\chi^2$ & - & 22.35& 24.62 \\
				\hline
				$\sigma$-level\footnote{The 	$\sigma$-level corresponds to the significance of the ADI.}& - & 5.99$\sigma$ & 4.76$\sigma$ \\
				\hline
			\end{tabular}
		}          
		\label{table5}
	\end{table}

	\subsection{TauREx setup}\label{Sect3.1}
	Once each planetary spectrum was obtained, we fitted it using the retrieval code TauREx3\footnote{\url{https://github.com/ucl-exoplanets/TauREx3_public}} \citep{al-refaie_taurex3}. This algorithm uses the nested sampling code Multinest \citep{multinest} to map the atmospheric forward model parameter space and find the best fit to our empirical spectra. %We employed linelists from ExoMol \citep{ExoMol}, HITEMP \citep{HITEMP} and HITRAN \citep{HITRAN}. 
	In our retrieval analysis we used 1500 live points and an evidence tolerance of 0.5. 
	
	The atmosphere of the two warm small planets was simulated by assuming an isothermal temperature-pressure (T/P) profile with molecular abundances constant as a function of altitude. These assumptions are acceptable since, due to the short wavelength covered by HST/WFC3, we are probing a  restricted range of the planetary T/P profile \citep{Tsiaras2018}. We note that this may not be the case anymore with next generation space telescopes \citep{Rocchetto_2016, Changeat_2019}.
	We calculated the equilibrium temperatures of the two planets using the following formula:
	\begin{equation}
	\rm T_{\rm {eq}}=\rm T_{\star} \left( \frac{ \rm R_{\star}}{2\,a} \right)^{1/2} (1-\rm{A})^{1/4}
	\end{equation}
	where $R_{\star}$ is the stellar radius, $a$ is the semi-major axis, $A$ is the geometric albedo.
	Assuming an albedo of 0.2 \citep{CrossfieldTrends}, we obtained a temperature of $835\pm20$~K and $518\pm12$~K for HD\,106315\,c and HD\,3167\,c, respectively. We then used a wide range of temperature priors $\pm{60\%}$ $T_{\rm {eq}}$ ($334$--$1336$~K for HD\,106315\,c, and $207$--$829$~K for HD\,3167\,c) to allow different temperatures around the expected $T_{\mathrm{eq}}$. The planetary radius is also fitted in the model ranging from $\pm 50\%$ of the values reported in Table~\ref{table1} (0.22-0.68 R$_{\rm J}$ for HD\,106315\,c, and 0.12-0.38 R$_{\rm J}$ for HD\,3167\,c).
	
	We simulated atmospheres with pressures between $10^{-2}$ and $10^6$~Pa, uniformly distributed in log-space across 100 plane-parallel layers. We considered the following trace-gases: H$_2$O \citep[Polyansky linelist,][]{polyansky_h2o}, CH$_4$ \citep[Exomol linelist,][]{CH4}, CO \citep[linelist from][]{li_co_2015}, CO$_2$ \citep[Hitemp linelist,][]{COandCO2}, NH$_3$ \citep[Exomol linelist,][]{ExoMol_NH3} and assumed the atmosphere to be H$_2$/He dominated. Each trace-gas abundance was allowed to vary between $10^{-12}$ and $10^{-1}$ in volume mixing ratios (log-uniform prior). We used absorption cross-sections at a resolution of 15000 and include Rayleigh scattering and collision induced absorption of H$_2$–H$_2$ and H$_2$–He \citep{abel_h2-h2,fletcher_h2-h2,abel_h2-he}. Clouds are modeled assuming a grey opacity model and cloud top pressure bounds are set between $10^{-2}$ and $10^6$~Pa. All priors are listed in Table~\ref{table5}. Recently, \citet{Kreidberg_2020} presented a transmission spectrum of HD\,106315\,c based on HST/WFC3, K2, and Spitzer observations. They chose to add N$_2$ in the retrieval analysis of HD\,106315\,c to compensate for invisible molecular opacities that could impact the mean molecular weight. The high equilibrium temperature of  HD\,106315\,c ($\sim$800~K) suggests indeed the favored presence of N$_2$. However, we note that no further constraints have been found regarding N$_2$ opacity in the posterior distributions presented in \citet{Kreidberg_2020} . Considering this result and for consistency with  HD\,3167\,c whose equilibrium temperature is lower ($\sim$500~K), we decided to consider NH$_3$ instead of N$_2$ in the retrieval analysis for both planets. This choice is mainly motivated by the low density of HD\,106315\,c ($\sim$600~kg/m$^3$) indicating, most likely, a primary light atmosphere. We therefore decided not to add N$_2$ to the analysis in order to maintain a primary mean molecular weight ($\mu\sim$2.3~amu). 
	
	To assign a significance to our detection, we used the ADI (Atmospheric Detectability Index) \citep{Tsiaras2018}. It is a positively defined Bayes Factor between the nominal atmospheric model and a flat-line model (a model which contains no active trace gases, Rayleigh scattering or collision induced absorption).
	%This value was then translated into a statistical significance \citep{Kass1995} by using Table 2 of \citet{Benneke2013}.
	We also computed two other Bayes factors in the same way as the ADI. The first one, $\Delta_{\rm E1}$ is used to compute the significance of a molecule detection using a Bayes factor between the nominal atmospheric model and the same model without the considered molecule. The second one, $\Delta_{\rm E2}$ compares a given model to a model containing only water, Rayleigh scattering and collision-induced absorption as the reference Bayesian's evidence. It is used to asses the necessity of a complex model to explain the atmosphere of the observed planet.
	These Bayes factors were then translated into a statistical significance \citep{Kass1995} by using Table 2 of \citet{Benneke2013}. Significances greater than 3.6 are considered `strong', 2.7-3.6 are `moderate', 2.1-2.7 `weak', and below 2.1 `insignificant'.
	\subsection{Results} \label{Sect3.2}
\begin{figure*}
		\centering
		\includegraphics[width=\linewidth]{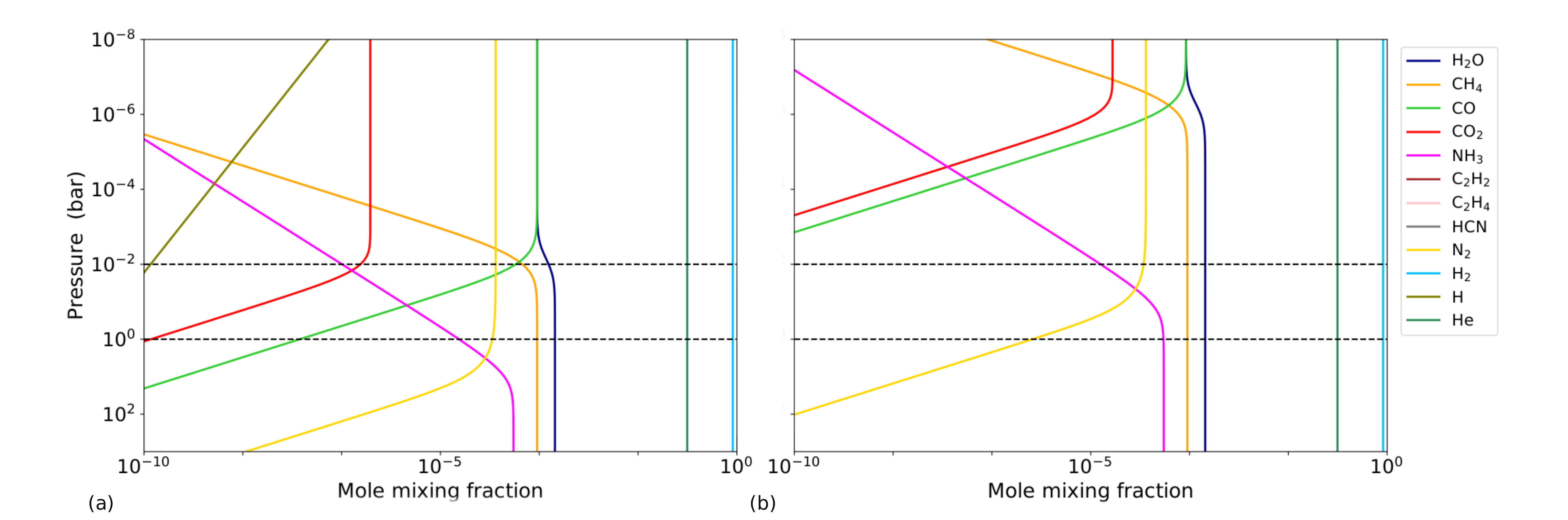}
		\caption{Thermochemical-equilibrium abundances for an atmosphere with an isothermal T/P profile and a temperature equal to that of HD\,106315\,c, T$_{\rm{eq}}$=835.0~K, (a panel), and to that of HD\,3167\,c, T$_{\rm{eq}}$=518.0~K, (b panel). These abundances' profiles have been calculated using the Reliable Analytic Thermochemical Equilibrium (RATE) Python open-source package \citep{Cubillos2019}, assuming a solar elemental  composition.  Panel a highlights that given the range of pressures probed by HST/WFC3 -marked by black dashed horizontal lines-, and the planet's equilibrium temperature the presence of N$_2$ should be favour over that of NH$_3$ in the atmosphere of HD\,106315\,c. Moreover, panel b shows as H$_2$O, CH$_4$, and NH$_3$ are the expected molecules in the atmosphere of HD\,3167\,c}.
		\label{abb_profile}
	\end{figure*}

\begin{figure*}
		\centering
		\begin{subfigure}[ ]{}	
			\includegraphics[width=0.45\linewidth]{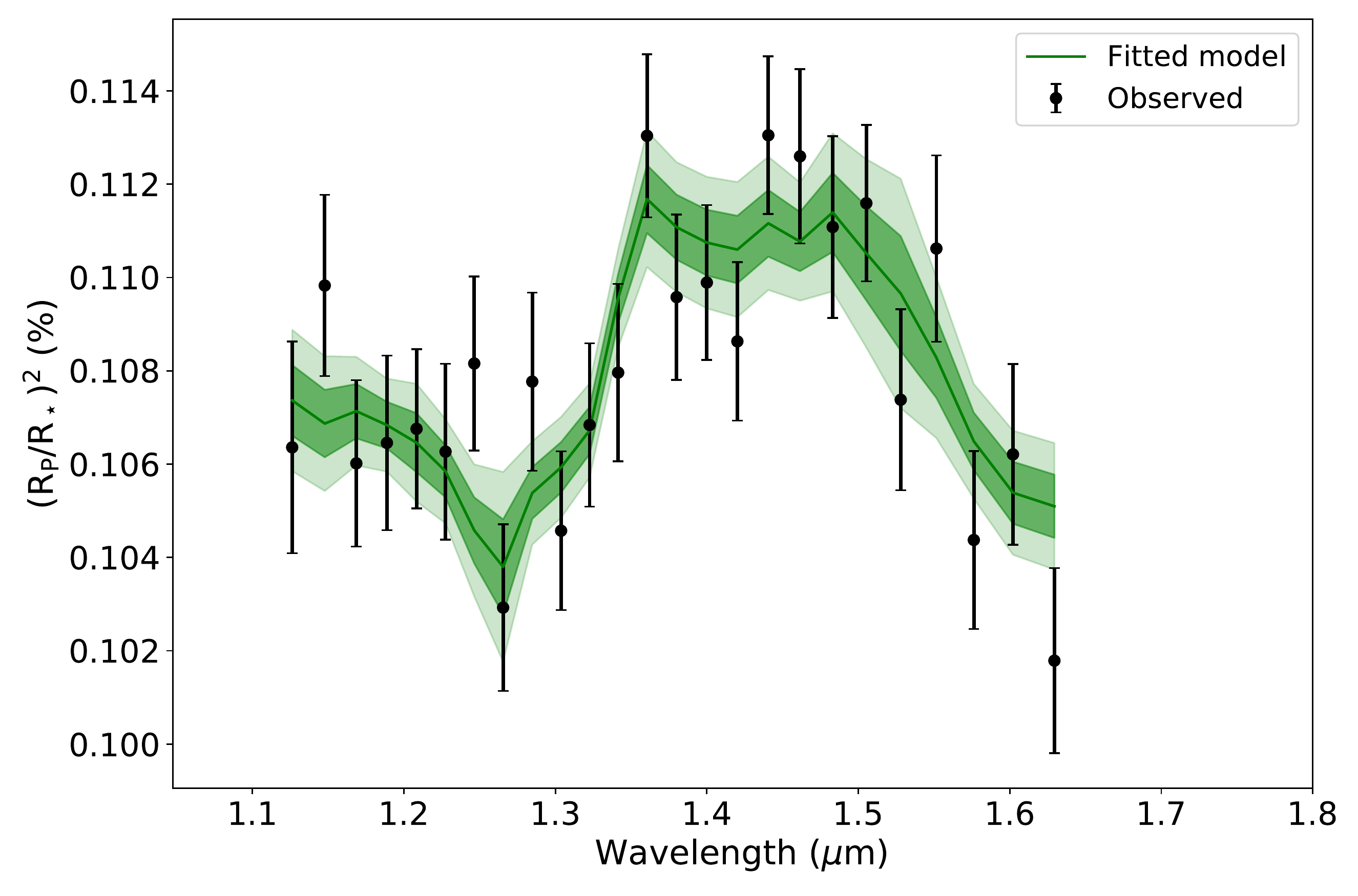}
		\end{subfigure} 
		~
		\begin{subfigure}[ ]{}
			\includegraphics[width=0.45\linewidth]{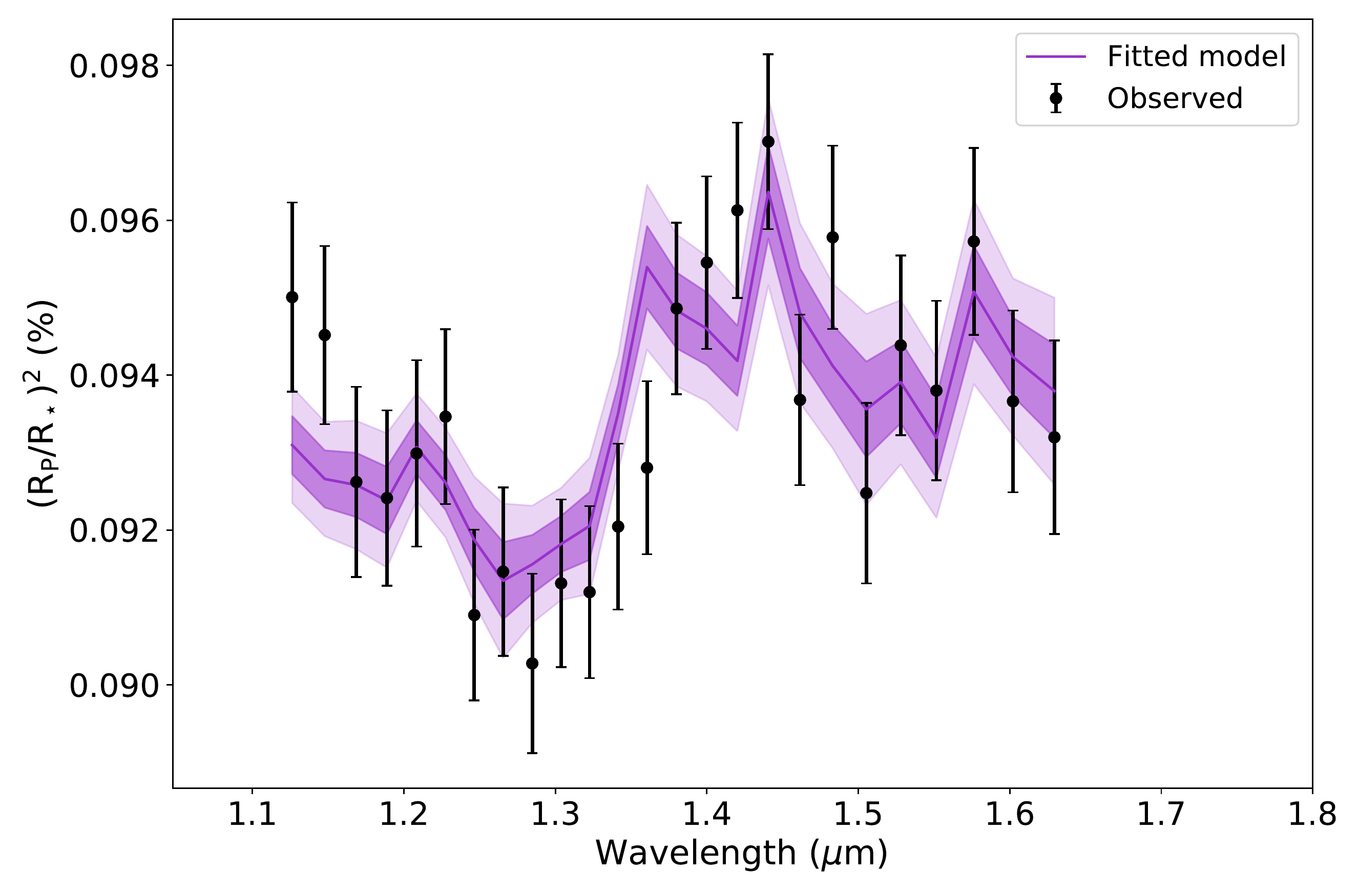}
		\end{subfigure}
		
		\begin{subfigure}[ ]{}	
			\includegraphics[width=0.45\linewidth]{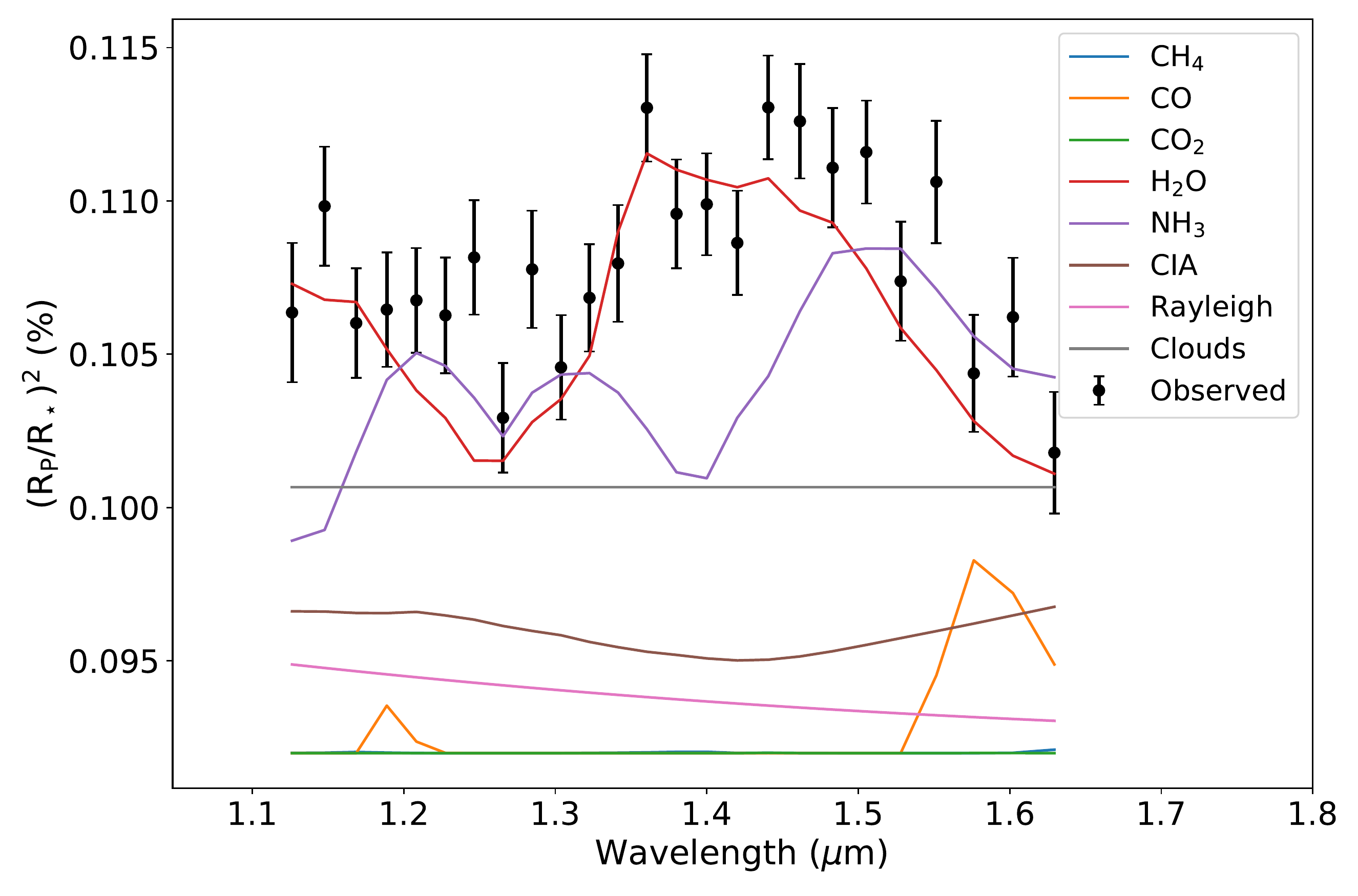}
		\end{subfigure} 
		~
		\begin{subfigure}[ ]{}
			\includegraphics[width=0.45\linewidth]{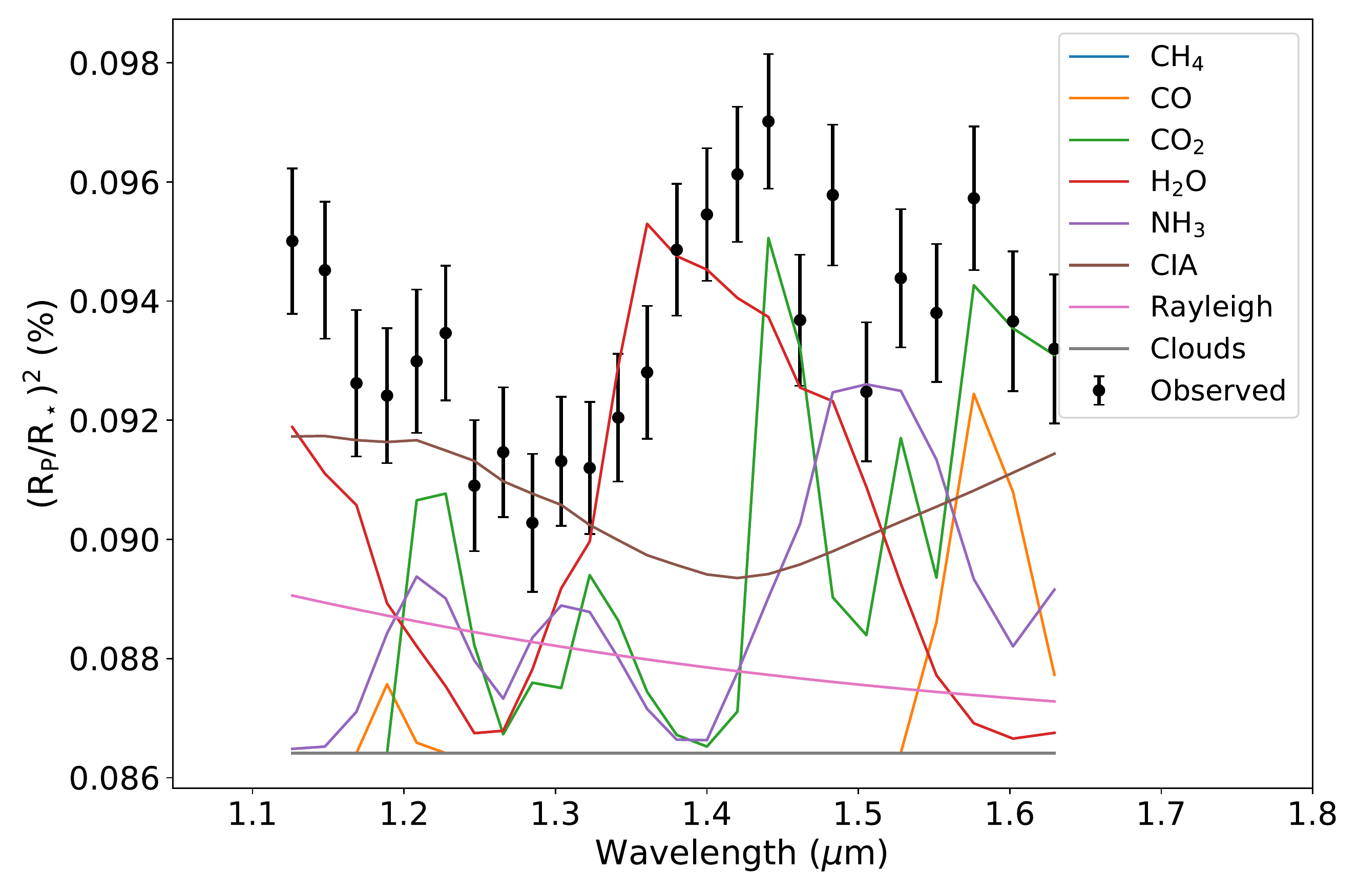}
		\end{subfigure}
		
		\caption{Best-fit atmospheric modeling results for HD\,106315\,c (a and c) and HD\,3167\,c (b and d).\protect\newline Top panels: best fit spectra, 1$\sigma$ and 2$\sigma$ uncertainty ranges. Bottom panels: contributions of active trace gases, Rayleigh scattering, collision induced absorption (CIA), and clouds. From c and d panels it is evident that some opacity contributions are very offset from the data. These correspond to molecules that do not contribute to the fit and are found to be unconstrained. }
		\label{fig6}
	\end{figure*}

	Table~\ref{table5} lists our full TauREx retrieval results for the two planets while posterior distributions are plotted in Figure~\ref{fig7} and Figure~\ref{fig8}. Retrieved best-fit spectra and corresponding best-fit molecular opacity contributions are shown in Figure~\ref{fig6}. For each opacity source, the contribution function is the transit depth that we would obtain if the molecule was alone in the atmosphere. Therefore, the opacity sources, like H$_2$O in HD\,3167\,c (Figure~\ref{fig6} d) are never fully dominant since there are always some residuals CIA, Rayleigh or other molecules that contribute to the model. Opacity contributions are represented for one solution, the one considered as the best one statistically speaking, i.e with the highest log evidence. The offset opacities correspond to molecules that do not contribute to the fit and are found to be unconstrained. Besides, the grey line in Figure~\ref{fig6} c and d represents the top cloud pressure retrieved by TauREx for the best fit solution. The signal is theoretically blocked by this layer and nothing can be observed at higher pressures. Opacities found below this line are unconstrained. Using the Bayesian log evidences, we computed the ADI, $\Delta_{\rm E1}$ and $\Delta_{\rm E2}$ as explained in \S~\ref{Sect3.1}.  
	For both planets, retrieval results are consistent with water absorption features detectable in the spectral band covered by the G141 grism. We note a significant detection of carbon-bearing species in the atmosphere of HD\,3167\,c consistent with CO$_2$ absorption features. This result is unexpected, indeed, considering the planetary equilibrium temperature, CH$_4$ features are more likely to be present than CO$_2$ \citep[see e.g., Figure~\ref{abb_profile} and][]{Venot2020}.
	%\citet{Venot2014} showed that in a planet like HD\,3167\,c, methane should be the dominant carbon-bearing molecule and the transition to a CO-dominated atmosphere should appear for $T \sim 600$~K at $10^{-5}$~bar.
	Other species like NH$_3$, CO and CH$_4$ have either unconstrained or low abundances. They could be present in both atmospheres, but spectra do not present significant absorption features. We note however that NH$_3$ abundance is better constrained in the atmosphere of HD\,106315\,c (see Figure~\ref{fig7}). Clouds top pressure is retrieved at different levels, $10^{3.7}$~Pa for HD\,106315\,c and $10^{5.3}$~Pa for HD\,3167\,c, corresponding to an upper bound (see the posterior distribution in Figure~\ref{fig7}, and \ref{fig8}). The presence of molecular features in our spectra suggests a clear atmosphere for both planets. If opaque clouds are present, they are located below the region probed by WFC3/G141 observations.
	
	\subsubsection{HD\,106315\,c}
\begin{figure*}[ht!]
		\centering
		\includegraphics[width=\linewidth]{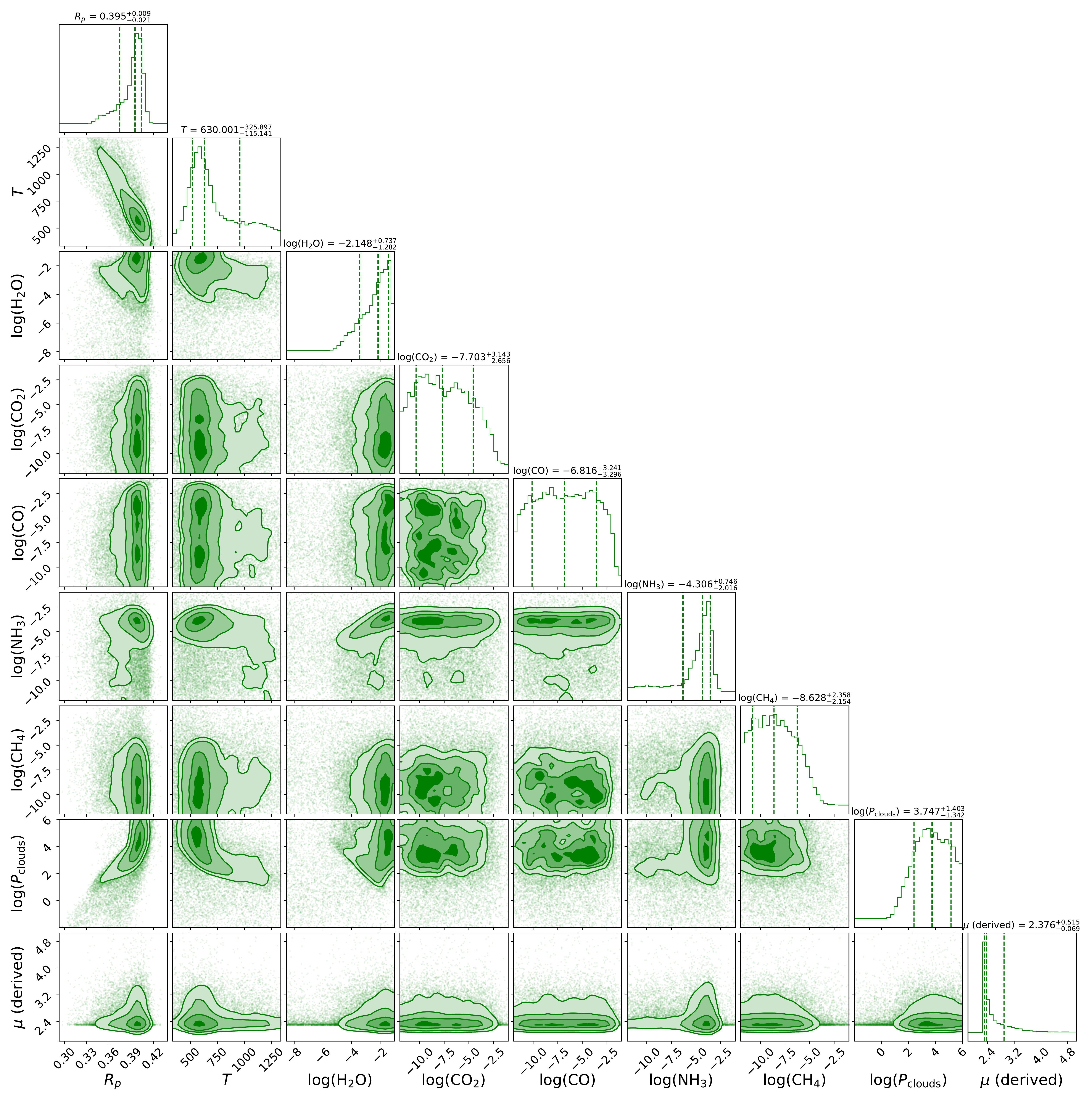}
		\caption{HD\,106315\,c atmospheric retrieval posterior distributions}
		\label{fig7}
	\end{figure*}

\begin{figure*}[ht!]
		\centering
		\includegraphics[width=\linewidth]{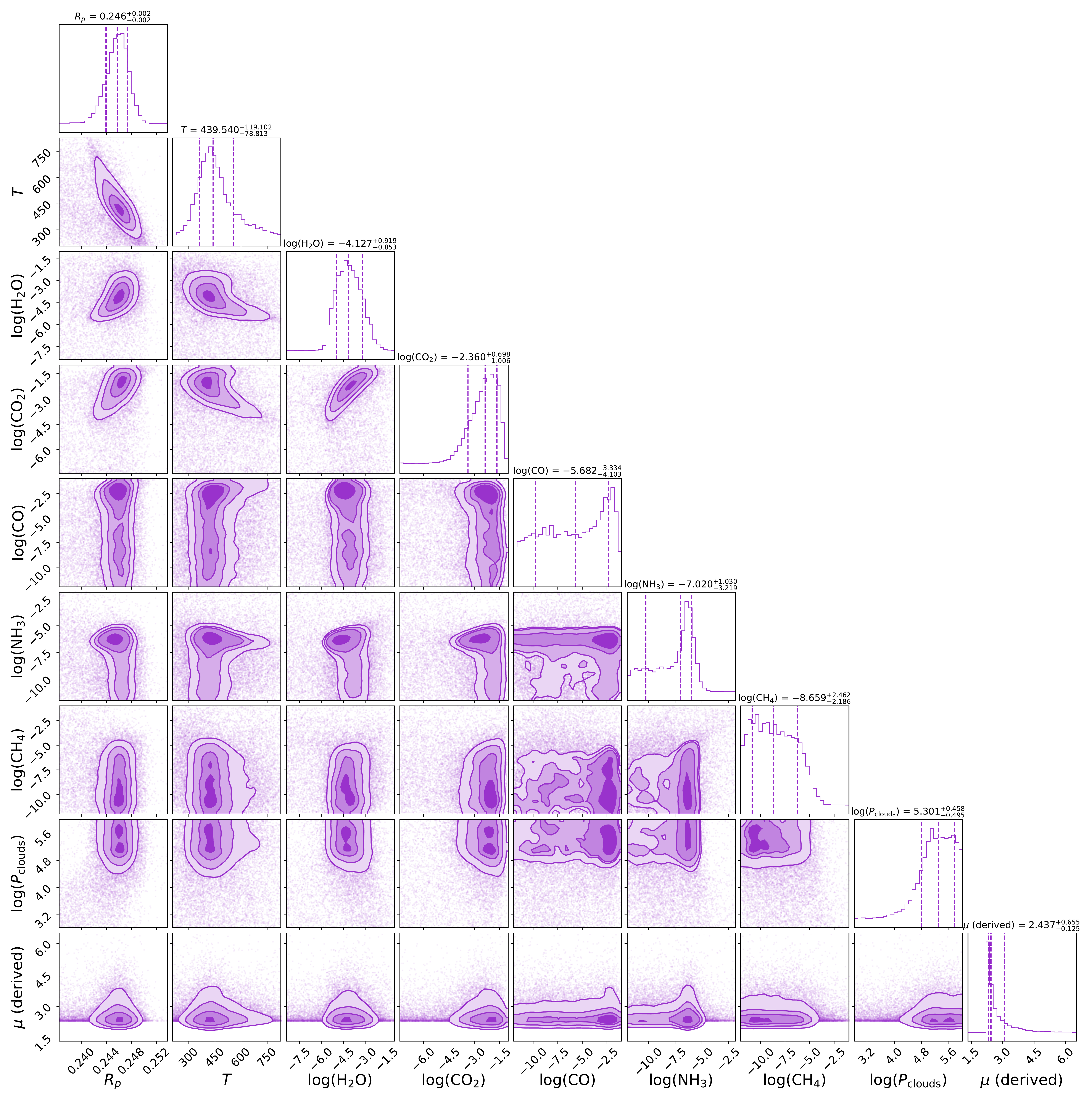}
		\caption{HD\,3167\,c atmospheric retrieval posterior distributions}
		\label{fig8}
	\end{figure*}

\begin{figure}
		\centering
		\includegraphics[width=\linewidth]{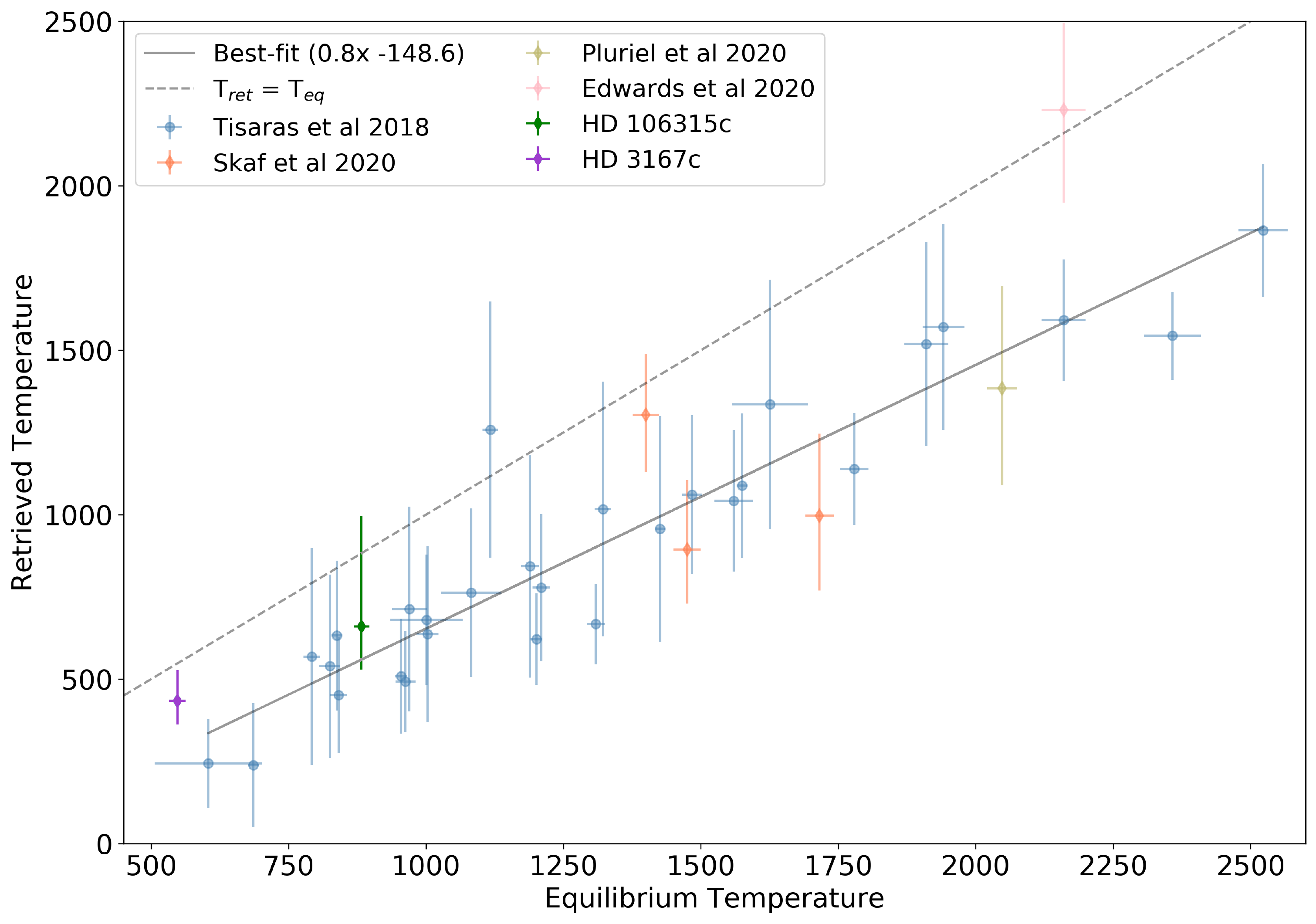}
		\caption{Trend between the retrieved and the equilibrium temperatures (or irradiation temperature) for the planets studied in \citet{Tsiaras2016}, \citet{Skaf2020}, and the two planets analysed in this work. For consistency with the above-mentioned two works, a zero albedo has been assumed to calculate the equilibrium temperature. For completeness, the planets studied in ARES I \citep{Edwards2020} and in ARES III \citep{Pluriel2020_ares} are shown too. }
		\label{Nour}
	\end{figure}

	According to the ADI, we retrieved a significant (5.99$\sigma$) atmosphere around the warm Neptune HD\,106315\,c with a notable water detection. H$_2$O is the only species that explains the absorption features between 1.3 and 1.5~$\rm{\mu}$m (Figure~\ref{fig6}). We obtained a temperature of $630^{+326}_{-115}$~K which is lower than the equilibrium temperature, but consistent within 1$\sigma$. This could be explained by the fact that we are probing the atmosphere in the terminator area, and we modeled the atmosphere in 1D using an isothermal profile  \citep{Caldas2019,MacDonald2020,Pluriel2020}. \citet{Skaf2020}, by analysing their three Hot Jupiters (WASP-127\,b, WASP-79\,b and WASP-62\,b) together with the exoplanets from \citet{Tsiaras2018}, highlighted the existence of a global trend between the equilibrium and the retrieved temperatures, with the retrieved temperatures showing almost always lower values. 
	%This behaviour is also discussed in \citet{MacDonald2020}.
	In Figure~\ref{Nour} we updated Figure~6 from \citet{Skaf2020} by adding the retrieved/equilibrium temperatures of the two Neptunes-like planets analysed in this work. We can see that HD\,106315\,c follows the global trend.  \\
	%The retrieved radius at the pressure level of 10 bars, $0.40^{+0.01}_{-0.02}$~R$_{\rm J}$ ($4.39^{+0.11}_{-0.22}$~R$_{\oplus}$), is lower than the value derived in this work (i.e. $4.98\pm 0.23$~R$_{\oplus}$, Table~\ref{table1}) and a little bit higher than the radius computed by \citet{Crossfield2017} ($3.95^{+0.42}_{-0.39}$~R$_{\oplus}$). On the other hand, it is consistent with those proposed in other studies by \citet{Barros2017} ($4.35 \pm 0.23$~R$_{\oplus}$) and \citet{Rodriguez2017} ($4.31^{+0.24}_{-0.27}$~R$_{\oplus}$). As expected by the scale height (H) formula -H is proportional to both temperature and the square of the radius-, the radius is anti-correlated with the atmospheric temperature (Figure~\ref{fig7}) and for a higher temperature solution ($1000$~K) the radius obtained should be less than $0.34$~R$_{\rm J}$ ($\sim3.73$~R$_{\oplus}$). {(\bf why are we talking about the radius? Is there a problem with it? The radius can be different depending on the instruments, the reduction techniques or the star variability. If we want to highlight differences with the literature we should try to explain why. My advise, if there is no massive inconsistencies don't talk about it as we are more interested in the atmosphere.)}
	The best-fit solution contains a notable amount of water, $\log_{10}[\mathrm{H_2O}]=-2.1^{+0.7}_{-1.3}$. Figure~\ref{fig7} shows that the right wing of the water's abundance Gaussian distribution is not complete. %but  indicating 
	This indicates that the abundance of H$_2$O could take even higher values ($\log_{10}[\mathrm{H_2O}]\sim-1$), but this is an unrealistic solution for a primary atmosphere, expected here for this Neptune-type planet. This is due to the limited coverage of HST/WFC3 G141. We note that the Bayes factor between a pure water model and the full chemical model $\Delta_{\rm E2}$ is equal to 6.07 (see Table~\ref{table8}) meaning that the complexity of the full chemical model is justified with a `strong' significance (3.91$\sigma$). %However a pure water  model could also be a good fit of our spectrum because the ADI .
	
	The temperature retrieved by TauRex ($\sim$600~K) is compatible with absorption from NH$_3$, and this strenghtens our choice to consider NH$_3$ as active gas instead of N$_2$. However, NH$_3$ contribution is debatable -- the detection is driven by a few points at 1.28, 1.55 $\rm{\mu}$m and 1.60 $\rm{\mu}$m, hence the weak abundance of $\log_{10}[\mathrm {NH_3}]=-4.3^{+0.7}_{-2.0}$. We note that a high temperature solution gives no constraint on NH$_3$ abundance whereas a lower temperature requires the molecule to be present (Figure~\ref{fig7}). NH$_3$ abundance is also correlated to the amount of H$_2$O. 
	Moreover, we can only put constraints on the higher abundance of CH$_4$: it could be found below $10^{-5}$. CO and CO$_2$ abundances are unconstrained.
	The model finds a clouds top pressure of $10^{3.7}$~Pa correlated to the amount of H$_2$O: the deeper the clouds are, the more water we have. The best-fit solution suggests a clear atmosphere with a significant amount of water. In order to give an estimation of the planetary C/O ratio, we employed the following formula re-adapted from \citet{C_O}: $C/O = (XCH_4  + XCO + XCO_2) /(XH_2O + XCO + 2CO_2 )$, where the numerator indicates all species containing
		C atoms, while the numerator indicates all
		other O-bearing species. As we obtained a constrained value only for the water abundance, we decided to explore the range of valid C/O by using not only the mean abundances, but also the upper/lower possible values allowed by the posteriors (see Table~\ref{table5}). In this way, we obtained a C/O ratio that could vary in the range (7.5 $\times$10$^{-9}$-0.60).

	\subsubsection{HD\,3167\,c}
	The ADI value found for HD\,3167\,c retrieval is lower than the one computed for HD\,106315\,c (Table~\ref{table5}), yet it corresponds to a 4.76$\sigma$ significance detection of an atmosphere around this sub-Neptune. The temperature retrieved by TauREx ($440^{+119}_{-79}$~K) is lesser then the equilibrium temperature obtained assuming an albedo equal to 0.2, but it is consistent within 1$\sigma$. 
	%The planet retrieved radius $0.242 \pm 0.002$~R$_{\rm J}$ ($2.66\pm 0.02$~R$_{\oplus}$) is consistent within 1-$\sigma$ with previous studies by \citet{Gandolfi2017} (2.740$^{+0.106}_{-0.100}$~R$_{\oplus}$), and slightly lower than the one computed by \citet{vanderburg} (2.85$^{+0.24}_{-0.15}$~R$_{\oplus}$). The radius is anti-correlated with the temperature (Figure~\ref{fig8}) and for a higher temperature solution ($\sim$700K), the radius obtained should be around $0.235$~R$_{\rm J}$ ($\sim 2.58~$R$_{\oplus}$).
	
	The main difference with HD\,106315\,c's atmosphere is the strong detection of CO$_2$, and more generally the presence of carbon-bearing species. Opacity source contributions in Figure~\ref{fig6} show both water and carbon dioxide features; these two species seem required to fit the data obtained by HST/WFC3 and their abundances are highly correlated (see Figure~\ref{fig8}). $\Delta_{\rm E2}$ is equal to 6.65 (Table~\ref{table8}) meaning that the full chemical model is statistically significant (4.07$\sigma$) compared to a pure water model. This is probably driven by the carbon dioxide detection that explains the absorption features at $1.20\;\upmu$m, $1.45\;\upmu$m, and $1.60\;\upmu$m. The best-fit solution contains a significant amount of carbon dioxide $\log_{10}[\mathrm{CO_{2}}]= -2.4^{+0.7}_{-1.0}$ and a lower amount of water $\log_{10}[\mathrm{H_2O}]=-4.1^{+0.9}_{-0.9}$. As explained in \S~\ref{Sect3.2} we would have expected CH$_4$ to be the main carbon-bearing species instead of CO$_2$.
	
	Looking at the posterior distributions in Figure~\ref{fig8}, we can constrain the higher limits of ammonia and methane abundances, which are below $10^{-5}$. The monoxide abundance posterior distribution is highly degenerate, hence the weak detection. Carbon dioxide and monoxide features are difficult to distinguish in WFC3/G141 observations because they have similar features between $1.5\;\upmu$m and $1.6\;\upmu$m, leading potentially to degeneracies between the two abundances. The amounts of H$_2$O and CO$_2$, as well as the planet temperature and radius are correlated. For less water and carbon dioxide, the model requires a higher temperature and lower radius at $10$~bar atmospheric pressure (see Figure~\ref{fig8}). The best-fit solution suggests a clear atmosphere with a top cloud pressure retrieved at $1$~bar. As for HD\,106315\,c, we derived a range of possible values in which the C/O ratio could vary, i.e. (0.49-0.85).

	\section{Discussion}
	\label{discussion}
	Considering the narrow wavelength coverage and the low data resolution, the results obtained here are to be considered carefully and put into perspective. The model we tested has 8 free parameters and 25 observation data points. Molecular abundances and temperatures retrieved by TauREx are sensitive to the inputs and bounds set up by the users. TauREx gives us a first insight into these exoplanets' atmospheres and, in particular for HST/WFC3, helps us to infer the presence of water. 
	To better constrain the molecular detections found in \S~\ref{Sect3.2}, we analysed different simulations (\textbf{A0-A5} and \textbf{B0-B7} in Table~\ref{table8}, for HD\,106315\,c and HD\,3167\,c, respectively) that include the expected molecules considering the wavelength coverage and the equilibrium temperature
	\textbf{A0} and \textbf{B0} are flat-line models that help us compute the ADI and \textbf{A2} and \textbf{B2}, pure water models are used to compute $\Delta_{\rm E2}$. 
	
	\begin{center}
		
		\begin{table*}
			\footnotesize
			\centering
			\caption{Comparison of the Bayesian log evidence for different models. The logarithm is taken to the base 10 (log $\rightarrow$ log$_{10}$).}
			\label{table8}
			\begin{tabular}{p{0.2cm} p{1.8cm} | c c c c | c p{1.2cm} c c c c} \hline\hline
				\multicolumn{12}{c}{\textbf{HD\,106315\,c}}\\\hline
				N$^{\circ}$ & Setup & Log E & ADI &$\Delta_{\rm E1}$ &$\Delta_{\rm E2}$ &T(K)&R$_{\rm P}$ (R$_{\rm J}$)&log[P$_{\rm clouds}$/1Pa] &log[H$_2$O] &log[NH$_3$]& log[CH$_4$] \\\hline
				
				A0 & No active gas & 210.94 & N/A & N/A & N/A & $798^{+356}_{-315}$ & 0.388$^{+ 0.027}_{-0.035}$ & 2.5$^{+2.5} _{-3.2}$ & N/A& N/A & N/A\\
				
				A1 & Full chemical &  226.91  & 15.97 & N/A & 6.07 &$630^{+326}_{-115} $ & 0.395$^{+0.009}_{-0.021}$ & $3.7^{+1.4}_{-1.3}$  & $ -2.1^{+0.7}_{-1.3}$ & $ -4.3^{+0.7}_{-2.0}$ &  $<-5$\\
				
				A2 & H$_2$O only &  220.84 & 9.52 & N/A & N/A &$859^{+66}_{-99}$ & 0.404$^{+0.002}_{-0.002}$ & N/A & $-5.1^{+ 0.3}_{-0.2}$ & N/A & N/A\\
				
				A3 & No H$_2$O &  212.70 &  1.76 & 14.21 & N/A & $417^{+156}_{-56}$ & $0.402^{+0.006}_{-0.011}$ & $4.1^{+1.3}_{-1.8}$  & N/A & $-3.4^{+1.0}_{-1.5}$ & $-3.0^{+1.0}_{-2.9}$\\
				
				A4 & No clouds & 226.98 & 16.04 & N/A & 6.14 & $546^{+93}_{-87}$ & $0.402^{+0.005}_{-0.007}$ & N/A & $-2.1^{+0.7}_{-1.5}$ & $-4.3^{+0.7}_{-1.0}$ & $<-5$ \\
				
				A5 & No NH$_3$ & 226.00  & 15.06 & 0.91 & 5.16 & $1004^{+223}_{-278}$ & 0.374$^{+0.022}_{-0.020}$ & $ 2.5^{+1.1}_{-0.9}$ & $-2.6^{+1.1}_{-1.3}$ & N/A & $<-5$ \\
				\hline\hline
				\multicolumn{12}{c}{\textbf{HD\,3167\,c}}\\ \hline
				N$^{\circ}$ & Setup & Log E & ADI &$\Delta_{\rm E1}$ &$\Delta_{\rm E2}$&T(K)&R$_{\rm P}$ (R$_{\rm J}$)&log[P$_{\rm clouds}$/1Pa] &log[H$_2$O] &log[CO$_2$]& log[CO] \\\hline
				
				B0 & No active gas & 225.84 & N/A & N/A & N/A & $ 473^{+225}_{-180}$ & 0.238$^{+ 0.010 }_{-0.016 }$ & $2.2^{+2.6}_{-2.6}$ & N/A & N/A & N/A \\
				
				B1 & Full chemical & 235.41 & 9.58 & N/A & 6.65  & $440^{+119}_{-79} $ &0.246$^{+0.002}_{-0.002}$ & $5.3^{+0.5}_{-0.5}$ & $-4.1^{+0.9}_{-0.9}$ & $-2.4^{+0.7}_{-1.0}$ & unconstrained\\ 
				
				B2 & H$_2$O only & 228.76 & 2.92 & N/A & N/A & $785^{+33}_{-73}$ &0.2425$^{+0.0009}_{-0.0006}$ &N/A & $-5.62^{+0.19}_{-0.18}$ & N/A & N/A \\ 
				
				B3 & No H$_2$O & 231.80 & 5.97 & 3.61 & N/A & $449^{+100}_{-88}$ &  0.246$^{+0.002}_{-0.002}$ & $5.1^{+0.6}_{-0.6}$ & N/A & $-1.9^{+0.5}_{-0.8}$ & unconstrained\\
				
				B4 & No clouds & 236.45 & 10.62 & N/A & 7.69 & $426^{+127}_{-75}$ &  0.246$^{+0.002}_{-0.002}$& N/A & $-4.2^{+0.9}_{-0.8}$ & $-2.4^{+0.7}_{-1.0}$ & unconstrained\\
				
				B5 & No CO$_2$ & 231.48 & 5.64 & 3.93 & 2.72 & $605^{+151}_{-237}$ & 0.245$^{+0.003}_{-0.003}$ & $5.2^{+0.5}_{-0.7}$ & $-4.8^{+1.4}_{-0.6}$ & N/A & $-1.9^{+0.5}_{-1.6}$\\
				
				B6 & No CO & 234.84 & 9.00 & 0.60 & 6.08 & $440^{+110}_{-82}$ & 0.246$^{+0.002}_{-0.002}$ &$5.3^{+0.5}_{-0.6}$ & $-3.9^{+1.1}_{-1.0}$ & $-2.1^{+0.7}_{-0.9}$ &N/A \\
				
				B7 & No CO$_2$, CO & 229.86 & 4.03 & 5.55 & 1.10& $732^{+60}_{-105}$ &  0.2423$^{+0.0013}_{-0.0009 }$ &$5.4^{+0.4}_{-0.4}$ & $-5.5^{+0.3}_{-0.2}$ & N/A &N/A \\
				
				\hline
			\end{tabular}%}
		\end{table*}
	\end{center}
	
	\subsection{Strength of H$_2$O detection}

\begin{figure}
		\centering
		\includegraphics[width=\columnwidth]{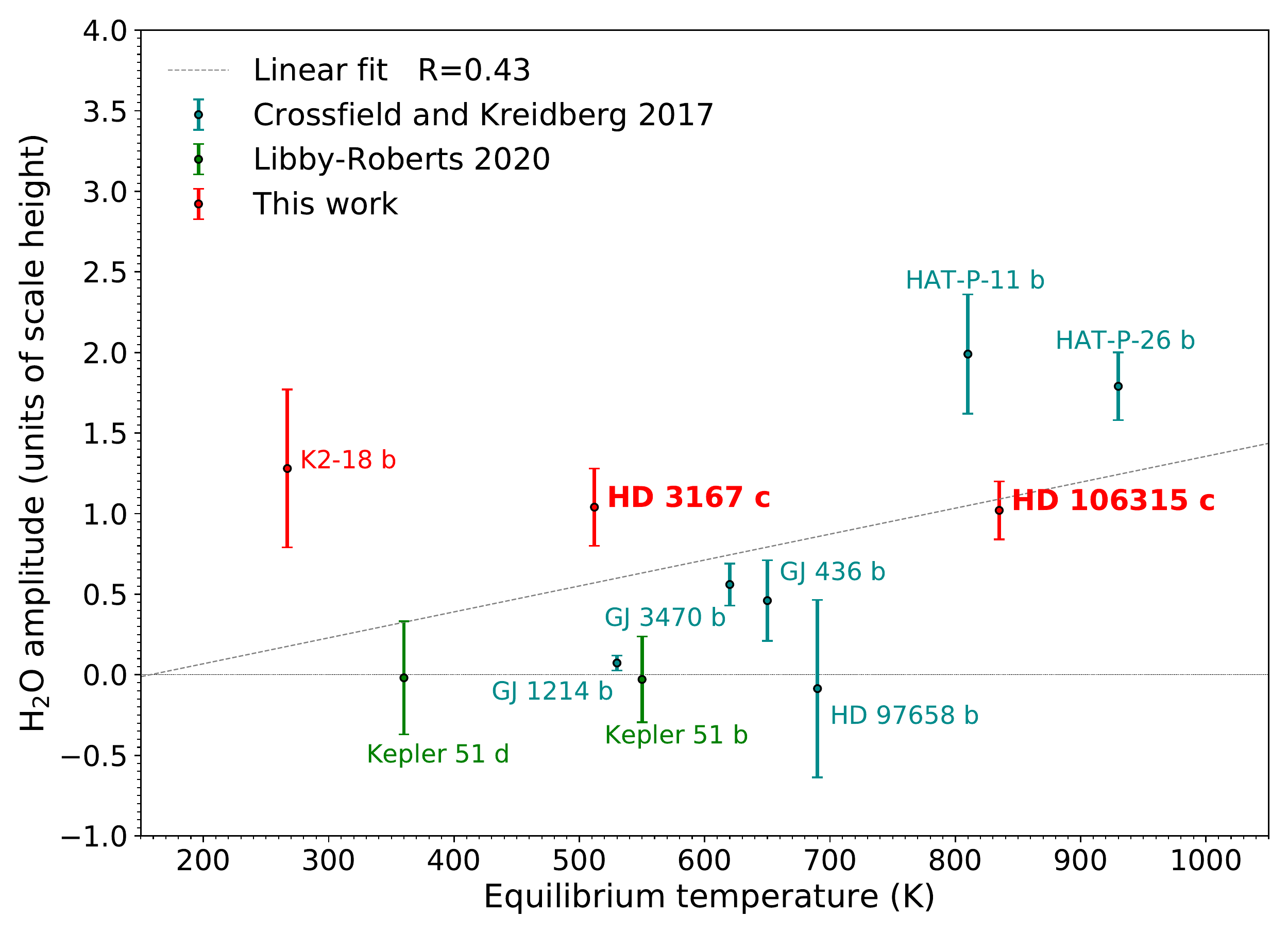}
		\caption{Normalized H$_2$O amplitude in units of scale height with respect to equilibrium temperatures. Blue points are from \citet{CrossfieldTrends} and green points are from \citet{Roberts2020}. Red points are computed using the method described in \citet{CrossfieldTrends} and spectra obtained in this work and from \citet{TsiarasK218b} for K2-18 b. The dotted line corresponds to a linear fit. The correlation coefficient was found to be lower than in \citet{CrossfieldTrends}, 0.43 compared to 0.83.}
		
		\label{amp_teq}
	\end{figure}

	For both planets, to assess the significance of H$_2$O detection, we removed this active gas from the full chemical model and we analysed the Bayes factor $\Delta_{\rm E1}$. It decreases from 226.91 (\textbf{A1}) to 212.70 (\textbf{A3}) (see Table~\ref{table8}) and from 235.41 (\textbf{B1}) to 231.80 (\textbf{B3}) for HD\,106315\,c and HD\,3167\,c, respectively. H$_2$O detection is statistically confirmed for both planets with a `strong' %\citep{Kass1995, Benneke2013} 
	significance (5.68$\sigma$) for  HD\,106315\,c, and a `moderate' one (3.17$\sigma$)  for HD\,3167\,c. 
	%Recently, \citet{Kreidberg_2020} presented a transmission spectrum of HD\,106315\,c based on HST/WFC3, K2, and Spitzer observations. 
	In the recent paper by \citet{Kreidberg_2020}, they reported a tentative detection (with a Bayes factor of 1.7 or 2.6, depending on prior assumptions) of water vapor with a small amplitude of 30~ppm. In this simultaneous and independent analysis, by using different algorithms both for the extraction of the transmission spectrum from the WFC3 data (with \verb+Iraclis+), and for the retrieval analysis (performed with TauREx3), we also detect the presence of water in the atmosphere of HD\,106315\,c with a high significance. Moreover our observed spectrum seems to be compatible with deeper H$_2$O features, which reinforces the detection. 
	To date, water has been detected on several Neptune and sub-Neptune planets which allows comparisons.  HD\,106315\,c could be compared to %HAT-P-26\,b (water detection's significance, hereafter $\sigma_{\rm{H_2O}}$, greater than 7$\sigma$, \citealt{MacDonald_2019,Wakeford2017}), 
	to HAT-P-11\,b (with a water detection's significance, hereafter $\sigma_{\rm{H_2O}}$, of 5.1$\sigma$, \citealt{Fraine2014}), and to GJ\,3470\,b ($\sigma_{\rm{H_2O}}$= 5.2$\sigma$, \citealt{Benneke2019}). While HD\,3167\,c has a lower water detection, appearing more similar to K2-18\,b ($\sigma_{\rm{H_2O}}$=3.6$\sigma$, \citealt{TsiarasK218b}, and $\sigma_{\rm{H_2O}}$=3.93$\sigma$, \citealt{benneke2019water}).
	\citet{CrossfieldTrends} studied  the water features amplitude of six warm Neptune planets and highlighted correlations with the equilibrium temperature and the mass fraction of hydrogen and helium. To verify the correlation of H$_2$O amplitude, in units of atmospheric scale height, with the equilibrium temperature we computed HD\,106315\,c, HD\,3167\,c and K2-18\,b water amplitude using HST/WFC3 spectra obtained here and in \citet{TsiarasK218b}. We used the same method described in \citet{CrossfieldTrends}. We fitted a carbon-free template of GJ\,1214\,b normalized in units of scale height \citep{Crossfield2011} to the observations using the Levenberg and Marquardt’s least squares method (L-M) \citep{Markwardt}. Then, we measured the amplitude taking the normalized average value from 1.34${\rm \mu}$m to 1.49${\rm \mu}$m and subtracting it from the average value outside this wavelength range. The scale height H=K$_{\rm B}$T$_{\rm eq}$/${\rm \mu}$g is computed assuming a hydrogen rich atmosphere (${\rm \mu}$=2.3~amu) and the equilibrium temperature is calculated for an albedo of 0.2. We find a water feature amplitude of 1.02$\pm$0.18 for HD 106315 c, of 1.04$\pm$0.24 for HD\,3167\,c, and of 1.28$\pm$0.49 for K2-18 b. We note that \citet{Kreidberg_2020} recent paper found a lower absorption feature, i.e 0.80$\pm$0.04 for HD 106315 c. We plot our values in Figure~\ref{amp_teq} along with the amplitudes computed in \citet{CrossfieldTrends} and the ones found in \citet{Roberts2020} for Kepler\,51\,b and Kepler\,51\,d. Finally, we fitted a linear relation and compared the Pearson correlation coefficient and the probability. We find a correlation coefficient of 0.43 and a p-value of 0.18. The strong correlation highlighted in \citet{CrossfieldTrends} is not found here, mostly because of K2-18\,b high water feature amplitude at low temperature.
		While removing K2-18\,b and Kepler\,51\,d amplitudes – to focus on planets with temperature between 500 and 1000~K as in \citet{CrossfieldTrends}- we find a correlation coefficient of 0.70 and p-value of 0.04 while they found a coefficient of 0.83 for a p-value equal to 0.04. A refinement of the scale height, HST/WFC3 water amplitude and correlations computations will be detailed in a follow-up paper focusing on intermediate size planets (R$_{\rm P} <$ 6~R$_{\rm \oplus}$) with consistent published spectra. 
	
	\subsection{Clear or cloudy atmospheres}
	In \S~\ref{Sect3.2}, we retrieved a clear atmosphere for both planets, but we expect species to condense and clouds to form on warm Neptune and sub-Neptune planets. The flat spectra of GJ\,436\,b \citep{knutsonGJ}, GJ\,1214\,b \citep{kreidbergGJ} and HD\,97658\,b \citep{knutsonHD} were interpreted as high cloud or haze at low pressure. We confirm the clear atmosphere by removing the cloud top pressure parameter from the full chemical model. ADIs of cloud free models (\textbf{A4} and \textbf{B4} in Table~\ref{table8}) are higher than ADIs of full chemical models including clouds (\textbf{A1} and \textbf{B1} in Table~\ref{table8}). Clouds do not impact retrieval results, even for HD\,106315\,c with a lower top clouds pressure, and this means that either the planet has a clear atmosphere or the clouds are located below the visible pressure where the atmosphere is opaque. %We note that a cloud free model implies a higher amount of water i.e 
	%10^{-1.9}$ for HD\,106315\,c and $10^{-4.3}$ for  HD\,3167\,c. 
	Looking at HD\,106315\,c's clouds top pressure correlations with H$_2$O abundance (see Figure~\ref{fig7}), a second mode appears meaning that clouds could be present in the region we are probing. %{(\bf we don't necessarily want to talk about this but for HD106 the cloud post look very much like a upper bound, however if you look into the detailed correlations with H2O, there is a beginning of a bi-modal, meaning that the clouds are in fact starting to appear in the pressure range we probe. We can keep this in mind and update if required by the referee.)}
	
	The TauREx retrieval does not bring any information on cloud composition and we must recall that the wavelength coverage is not wide enough to constrain cloud chemistry. All things considered, models have predicted that for hot atmospheres (900 to 1300~K) we could find condensates like KCl, ZnS and Na$_2$S, and for colder atmospheres (400 to 600~K) KCl and NH$_4$H$_2$PO$_4$ \citep{lodders, morley}.  GJ\,1214\,b (6.26 $\pm$ 0.86 M$_\oplus$, 2.85 $\pm$ 0.20 R$_\oplus$, \citealt{Harpsoe2013}), K2-18 b (8.92 $\pm$ 1.7 M$_\oplus$, 2.37 $\pm$ 0.22 R$_\oplus$, \citealt{Sarkis2018} ) and HD\,3167\,c (this paper) have a similar mass and radius, and yet present very different atmospheric properties. The equilibrium temperature is lower for K2-18\,b \citep[284$\pm$15~K,][]{Sarkis2018}, but presents water detection. GJ\,1214\,b has a similar equilibrium temperature \citep[547$_{-8}^{+7}$,][]{Kundurthy2011}, but exhibits a flat spectrum suggesting the presence of clouds. 
	
	\subsection{NH$_3$ in HD\,106315\,c's atmosphere}
	HD\,106315\,c's best fit solution includes a small amount of NH3, i.e $\log_{10}[\mathrm {NH_3}]=-4.3^{+0.7}_{-2.0}$. Looking at the posteriors distribution (Figure~\ref{fig7}), NH$_3$ abundance converges toward a solution. To confirm this detection, we removed this gas from the full chemical model and computed $\Delta_{\rm E1}$ (see \textbf{A5} in Table~\ref{table8}). The difference is 0.91 meaning that NH$_3$'s detection has to be considered `not-significant' (1.97$\sigma$). %\citep{Kass1995, Benneke2013}. 
	However, we observe some differences, the temperature rises to 1004~K with less constraints and consequently, the radius decreases to 0.374~R$_{\rm J}$. Clouds are found at a higher level $10^{2.5}$~Pa. The cloud deck compensates for NH$_3$ features by cutting H$_2$O ones and shrinking the spectrum. From this analysis, we conclude that HD\,106315\,c can be surrounded by either %We can conclude in 
	a primary clear atmosphere with H$_2$O and traces of NH$_3$ or by a primary atmosphere with H$_2$O and deep clouds. \\
	As mentioned in \S~\ref{Sect3.1} the high equilibrium temperature of HD\,106315\,c should have favor the presence of N$_2$ instead of NH$_3$ (see e.g., Figure~\ref{abb_profile}). NH$_3$ is expected to disappear above 500-550~K. However, we retrieve at the terminator, so we should expect a lower temperature (closer to this 500~K limit) and more NH$_3$. Moreover, N$_2$ is an inactive gas, with no feature in WFC3, which means that the ‘free’ retrieval we perform – the retrieval in ‘free’ mode is used to retrieve the abundance for active molecules that have features in the spectrum- will not pick up this molecule except if it influences the mean molecular weight. To test this, we added N$_2$ in the analysis to see the possible consequences that this molecule could have had on the mean molecular weight.  We assumed an initial N$_2$ abundance of 10$^{-4}$, compatible with the one expected by thermochemical-equilibrium condition (see Figure~\ref{abb_profile}), and we allowed it to vary between $10^{-12}$ and $10^{-1}$ in volume mixing ratios (log-uniform prior) -as for the other molecules. The inclusion of  N$_2$ does  not  affect the mean molecular weight, a simple clouds model added to H$_2$O and NH$_3$ features are enough to fit the spectrum, there is no need to add extra molecular weight to shrink the spectrum. Moreover, NH$_3$ detection remains around 10$^{-4}$ (see Figure~\ref{FigA3}).
	
	\subsection{CO$_2$ in HD\,3167\,c's atmosphere}
	HD\,3167\,c best fit solution includes an important amount of CO$_2$ (i.e $\log_{10}[\mathrm {CO_2}]=-2.4^{+0.7}_{-1.0}$). This detection is supported by the data points from $\sim$1.5 to 1.6~$\mu$m, but water seems to explains better the absorption features around 1.4~$\mu$m (see Figure~\ref{fig6}). We removed CO$_2$ from the full chemical and compared log evidences, it decreases from 235.41 (\textbf{B1}, Table~\ref{table8} ) to 231.48 (\textbf{B5}) corresponding to a 3.28$\sigma$ ``moderate" detection. The ADI decreases as well to 5.64. We note that CO is now compensating for CO$_2$ features and its log abundance increases to $\log_{10}[\mathrm{CO}]=-1.9^{+0.5}_{-1.6}$. This value is too high for a realistic primary hydrogen-rich atmosphere that we expect for this planet. We successively removed CO from the full chemical model, but it does not impact the retrieval results (\textbf{B6} in Table~\ref{table8}) and $\Delta_{\rm E1}$ is below 1 (`not significant'). Finally, we removed both CO and CO$_2$ to asses the detection of those carbon-bearing species (\textbf{B7}). The difference in log evidences is now equal to $\Delta_{\rm E1}$=5.55 and corresponds to more than 3$\sigma$ carbon detection. This test does not impact the abundance of water nor the top cloud pressure, but constrains better the abundance of ammonia to $10^{-6.4}$. We note that CH$_4$ does not compensate the lack of the other carbon-bearing species, it's abundance remains constrained below $10^{-5}$.  The temperature increases to keep a primary atmosphere hypothesis and an extended clear atmosphere. 
	
	This unexpected detection of carbon bearing species could be explained by noise or systematic effects that were not removed during the white light curve fitting step (see Section \ref{Sect2.4}). It could also be the result of phenomena that our
	1D equilibrium chemistry modeling cannot reproduce, e.g. 3D transport
	cross-terminator. An other interpretation could be the actual presence of CO$_2$ in the atmosphere of HD\,3167\,c due for example to a very high metallicity, enhanced over that of the host star which is consistent with solar metallicity ([Fe/H]=0.03$\pm$0.03~dex, \citealt{Gandolfi2017}). It is known indeed that the abundance of CO$_2$ scales quadratically with metallicity \citep[see e.g.][]{Moses_2014}, and other examples of overabundance of CO$_2$ interpreted as caused by an high metallicity can be found in the literature \citep[see e.g.][]{GJ436b_2011}. However, if we use the water abundance as a proxy of metallicity \citep[see e.g.,][]{Kreidberg_metallicity} we infer a solar or sub-solar metallicity for HD\,3167\,c, which would be in tension with the possibility that CO$_2$ could be present due to high metallicity. More observations are thus necessary to better constraint a possible presence of CO$_2$ in the atmosphere of HD\,3167\,c.

	%\subsection{Future characterisations}
	%Ariel JWST ? simulations ?
	\subsection{Inferences from the Mass and Radius}
	
	There is a strong degeneracy in exoplanet interiors as there are many compositional models that are compatible with an observed mass and radius. However, by combining the mass, radius, and the spectroscopic results of our study we can get an inference for the interior composition of HD\,106315\,c and HD\,3167\,c. Our discovery of icy constituents, such as $\rm H_{2}O$ in both planetary atmospheres (and maybe $\rm NH_{3}$ in the envelope of HD\,106315\,c) indicate an ice-rich embryo. Curiously, the mass and radius of HD\,106315\,c and HD\,3167\,c are also consistent with an ice-rich core which we explain below.
	
\begin{figure}[!ht]
		\includegraphics[width=\linewidth]{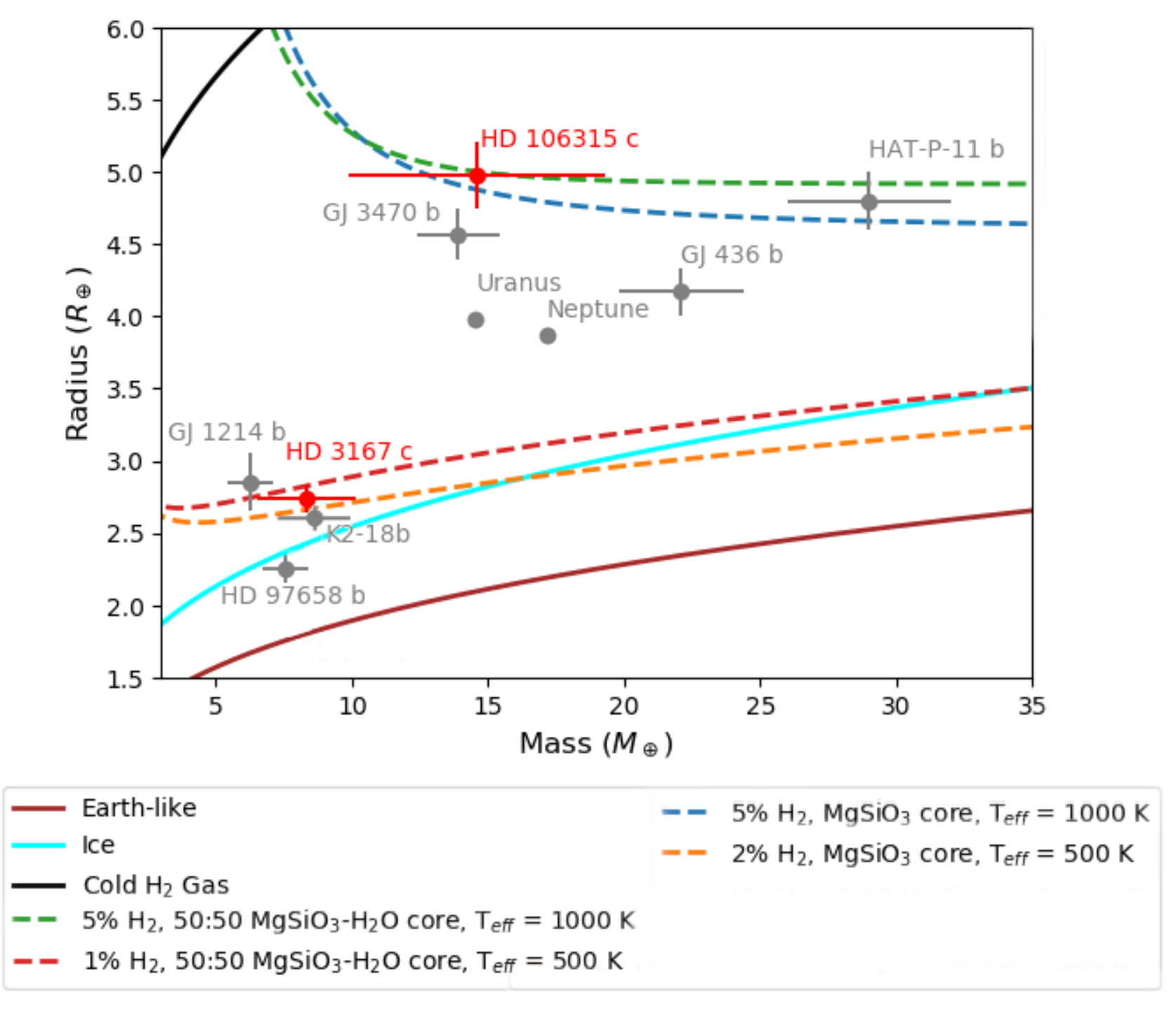}
		\caption{The mass and radius of HD\,106315\,c and HD\,3167\,c (from Table~\ref{table1}) plotted against other planets with size between 1.5-4~R$_{\oplus}$ and published atmospheric characterization studies (see Table~\ref{table0}) -GJ\,3470\,b \citep{Awiphan2016}, GJ\,436\,b \citep{Maciejewski2014}, GJ\,1214\,b \citep{Harpsoe2013}, HD\,97658\,b \citep{VanGrootel2014}, HAT-P-11\,b \citep{Stassun2017}, K2-18\,b \citep{benneke2019water}- and Uranus and Neptune (\url{https://nssdc.gsfc.nasa.gov/planetary/factsheet/}). The mass and radius models are from \citet{Zeng2013} and \citet{Zeng2016}. } 
		\label{fig:mass_radius}
	\end{figure}

	For the following results we adopted the planetary models from \citet{Zeng2013}, \citet{Zeng2016}, and \citet{Zeng2019}. Based on the mass and radius of HD\,106315\,c and HD\,3167\,c they are both consistent with icy cores with hydrogen envelopes $\sim 5~ \rm wt. \%$ and $0.3-1~ \rm wt. \%$ of their total planetary masses respectively. We show these results in Figure~\ref{fig:mass_radius}. Nevertheless, there is still enough uncertainty in the results that a silicate embryo engulfed by a hydrogen atmosphere is still plausible for both planets. Certainly, with improved mass and radius measurements, together with more accurate spectroscopic observations, the interior structure of exoplanets such as HD\,106315\,c and HD\,3167\,c will get further constrained. We discuss the implications of this in \S~\ref{future}. 
	
	Besides, \citet{Mousis_2020} recent publication, showed that close-in planets could have water-rich hydrospheres in super-critical state. Their model suggests that intermediate-size planets could be hydrogen/helium-free and their interiors would simply vary from one another depending on the water content.
	
		\subsection{Comparison with previous results} 
\begin{figure*}[!ht]
		\includegraphics[width=\linewidth]{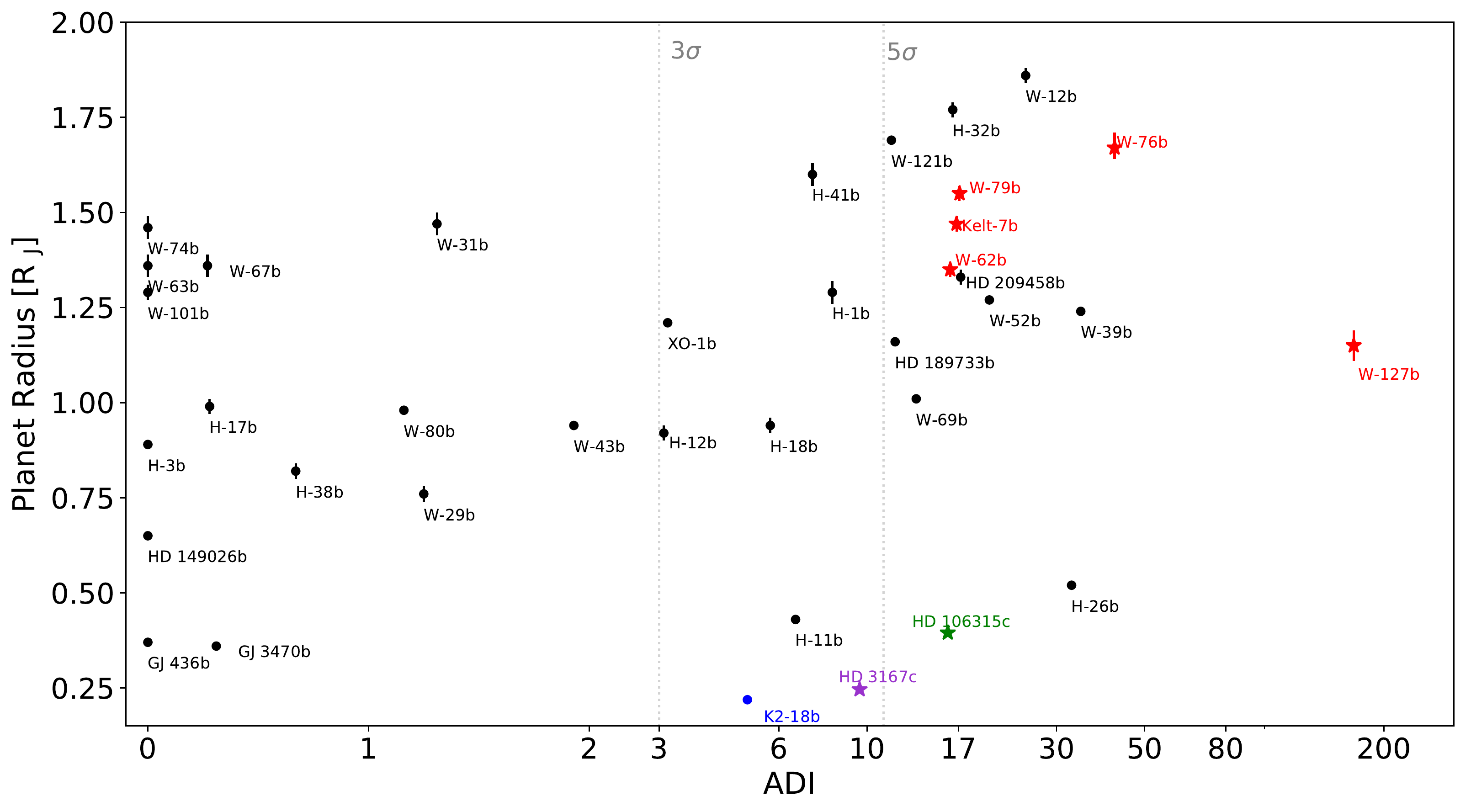}
		\caption{Exoplanetary radii as a function of the ADI index (in logarithmic scale) for the targets analysed in this work (in green and in violet), in ARES I-II-III (in red), in \citet{Tsiaras2016} (in black), and in \citet{TsiarasK218b} (in blue).} 
		\label{fig9}
	\end{figure*}

		This paper is the result of work carried out during the ARES Summer School, where we used algorithms and data available to the public, thus allowing our results to be tested and reproduced. This is the fourth paper output of this summer school. In the first work ARES I
		%\ion{ARES}{I} 
		\citep{Edwards2020} and in the third one ARES III
		%\ion{ARES}{III} 
		\citep{Pluriel2020_ares} we analysed the transmission and the emission spectra of WASP-76\,b and Kelt-7\,b respectively, while in the second one ARES II
		%\ion{ARES}{II} 
		\citep{Skaf2020}, the atmospheric study of WASP-42\,b, WASP-79\,b, and WASP-127\,b was performed.
		In this work, we used the ADI as a significance index to make the approach in our work uniform with these previous papers, with \citet{Tsiaras2018}, and with \citet{TsiarasK218b}. Figure~\ref{fig9}, shows the gaseous exoplanets studied by \citet{Tsiaras2018} (in black), K2-18b examined in \citet{TsiarasK218b} (in blue), the hot Jupiters analysed in ARES I, ARES II, ARES III
		%\ion{ARES}{I}, \ion{ARES}{II}, and \ion{ARES}{III} 
		- for consistency with other works here we plot the ADI obtained from the analysis of WASP-76\,b's and Kelt-7\,b's transmission spectra - (in red), and finally the Neptune-like planets, HD\,106315\,c (in green) and HD\,3167\,c (in violet), studied in this paper. 
		From this figure, it emerges that, even if the two exoplanets characterized in this paper have smaller radii than most of the other targets, their ADI is not smaller. Our study, together with  \citet{TsiarasK218b}, shows that even smaller planets' atmospheres can be characterized with high significance. This opens the way for the atmospheric study of planets with smaller radii than the hot Jupiter targets which have mostly been analyzed so far.

	\subsection{Future Characterization}
	\label{future}

	\begin{figure*}
		\centering
		\begin{subfigure}[ ]{}	
			\includegraphics[width=0.45\linewidth]{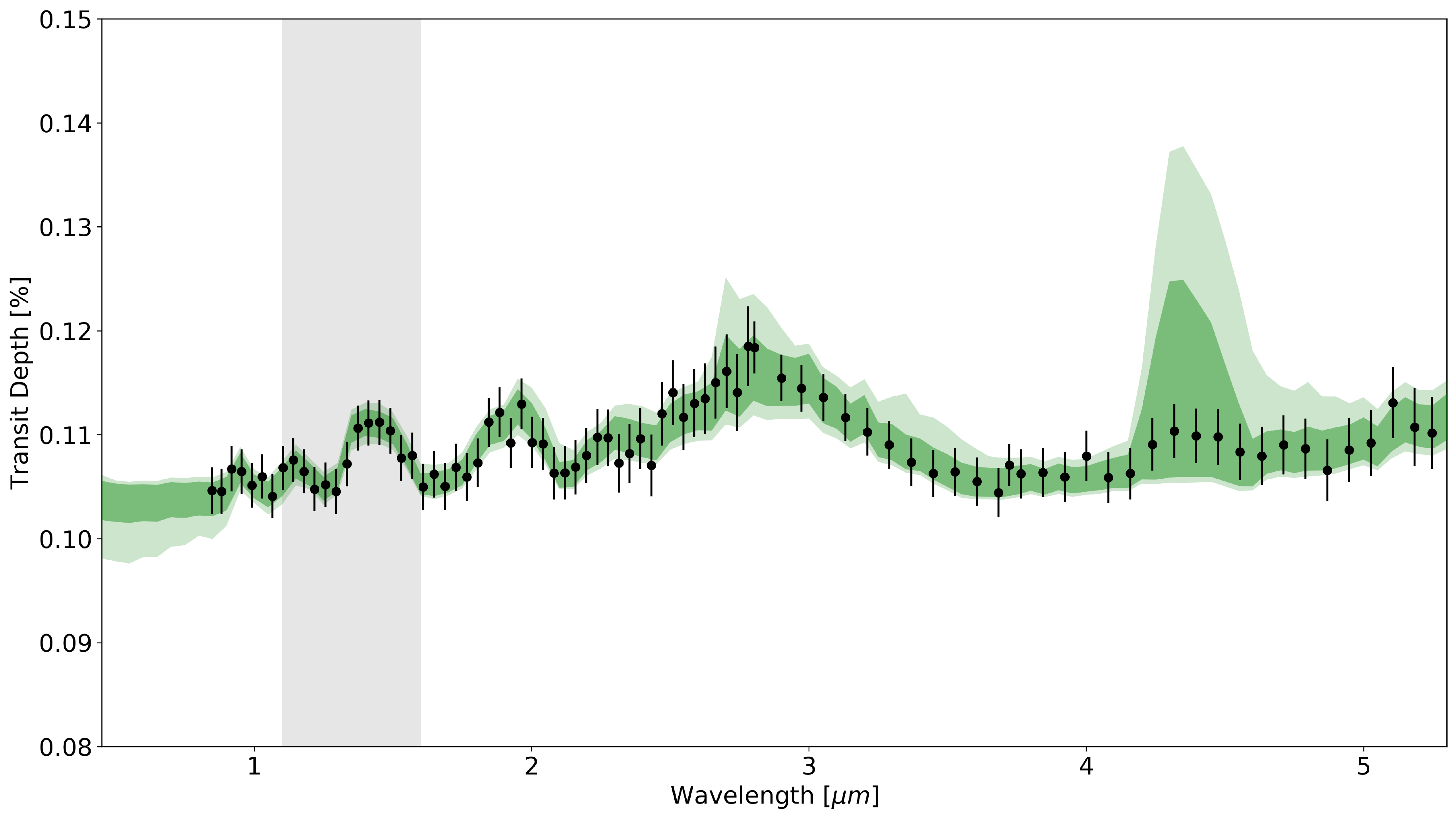}
		\end{subfigure} 
		~
		\begin{subfigure}[ ]{}
			\includegraphics[width=0.45\linewidth]{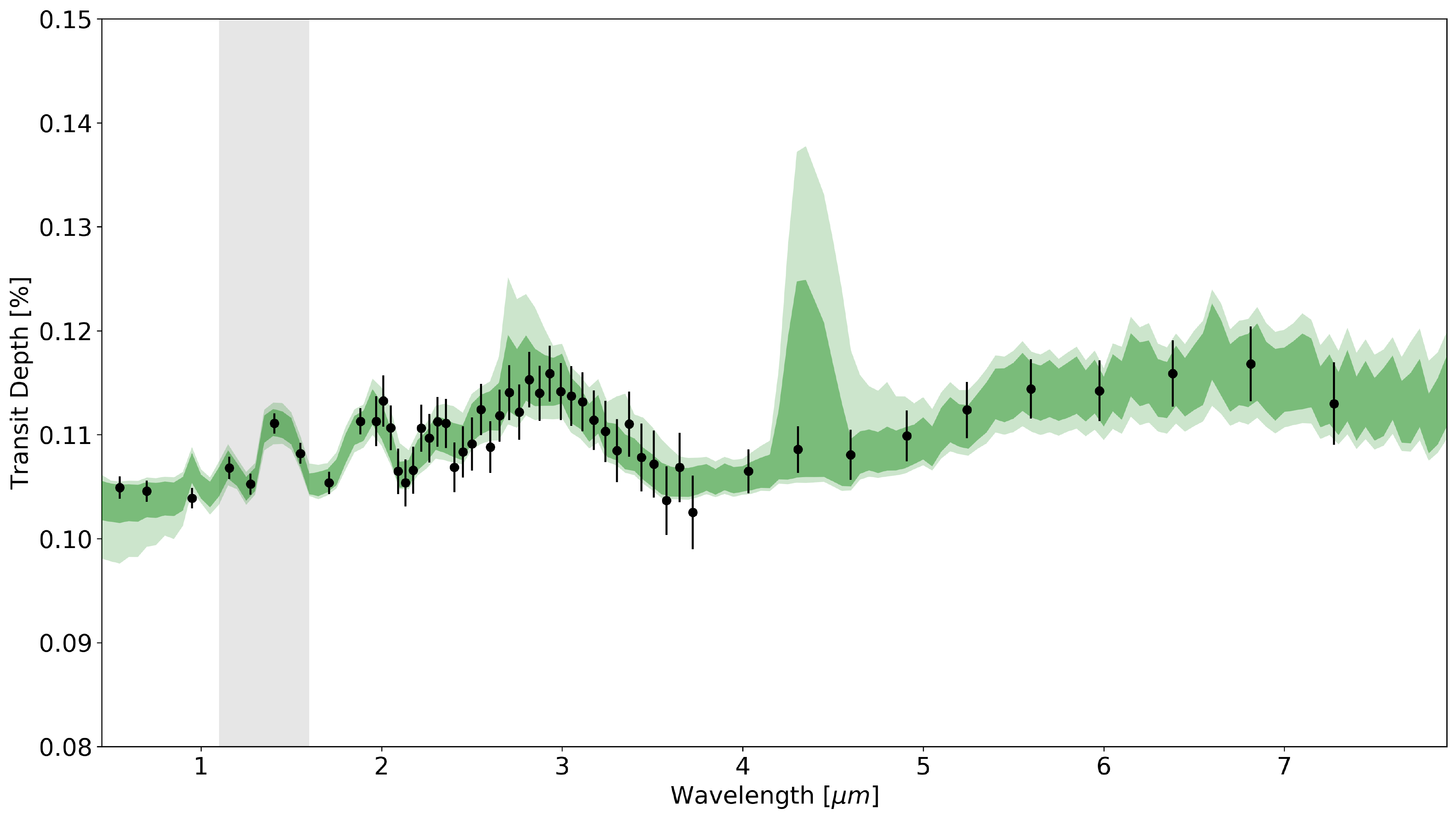}
		\end{subfigure}
		
		\begin{subfigure}[ ]{}	
			\includegraphics[width=0.45\linewidth]{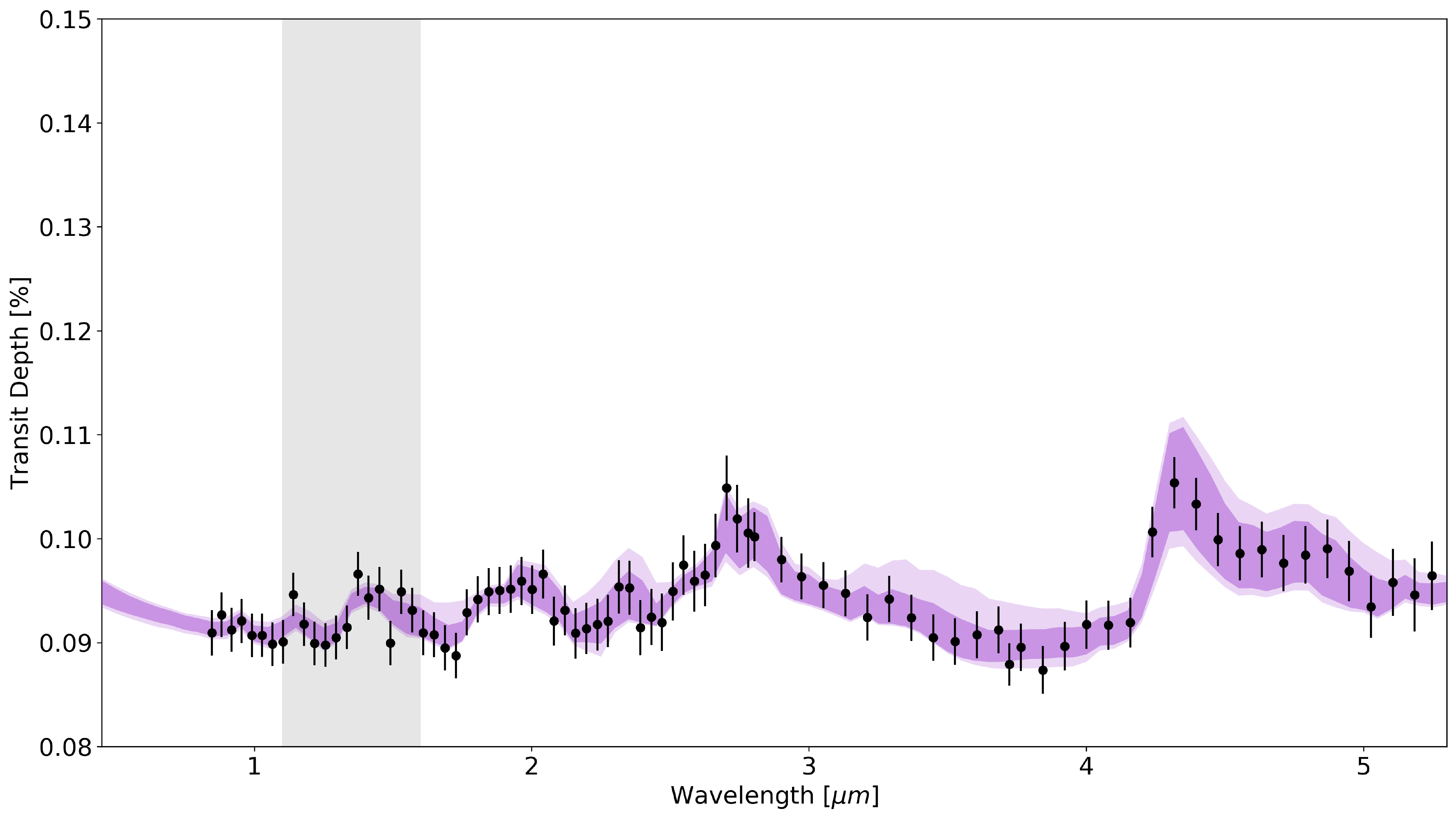}
		\end{subfigure} 
		~
		\begin{subfigure}[ ]{}
			\includegraphics[width=0.45\linewidth]{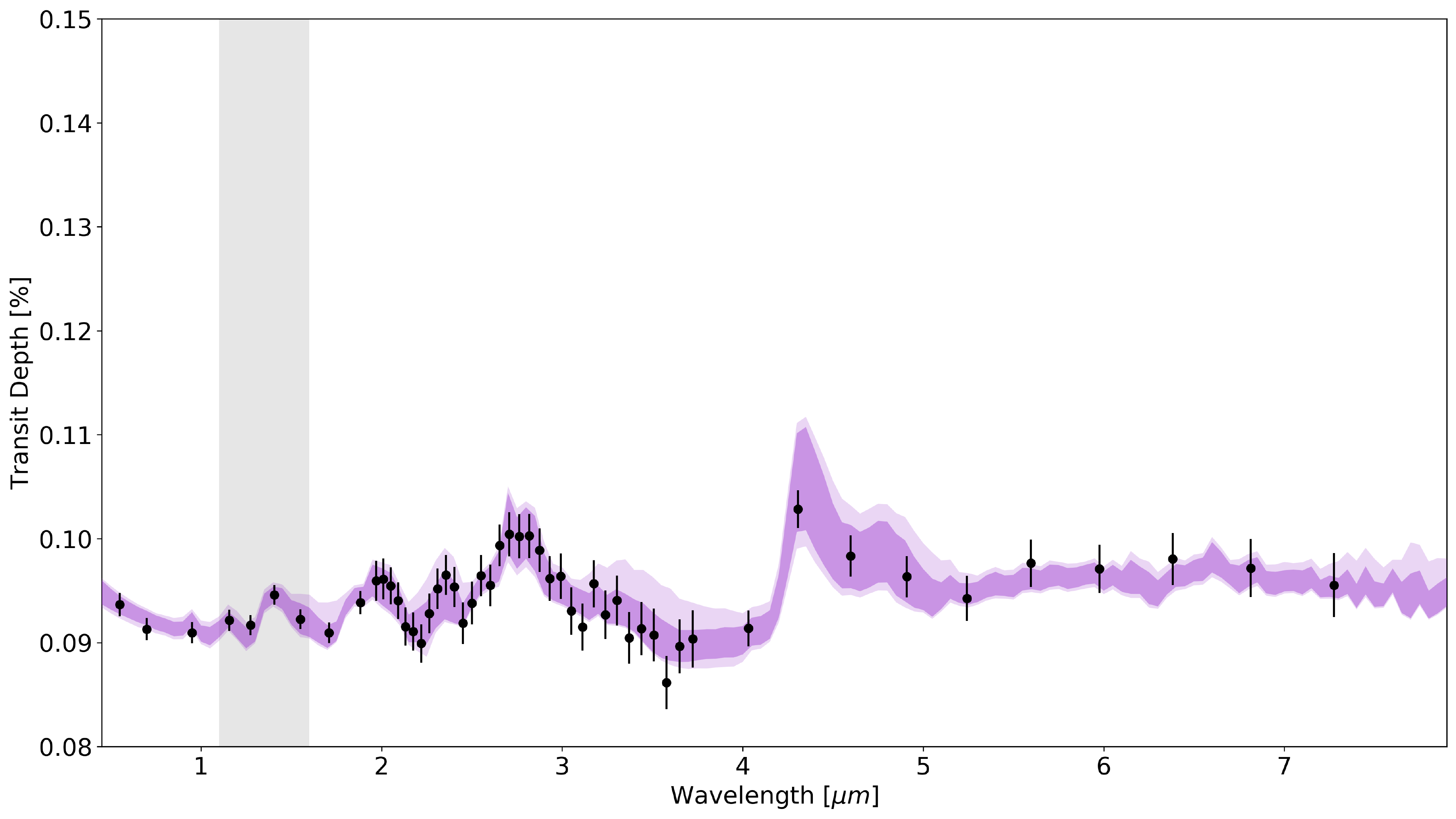}
		\end{subfigure}
		
		\caption{Simulated JWST (a and c) and Ariel (b and d) observations of the best-fit solutions, i.e. the full chemical scenario, retrieved in this work. For Ariel, 10 transits have been assumed for each planet, while JWST simulations have been performed using a single transit with NIRISS GR700XD as well as an observation with NIRSpec G395H. HD\,106315\,c and HD\,3167\,c are shown in green (a and b) and in purple (c and d), respectively.}
		\label{JWST}
	\end{figure*}
	It is evident that in the future the exoplanetary field will be based on the detailed characterisation of exo-atmospheres. In this scenario, NASA’s upcoming JWST telescope will play an important role; its large aperture, high sensitivity and wide spectral range will allow the detection of molecular species in the atmospheres of planets with different masses: from super Earths to super-Jovians. Scheduled to launch in the late 2020s, the ESA Ariel space mission will enable atmospheric characterisation of a large sample ($\sim$1000) of exoplanets in order to address how the chemical composition of an exoplanet is linked to its formation/evolution environment \citep{Tinetti2018,Ariel_mrs}.
	With this prospect in mind, HD\,106315\,c and HD\,3167\,c represent suitable targets for both these space-borne instruments and so we used the Ariel Radiometric Model (ArielRad) \citep{Mugnai} to simulate observations by Ariel. For each planet, we took the best-fit solution from the HST/WFC3 analysis to model Ariel observations at its native resolution (i.e. the TIER~3 resolution); we considered ten Ariel transits. In addition, we simulated JWST observations using ExoWebb \cite{exowebb}, assuming the collection of one single transit using NIRISS GR700XD plus a transit with NIRSpec G395M. 
	Figure~\ref{JWST}, shows, for the two planets, the results of our simulations for both JWST (left panels, a and c) and Ariel (right panels, b and d). It highlights the increased wavelength coverage and data quality that will be obtained with both Ariel and JWST. The power of having a broad wavelength coverage is that we can probe multiple absorption bands for each molecule. This helps break degeneracies due to overlapping features and always molecular compositions to be more readily constrained. Additionally, these future missions could shore up the detections of both NH$_3$ (for HD\,106315\,c), and CO$_2$ (for HD\,3167\,c): the larger the spectral range covered, the more absorption bands may be present. Namely, on one hand, JWST and Ariel could highlights the CO$_2$ absorption features between $\sim$1.7-2.0~$\mu$m and 4.0-5.32~$\mu$m; on the other hand NH$_3$ presents strong absorption features at longer wavelengths compared to the one probed with WFC3.

	\section{Summary and conclusion}\label{Conclusion}
	We presented here the analysis of HST/WFC3 spatially scanned observations of the Neptune-type HD\,106315\,c, and of the sub-Neptune HD\,3167\,c resulting in the detection of water vapor in both atmospheres. Starting from the raw data, and using the routine \verb+Iraclis+, we extracted a transmission spectrum for both planets. We then interpreted it through the use of the Bayesian spectral retrieval algorithm TauREx3. We found a statistically significant atmosphere surrounding the two planets and evaluated the strength of our detection through the ADI metric.
	
	%Retrieval models show evidence of water vapor in both the atmospheres.
	%From the TauREx HD\,106315\,c analysis, 
	From the TauREx analysis, we retrieved a `strong' detection of H$_2$O ($\log_{10}[\mathrm{H_2O}]=-2.1^{+0.7}_{-1.3}$, $\Delta_{\rm E1}$=14.21) and a `possible evidence' of NH$_3$ ($\log_{10}[\mathrm {NH_3}]=-4.3^{+0.7}_{-2.0}$, $\Delta_{\rm E1}$=0.91, even if it is not significant) in the atmosphere of HD\,106315\,c. When removing ammonia, a deep cloud deck is required to fit the spectrum. We can only put an upper bound on methane abundance ($10^{-5}$), while carbon dioxide and monoxide abundances are unconstrained.
	
	The HD\,3167\,c analysis resulted in both a water vapor ($\log_{10}[\mathrm{H_2O}]=-4.1^{+0.9}_{-0.9}$, $\Delta_{\rm E1}$=3.61) and a carbon dioxide ($\log_{10}[\mathrm {CO_2}]=-2.4^{+0.7}_{-1.0}$, $\Delta_{\rm E1}$=3.93) `moderate' detection. As CO$_2$ is not explained by 1D equilibrium chemistry models, its presence could be due to noise and highlights the limitations of our data quality. More precise constraints on the chemical abundances could be given if 3D models were employed instead of 1D ones. The shortcomings of retrieval analyses performed with 1D forward models have been highlighted already in previous papers \citep[see e.g.][]{Caldas2019}. On the contrary, if we assume a  high metallicity, CO$_2$ could actually be present in the atmosphere of HD\,3167\,c  (an increase in metallicity by a factor of x tends to increase the abundance of CO$_2$ by a factor of x$^2$, see \citealt[see e.g.][]{Moses_2014}), and what we are seeing could not be due to noise or to systematics. Thus, further observations are needed to establish whether the CO$_2$ might actually be present in the atmosphere of this exoplanet.
	
	\par
	%\medskipl
	The future is bright for atmospheric studies of exoplanets thanks to both space-based and ground-based facilities. On one hand, \citet{Cowan}, \citet{ Greene2016}, \citet{Tinetti2018}, and \citet{twinkle} have shown the potential of the JWST, Twinkle, and Ariel space missions to characterize exo-atmospheres. On the other, ground-based instruments such as the European Extremely Large Telescope (E-ELT), -and in particular the Mid-Infrared E-ELT Imager and Spectrograph (METIS) instrument \citep{Brandl2018}-, the Thirty Meter Telescope (TMT, \citealt{Skidmore2018}), and the Giant Magellan Telescope (GMT, \citealt{Fanson2018}), will become available. This will lead to the systematic study of thousands of exoplanets' day sides and terminators both at high-(HRS, from the ground) and at low-(LRS, from the space) spectral resolution. By combining HRS with LRS, and thus probing different regions of the exoplanetary atmospheres (higher atmospheric altitudes with HRS, lower atmospheric altitudes with LRS), we will better understand the atmospheric compositions, and thus be able to apply more constraints on their formation and evolution. Given the brightness of their respective host stars, and the large scale heights we computed (H$\sim$518 $\pm$ 174~km and H$\sim$171 $\pm$ 40~km for HD\,106315\,c and HD\,3167\,c, respectively), the two Neptune-like planets we studied in this paper are suitable targets for these upcoming instruments.
	
	\medskip
	\vspace{1cm}
	
	\noindent{\textbf{Acknowledgments:}}
	We want to thank the anonymous referee for the constructive comments which helped improve the quality of the manuscript.
	This work was  realised as part of ``ARES Ariel School" in Biarritz in 2019. The school was organised by JPB, AT and IW with the financial support of CNES.
	JPB acknowledge the support of the University of Tasmania through the UTAS Foundation and the endowed Warren Chair in Astronomy, Rodolphe Cledassou, Pascale Danto and Michel Viso (CNES). WP, TZ, and AYJ have received funding from the European Research Council (ERC) under the European Union's Horizon 2020 research and innovation programme (grant agreement n$^\circ$ 679030/WHIPLASH and n$^\circ$ 758892/ExoAI). SW was supported through the STFC UCL CDT in Data Intensive Science (grant number ST/P006736/1). GG acknowledges the financial support of the 2017 PhD fellowship programme of INAF. RB is a PhD fellow of the Research Foundation -- Flanders (FWO). 
	DB acknowledges financial support from the ANR project "e-PYTHEAS" (ANR-16-CE31-0005-01). LVM and DMG acknowledge the financial support of the Ariel ASI grant n. 2018-22-HH.0.
	BE, QC, MM, AT and IW acknowledge funding from the European Research Council (ERC) under the European Union's Horizon 2020 research and innovation programme grant ExoAI (GA No. 758892) and the STFC grants ST/P000282/1, ST/P002153/1, ST/S002634/1 and ST/T001836/1.
	NS acknowledges the support of the IRIS-OCAV, PSL.
	MP acknowledges support by the European Research Council under Grant Agreement ATMO 757858 and by the CNES.
	OV thank the CNRS/INSU Programme National de Plan\'etologie (PNP) and CNES for funding support.

	\vspace{3mm}
	\software {PASTIS \citep{Pastis2014}, Iraclis \citep{Iraclis}, PyLightcurve \citep{tsiaras_plc}, RATE \citep{Cubillos2019}, TauREx3 \citep{al-refaie_taurex3}, Multinest \citep{multinest}, ArielRad \citep{Mugnai}, ExoWebb \citep{exowebb}, emcee \citep{emcee}, Astropy \citep{astropy}, h5py \citep{hdf5_collette}, Matplotlib \citep{Hunter_matplotlib}, Numpy \citep{oliphant_numpy}, lmfit \citep{limfit}.}
	
	\bibliographystyle{yahapj}
	\bibliography{ms1}

\begin{thebibliography}{}
\providecommand\natexlab[1]{#1}
\providecommand\JournalTitle[1]{#1}

\bibitem[{Abel {et~al.}(2011)Abel, Frommhold, Li, \& Hunt}]{abel_h2-h2}
Abel, M., Frommhold, L., Li, X., \& Hunt, K.~L. 2011, \JournalTitle{The Journal
  of Physical Chemistry A}, 115, 6805

\bibitem[{Abel {et~al.}(2012)Abel, Frommhold, Li, \& Hunt}]{abel_h2-he}
---. 2012, \JournalTitle{The Journal of chemical physics}, 136, 044319

\bibitem[{{Al-Refaie} {et~al.}(2019){Al-Refaie}, {Changeat}, {Waldmann}, \&
  {Tinetti}}]{al-refaie_taurex3}
{Al-Refaie}, A.~F., {Changeat}, Q., {Waldmann}, I.~P., \& {Tinetti}, G. 2019,
  \JournalTitle{arXiv e-prints}, arXiv:1912.07759

\bibitem[{{Allart} {et~al.}(2018){Allart}, {Bourrier}, {Lovis}, {Ehrenreich},
  {Spake}, {Wyttenbach}, {Pino}, {Pepe}, {Sing}, \& {Lecavelier des
  Etangs}}]{Allart2018}
{Allart}, R., {Bourrier}, V., {Lovis}, C., {et~al.} 2018, \JournalTitle{arXiv
  e-prints}, \href{http://arxiv.org/abs/1812.02189}{{\sffamily arXiv:1812.02189
  [astro-ph.EP]}}

\bibitem[{{Astropy Collaboration} {et~al.}(2018){Astropy Collaboration},
  {Price-Whelan}, {Sip{\H{o}}cz}, {G{\"u}nther}, {Lim}, {Crawford}, {Conseil},
  {Shupe}, {Craig}, {Dencheva}, {Ginsburg}, {Vand erPlas}, {Bradley},
  {P{\'e}rez-Su{\'a}rez}, {de Val-Borro}, {Aldcroft}, {Cruz}, {Robitaille},
  {Tollerud}, {Ardelean}, {Babej}, {Bach}, {Bachetti}, {Bakanov}, {Bamford},
  {Barentsen}, {Barmby}, {Baumbach}, {Berry}, {Biscani}, {Boquien}, {Bostroem},
  {Bouma}, {Brammer}, {Bray}, {Breytenbach}, {Buddelmeijer}, {Burke},
  {Calderone}, {Cano Rodr{\'\i}guez}, {Cara}, {Cardoso}, {Cheedella}, {Copin},
  {Corrales}, {Crichton}, {D'Avella}, {Deil}, {Depagne}, {Dietrich}, {Donath},
  {Droettboom}, {Earl}, {Erben}, {Fabbro}, {Ferreira}, {Finethy}, {Fox},
  {Garrison}, {Gibbons}, {Goldstein}, {Gommers}, {Greco}, {Greenfield},
  {Groener}, {Grollier}, {Hagen}, {Hirst}, {Homeier}, {Horton}, {Hosseinzadeh},
  {Hu}, {Hunkeler}, {Ivezi{\'c}}, {Jain}, {Jenness}, {Kanarek}, {Kendrew},
  {Kern}, {Kerzendorf}, {Khvalko}, {King}, {Kirkby}, {Kulkarni}, {Kumar},
  {Lee}, {Lenz}, {Littlefair}, {Ma}, {Macleod}, {Mastropietro}, {McCully},
  {Montagnac}, {Morris}, {Mueller}, {Mumford}, {Muna}, {Murphy}, {Nelson},
  {Nguyen}, {Ninan}, {N{\"o}the}, {Ogaz}, {Oh}, {Parejko}, {Parley}, {Pascual},
  {Patil}, {Patil}, {Plunkett}, {Prochaska}, {Rastogi}, {Reddy Janga},
  {Sabater}, {Sakurikar}, {Seifert}, {Sherbert}, {Sherwood-Taylor}, {Shih},
  {Sick}, {Silbiger}, {Singanamalla}, {Singer}, {Sladen}, {Sooley},
  {Sornarajah}, {Streicher}, {Teuben}, {Thomas}, {Tremblay}, {Turner},
  {Terr{\'o}n}, {van Kerkwijk}, {de la Vega}, {Watkins}, {Weaver}, {Whitmore},
  {Woillez}, {Zabalza}, \& {Astropy Contributors}}]{astropy}
{Astropy Collaboration}, {Price-Whelan}, A.~M., {Sip{\H{o}}cz}, B.~M., {et~al.}
  2018, \href{http://dx.doi.org/10.3847/1538-3881/aabc4f}{\JournalTitle{\aj},
  156, 123}

\bibitem[{{Awiphan} {et~al.}(2016){Awiphan}, {Kerins}, {Pichadee},
  {Komonjinda}, {Dhillon}, {Rujopakarn}, {Poshyachinda}, {Marsh}, {Reichart},
  {Ivarsen}, \& {Haislip}}]{Awiphan2016}
{Awiphan}, S., {Kerins}, E., {Pichadee}, S., {et~al.} 2016,
  \href{http://dx.doi.org/10.1093/mnras/stw2148}{\JournalTitle{\mnras}, 463,
  2574}

\bibitem[{{Barros} {et~al.}(2017){Barros}, {Gosselin}, {Lillo-Box}, {Bayliss},
  {Delgado Mena}, {Brugger}, {Santerne}, {Armstrong}, {Adibekyan}, {Armstrong},
  {Barrado}, {Bento}, {Boisse}, {Bonomo}, {Bouchy}, {Brown}, {Cochran},
  {Collier Cameron}, {Deleuil}, {Demangeon}, {D{\'\i}az}, {Doyle}, {Dumusque},
  {Ehrenreich}, {Espinoza}, {Faedi}, {Faria}, {Figueira}, {Foxell},
  {H{\'e}brard}, {Hojjatpanah}, {Jackman}, {Lendl}, {Ligi}, {Lovis}, {Melo},
  {Mousis}, {Neal}, {Osborn}, {Pollacco}, {Santos}, {Sefako}, {Shporer},
  {Sousa}, {Triaud}, {Udry}, {Vigan}, \& {Wyttenbach}}]{Barros2017}
{Barros}, S.~C.~C., {Gosselin}, H., {Lillo-Box}, J., {et~al.} 2017,
  \href{http://dx.doi.org/10.1051/0004-6361/201731276}{\JournalTitle{\aap},
  608, A25}

\bibitem[{{Batalha} {et~al.}(2013){Batalha}, {Rowe}, {Bryson}, {Barclay},
  {Burke}, {Caldwell}, {Christiansen}, {Mullally}, {Thompson}, {Brown},
  {Dupree}, {Fabrycky}, {Ford}, {Fortney}, {Gilliland}, {Isaacson}, {Latham},
  {Marcy}, {Quinn}, {Ragozzine}, {Shporer}, {Borucki}, {Ciardi}, {Gautier},
  {Haas}, {Jenkins}, {Koch}, {Lissauer}, {Rapin}, {Basri}, {Boss}, {Buchhave},
  {Carter}, {Charbonneau}, {Christensen-Dalsgaard}, {Clarke}, {Cochran},
  {Demory}, {Desert}, {Devore}, {Doyle}, {Esquerdo}, {Everett}, {Fressin},
  {Geary}, {Girouard}, {Gould}, {Hall}, {Holman}, {Howard}, {Howell},
  {Ibrahim}, {Kinemuchi}, {Kjeldsen}, {Klaus}, {Li}, {Lucas}, {Meibom},
  {Morris}, {Pr{\v{s}}a}, {Quintana}, {Sanderfer}, {Sasselov}, {Seader},
  {Smith}, {Steffen}, {Still}, {Stumpe}, {Tarter}, {Tenenbaum}, {Torres},
  {Twicken}, {Uddin}, {Van Cleve}, {Walkowicz}, \& {Welsh}}]{Batalha2013}
{Batalha}, N.~M., {Rowe}, J.~F., {Bryson}, S.~T., {et~al.} 2013,
  \href{http://dx.doi.org/10.1088/0067-0049/204/2/24}{\JournalTitle{\apjs},
  204, 24}

\bibitem[{{Benneke} \& {Seager}(2013)}]{Benneke2013}
{Benneke}, B., \& {Seager}, S. 2013,
  \href{http://dx.doi.org/10.1088/0004-637X/778/2/153}{\JournalTitle{\apj},
  778, 153}

\bibitem[{{Benneke} {et~al.}(2019{\natexlab{a}}){Benneke}, {Knutson},
  {Lothringer}, {Crossfield}, {Moses}, {Morley}, {Kreidberg}, {Fulton},
  {Dragomir}, {Howard}, {Wong}, {D{\'e}sert}, {McCullough}, {Kempton},
  {Fortney}, {Gilliland }, {Deming}, \& {Kammer}}]{Benneke2019}
{Benneke}, B., {Knutson}, H.~A., {Lothringer}, J., {et~al.} 2019{\natexlab{a}},
  \href{http://dx.doi.org/10.1038/s41550-019-0800-5}{\JournalTitle{Nature
  Astronomy}, 3, 813}

\bibitem[{{Benneke} {et~al.}(2019{\natexlab{b}}){Benneke}, {Wong}, {Piaulet},
  {Knutson}, {Lothringer}, {Morley}, {Crossfield}, {Gao}, {Greene}, {Dressing},
  {Dragomir}, {Howard}, {McCullough}, {Kempton}, {Fortney}, \&
  {Fraine}}]{benneke2019water}
{Benneke}, B., {Wong}, I., {Piaulet}, C., {et~al.} 2019{\natexlab{b}},
  \href{http://dx.doi.org/10.3847/2041-8213/ab59dc}{\JournalTitle{\apjl}, 887,
  L14}

\bibitem[{{Borucki} {et~al.}(2011){Borucki}, {Koch}, {Basri}, {Batalha},
  {Brown}, {Bryson}, {Caldwell}, {Christensen-Dalsgaard}, {Cochran}, {DeVore},
  {Dunham}, {Gautier}, {Geary}, {Gilliland}, {Gould}, {Howell}, {Jenkins},
  {Latham}, {Lissauer}, {Marcy}, {Rowe}, {Sasselov}, {Boss}, {Charbonneau},
  {Ciardi}, {Doyle}, {Dupree}, {Ford}, {Fortney}, {Holman}, {Seager},
  {Steffen}, {Tarter}, {Welsh}, {Allen}, {Buchhave}, {Christiansen}, {Clarke},
  {Das}, {D{\'e}sert}, {Endl}, {Fabrycky}, {Fressin}, {Haas}, {Horch},
  {Howard}, {Isaacson}, {Kjeldsen}, {Kolodziejczak}, {Kulesa}, {Li}, {Lucas},
  {Machalek}, {McCarthy}, {MacQueen}, {Meibom}, {Miquel}, {Prsa}, {Quinn},
  {Quintana}, {Ragozzine}, {Sherry}, {Shporer}, {Tenenbaum}, {Torres},
  {Twicken}, {Van Cleve}, {Walkowicz}, {Witteborn}, \& {Still}}]{Borucki2011}
{Borucki}, W.~J., {Koch}, D.~G., {Basri}, G., {et~al.} 2011,
  \href{http://dx.doi.org/10.1088/0004-637X/736/1/19}{\JournalTitle{\apj}, 736,
  19}

\bibitem[{{Bourrier} {et~al.}(2016){Bourrier}, {Lecavelier des Etangs},
  {Ehrenreich}, {Tanaka}, \& {Vidotto}}]{Bourrier2016}
{Bourrier}, V., {Lecavelier des Etangs}, A., {Ehrenreich}, D., {Tanaka}, Y.~A.,
  \& {Vidotto}, A.~A. 2016,
  \href{http://dx.doi.org/10.1051/0004-6361/201628362}{\JournalTitle{\aap},
  591, A121}

\bibitem[{{Bourrier} {et~al.}(2018){Bourrier}, {Lecavelier des Etangs},
  {Ehrenreich}, {Sanz-Forcada}, {Allart}, {Ballester}, {Buchhave}, {Cohen},
  {Deming}, {Evans}, {Garc{\'\i}a Mu{\~n}oz}, {Henry}, {Kataria}, {Lavvas},
  {Lewis}, {L{\'o}pez-Morales}, {Marley}, {Sing}, \& {Wakeford}}]{Bourrier2018}
{Bourrier}, V., {Lecavelier des Etangs}, A., {Ehrenreich}, D., {et~al.} 2018,
  \href{http://dx.doi.org/10.1051/0004-6361/201833675}{\JournalTitle{\aap},
  620, A147}

\bibitem[{{Brandl} {et~al.}(2018){Brandl}, {Absil}, {Ag{\'o}cs}, {Baccichet},
  {Bertram}, {Bettonvil}, {van Boekel}, {Burtscher}, {van Dishoeck}, {Feldt},
  {Garcia}, {Glasse}, {Glauser}, {G{\"u}del}, {Haupt}, {Kenworthy}, {Labadie},
  {Laun}, {Lesman}, {Pantin}, {Quanz}, {Snellen}, {Siebenmorgen}, \& {van
  Winckel}}]{Brandl2018}
{Brandl}, B.~R., {Absil}, O., {Ag{\'o}cs}, T., {et~al.} 2018,
  \href{http://dx.doi.org/10.1117/12.2311492}{in Society of Photo-Optical
  Instrumentation Engineers (SPIE) Conference Series, Vol. 10702, \procspie},
  107021U

\bibitem[{{Brown} {et~al.}(2001){Brown}, {Charbonneau}, {Gilliland}, {Noyes},
  \& {Burrows}}]{Brown2001}
{Brown}, T.~M., {Charbonneau}, D., {Gilliland}, R.~L., {Noyes}, R.~W., \&
  {Burrows}, A. 2001,
  \href{http://dx.doi.org/10.1086/320580}{\JournalTitle{\apj}, 552, 699}

\bibitem[{{Caldas} {et~al.}(2019){Caldas}, {Leconte}, {Selsis}, {Waldmann},
  {Bord{\'e}}, {Rocchetto}, \& {Charnay}}]{Caldas2019}
{Caldas}, A., {Leconte}, J., {Selsis}, F., {et~al.} 2019,
  \href{http://dx.doi.org/10.1051/0004-6361/201834384}{\JournalTitle{\aap},
  623, A161}

\bibitem[{Chachan {et~al.}(2019)Chachan, Knutson, Gao, Kataria, Wong, Henry,
  Benneke, Zhang, Barstow, Bean, Mikal-Evans, Lewis, Mansfield,
  L{\'{o}}pez-Morales, Nikolov, Sing, \& Wakeford}]{Chachan_2019}
Chachan, Y., Knutson, H.~A., Gao, P., {et~al.} 2019,
  \href{http://dx.doi.org/10.3847/1538-3881/ab4e9a}{\JournalTitle{The
  Astronomical Journal}, 158, 244}

\bibitem[{Changeat {et~al.}(2019)Changeat, Edwards, Waldmann, \&
  Tinetti}]{Changeat_2019}
Changeat, Q., Edwards, B., Waldmann, I.~P., \& Tinetti, G. 2019,
  \href{http://dx.doi.org/10.3847/1538-4357/ab4a14}{\JournalTitle{The
  Astrophysical Journal}, 886, 39}

\bibitem[{{Christiansen} {et~al.}(2017){Christiansen}, {Vanderburg}, {Burt},
  {Fulton}, {Batygin}, {Benneke}, {Brewer}, {Charbonneau}, {Ciardi}, {Collier
  Cameron}, {Coughlin}, {Crossfield}, {Dressing}, {Greene}, {Howard}, {Latham},
  {Molinari}, {Mortier}, {Mullally}, {Pepe}, {Rice}, {Sinukoff}, {Sozzetti},
  {Thompson}, {Udry}, {Vogt}, {Barman}, {Batalha}, {Bouchy}, {Buchhave},
  {Butler}, {Cosentino}, {Dupuy}, {Ehrenreich}, {Fiorenzano}, {Hansen},
  {Henning}, {Hirsch}, {Holden}, {Isaacson}, {Johnson}, {Knutson}, {Kosiarek},
  {L{\'o}pez-Morales}, {Lovis}, {Malavolta}, {Mayor}, {Micela}, {Motalebi},
  {Petigura}, {Phillips}, {Piotto}, {Rogers}, {Sasselov}, {Schlieder},
  {S{\'e}gransan}, {Watson}, \& {Weiss}}]{Christiansen2017}
{Christiansen}, J.~L., {Vanderburg}, A., {Burt}, J., {et~al.} 2017,
  \href{http://dx.doi.org/10.3847/1538-3881/aa832d}{\JournalTitle{\aj}, 154,
  122}

\bibitem[{{Claret}(2000)}]{2000A&A...363.1081C}
{Claret}, A. 2000, \JournalTitle{A\&A}, 363, 1081

\bibitem[{Collette(2013)}]{hdf5_collette}
Collette, A. 2013, Python and HDF5 (O'Reilly)

\bibitem[{Cowan {et~al.}(2015)Cowan, Greene, Angerhausen, Batalha, Clampin,
  Colón, Crossfield, Fortney, Gaudi, Harrington, Iro, Lillie, Linsky,
  Lopez-Morales, Mandell, \& Stevenson}]{Cowan}
Cowan, N.~B., Greene, T., Angerhausen, D., {et~al.} 2015,
  \href{http://dx.doi.org/10.1086/680855}{\JournalTitle{Astrophysics - Earth
  and Planetary Astrophysics}, 127, 311}

\bibitem[{Crossfield {et~al.}(2011)Crossfield, Barman, \&
  Hansen}]{Crossfield2011}
Crossfield, I. J.~M., Barman, T., \& Hansen, B. M.~S. 2011,
  \href{http://dx.doi.org/10.1088/0004-637X/736/2/132}{\JournalTitle{\apj},
  736, 17}

\bibitem[{{Crossfield} \& {Kreidberg}(2017)}]{CrossfieldTrends}
{Crossfield}, I.~J.~M., \& {Kreidberg}, L. 2017,
  \href{http://dx.doi.org/10.3847/1538-3881/aa9279}{\JournalTitle{\aj}, 154, 6}

\bibitem[{{Crossfield} {et~al.}(2017){Crossfield}, {Ciardi}, {Isaacson},
  {Howard}, {Petigura}, {Weiss}, {Fulton}, {Sinukoff}, {Schlieder}, {Mawet},
  {Ruane}, {de Pater}, {de Kleer}, {Davies}, {Christiansen}, {Dressing},
  {Hirsch}, {Benneke}, {Crepp}, {Kosiarek}, {Livingston}, {Gonzales},
  {Beichman}, \& {Knutson}}]{Crossfield2017}
{Crossfield}, I.~J.~M., {Ciardi}, D.~R., {Isaacson}, H., {et~al.} 2017,
  \href{http://dx.doi.org/10.3847/1538-3881/aa6e01}{\JournalTitle{\aj}, 153,
  255}

\bibitem[{{Cubillos} {et~al.}(2019){Cubillos}, {Blecic}, \&
  {Dobbs-Dixon}}]{Cubillos2019}
{Cubillos}, P.~E., {Blecic}, J., \& {Dobbs-Dixon}, I. 2019,
  \href{http://dx.doi.org/10.3847/1538-4357/aafda2}{\JournalTitle{\apj}, 872,
  111}

\bibitem[{{D{\'\i}az} {et~al.}(2014){D{\'\i}az}, {Almenara}, {Santerne},
  {Moutou}, {Lethuillier}, \& {Deleuil}}]{Pastis2014}
{D{\'\i}az}, R.~F., {Almenara}, J.~M., {Santerne}, A., {et~al.} 2014,
  \href{http://dx.doi.org/10.1093/mnras/stu601}{\JournalTitle{\mnras}, 441,
  983}

\bibitem[{{Dressing} \& {Charbonneau}(2013)}]{Dressing2013}
{Dressing}, C.~D., \& {Charbonneau}, D. 2013,
  \href{http://dx.doi.org/10.1088/0004-637X/767/1/95}{\JournalTitle{\apj}, 767,
  95}

\bibitem[{Edwards {et~al.}(2020)Edwards, Al-Refaie, Lagage, \&
  Gastaud}]{exowebb}
Edwards, B., Al-Refaie, A., Lagage, P., \& Gastaud, R. 2020, \JournalTitle{in
  prep}

\bibitem[{{Edwards} {et~al.}(2019{\natexlab{a}}){Edwards}, {Mugnai}, {Tinetti},
  {Pascale}, \& {Sarkar}}]{Ariel_mrs}
{Edwards}, B., {Mugnai}, L., {Tinetti}, G., {Pascale}, E., \& {Sarkar}, S.
  2019{\natexlab{a}},
  \href{http://dx.doi.org/10.3847/1538-3881/ab1cb9}{\JournalTitle{\aj}, 157,
  242}

\bibitem[{{Edwards} {et~al.}(2019{\natexlab{b}}){Edwards}, {Rice}, {Zingales},
  {Tessenyi}, {Waldmann}, {Tinetti}, {Pascale}, {Savini}, \&
  {Sarkar}}]{twinkle}
{Edwards}, B., {Rice}, M., {Zingales}, T., {et~al.} 2019{\natexlab{b}},
  \href{http://dx.doi.org/10.1007/s10686-018-9611-4}{\JournalTitle{Experimental
  Astronomy}, 47, 29}

\bibitem[{{Edwards} {et~al.}(2020){Edwards}, {Changeat}, {Baeyens}, {Tsiaras},
  {Al-Refaie}, {Taylor}, {Yip}, {Bieger}, {Blain}, {Gressier}, {Guilluy},
  {Jaziri}, {Kiefer}, {Modirrousta-Galian}, {Morvan}, {Mugnai}, {Pluriel},
  {Poveda}, {Skaf}, {Whiteford}, {Wright}, {Zingales}, {Charnay}, {Drossart},
  {Leconte}, {Venot}, {Waldmann}, \& {Beaulieu}}]{Edwards2020}
{Edwards}, B., {Changeat}, Q., {Baeyens}, R., {et~al.} 2020,
  \href{http://dx.doi.org/10.3847/1538-3881/ab9225}{\JournalTitle{\aj}, 160, 8}

\bibitem[{{Fanson} {et~al.}(2018){Fanson}, {McCarthy}, {Bernstein}, {Angeli},
  {Ashby}, {Bigelow}, {Bouchez}, {Burgett}, {Chauvin}, {Contos}, {Figueroa},
  {Gray}, {Groark}, {Laskin}, {Millan-Gabet}, {Rakich}, {Sandoval}, {Pi}, \&
  {Wheeler}}]{Fanson2018}
{Fanson}, J., {McCarthy}, P.~J., {Bernstein}, R., {et~al.} 2018,
  \href{http://dx.doi.org/10.1117/12.2313340}{in Society of Photo-Optical
  Instrumentation Engineers (SPIE) Conference Series, Vol. 10700, \procspie},
  1070012

\bibitem[{{Feroz} {et~al.}(2009){Feroz}, {Hobson}, \& {Bridges}}]{multinest}
{Feroz}, F., {Hobson}, M.~P., \& {Bridges}, M. 2009,
  \href{http://dx.doi.org/10.1111/j.1365-2966.2009.14548.x}{\JournalTitle{\mnras},
  398, 1601}

\bibitem[{{Fisher} \& {Heng}(2018)}]{Fisher2018}
{Fisher}, C., \& {Heng}, K. 2018,
  \href{http://dx.doi.org/10.1093/mnras/sty2550}{\JournalTitle{\mnras}, 481,
  4698}

\bibitem[{Fletcher {et~al.}(2018)Fletcher, Gustafsson, \&
  Orton}]{fletcher_h2-h2}
Fletcher, L.~N., Gustafsson, M., \& Orton, G.~S. 2018, \JournalTitle{The
  Astrophysical Journal Supplement Series}, 235, 24

\bibitem[{{Foreman-Mackey} {et~al.}(2013){Foreman-Mackey}, {Hogg}, {Lang}, \&
  {Goodman}}]{emcee}
{Foreman-Mackey}, D., {Hogg}, D.~W., {Lang}, D., \& {Goodman}, J. 2013,
  \href{http://dx.doi.org/10.1086/670067}{\JournalTitle{\pasp}, 125, 306}

\bibitem[{{Fraine} {et~al.}(2014){Fraine}, {Deming}, {Benneke}, {Knutson},
  {Jord{\'a}n}, {Espinoza}, {Madhusudhan}, {Wilkins}, \&
  {Todorov}}]{Fraine2014}
{Fraine}, J., {Deming}, D., {Benneke}, B., {et~al.} 2014,
  \href{http://dx.doi.org/10.1038/nature13785}{\JournalTitle{\nat}, 513, 526}

\bibitem[{{Fressin} {et~al.}(2013){Fressin}, {Torres}, {Charbonneau}, {Bryson},
  {Christiansen}, {Dressing}, {Jenkins}, {Walkowicz}, \&
  {Batalha}}]{Fressin2013}
{Fressin}, F., {Torres}, G., {Charbonneau}, D., {et~al.} 2013,
  \href{http://dx.doi.org/10.1088/0004-637X/766/2/81}{\JournalTitle{\apj}, 766,
  81}

\bibitem[{Fulton {et~al.}(2017)Fulton, Petigura, Howard, Isaacson, Marcy,
  Cargile, Hebb, Weiss, Johnson, Morton, Sinukoff, Crossfield, \&
  Hirsch}]{Fulton2017}
Fulton, B., Petigura, E., Howard, A., {et~al.} 2017,
  \href{http://dx.doi.org/10.3847/1538-3881/aa80eb}{\JournalTitle{The
  Astronomical Journal}, 154}

\bibitem[{{Fulton} \& {Petigura}(2018)}]{FultonePetigura2018}
{Fulton}, B.~J., \& {Petigura}, E.~A. 2018,
  \href{http://dx.doi.org/10.3847/1538-3881/aae828}{\JournalTitle{\aj}, 156,
  264}

\bibitem[{{Gandolfi} {et~al.}(2017){Gandolfi}, {Barrag{\'a}n}, {Hatzes},
  {Fridlund}, {Fossati}, {Donati}, {Johnson}, {Nowak}, {Prieto-Arranz},
  {Albrecht}, {Dai}, {Deeg}, {Endl}, {Grziwa}, {Hjorth}, {Korth}, {Nespral},
  {Saario}, {Smith}, {Antoniciello}, {Alarcon}, {Bedell}, {Blay}, {Brems},
  {Cabrera}, {Csizmadia}, {Cusano}, {Cochran}, {Eigm{\"u}ller}, {Erikson},
  {Gonz{\'a}lez Hern{\'a}ndez}, {Guenther}, {Hirano}, {Su{\'a}rez
  Mascare{\~n}o}, {Narita}, {Palle}, {Parviainen}, {P{\"a}tzold}, {Persson},
  {Rauer}, {Saviane}, {Schmidtobreick}, {Van Eylen}, {Winn}, \&
  {Zakhozhay}}]{Gandolfi2017}
{Gandolfi}, D., {Barrag{\'a}n}, O., {Hatzes}, A.~P., {et~al.} 2017,
  \href{http://dx.doi.org/10.3847/1538-3881/aa832a}{\JournalTitle{\aj}, 154,
  123}

\bibitem[{{Greene} {et~al.}(2016){Greene}, {Line}, {Montero}, {Fortney},
  {Lustig-Yaeger}, \& {Luther}}]{Greene2016}
{Greene}, T.~P., {Line}, M.~R., {Montero}, C., {et~al.} 2016,
  \href{http://dx.doi.org/10.3847/0004-637X/817/1/17}{\JournalTitle{\apj}, 817,
  17}

\bibitem[{{Harps{\o}e} {et~al.}(2013){Harps{\o}e}, {Hardis}, {Hinse},
  {J{\o}rgensen}, {Mancini}, {Southworth}, {Alsubai}, {Bozza}, {Browne},
  {Burgdorf}, {Calchi Novati}, {Dodds}, {Dominik}, {Fang}, {Finet}, {Gerner},
  {Gu}, {Hundertmark}, {Jessen-Hansen}, {Kains}, {Kerins}, {Kjeldsen},
  {Liebig}, {Lund}, {Lundkvist}, {Mathiasen}, {Nesvorn{\'y}}, {Nikolov},
  {Penny}, {Proft}, {Rahvar}, {Ricci}, {Sahu}, {Scarpetta}, {Sch{\"a}fer},
  {Sch{\"o}nebeck}, {Snodgrass}, {Skottfelt}, {Surdej}, {Tregloan-Reed}, \&
  {Wertz}}]{Harpsoe2013}
{Harps{\o}e}, K.~B.~W., {Hardis}, S., {Hinse}, T.~C., {et~al.} 2013,
  \href{http://dx.doi.org/10.1051/0004-6361/201219996}{\JournalTitle{\aap},
  549, A10}

\bibitem[{{Houk} \& {Swift}(1999)}]{Houk1999}
{Houk}, N., \& {Swift}, C. 1999, \JournalTitle{Michigan Spectral Survey}, 5, 0

\bibitem[{{Howard} {et~al.}(2012){Howard}, {Marcy}, {Bryson}, {Jenkins},
  {Rowe}, {Batalha}, {Borucki}, {Koch}, {Dunham}, {Gautier}, {Van Cleve},
  {Cochran}, {Latham}, {Lissauer}, {Torres}, {Brown}, {Gilliland}, {Buchhave},
  {Caldwell}, {Christensen-Dalsgaard}, {Ciardi}, {Fressin}, {Haas}, {Howell},
  {Kjeldsen}, {Seager}, {Rogers}, {Sasselov}, {Steffen}, {Basri},
  {Charbonneau}, {Christiansen}, {Clarke}, {Dupree}, {Fabrycky}, {Fischer},
  {Ford}, {Fortney}, {Tarter}, {Girouard}, {Holman}, {Johnson}, {Klaus},
  {Machalek}, {Moorhead}, {Morehead}, {Ragozzine}, {Tenenbaum}, {Twicken},
  {Quinn}, {Isaacson}, {Shporer}, {Lucas}, {Walkowicz}, {Welsh}, {Boss},
  {Devore}, {Gould}, {Smith}, {Morris}, {Prsa}, {Morton}, {Still}, {Thompson},
  {Mullally}, {Endl}, \& {MacQueen}}]{Howard2012}
{Howard}, A.~W., {Marcy}, G.~W., {Bryson}, S.~T., {et~al.} 2012,
  \href{http://dx.doi.org/10.1088/0067-0049/201/2/15}{\JournalTitle{\apjs},
  201, 15}

\bibitem[{{Howarth}(2011)}]{2011MNRAS.413.1515H}
{Howarth}, I.~D. 2011,
  \href{http://dx.doi.org/doi:10.1111/j.1365-2966.2011.18122.x}{\JournalTitle{MNRAS},
  413, 1515}

\bibitem[{Hunter(2007)}]{Hunter_matplotlib}
Hunter, J.~D. 2007,
  \href{http://dx.doi.org/10.1109/MCSE.2007.55}{\JournalTitle{Computing in
  Science \& Engineering}, 9, 90}

\bibitem[{{Kass} \& {Raftery}(1995)}]{Kass1995}
{Kass}, R.~E., \& {Raftery}, A.~E. 1995,
  \href{http://dx.doi.org/10.1080/01621459.1995.10476572}{\JournalTitle{Journal
  of the American Statistical Association}, 90:430, 773}

\bibitem[{{Knutson} {et~al.}(2014{\natexlab{a}}){Knutson}, {Benneke}, {Deming},
  \& {Homeier}}]{knutsonGJ}
{Knutson}, H.~A., {Benneke}, B., {Deming}, \& {Homeier}, D. 2014{\natexlab{a}},
  \href{http://dx.doi.org/10.1038/nature12887}{\JournalTitle{\nat}, 505, 66}

\bibitem[{{Knutson} {et~al.}(2014{\natexlab{b}}){Knutson}, {Dragomir},
  {Kreidberg}, {Kempton}, {McCullough}, {Fortney}, {Bean}, {Gillon}, {Homeier},
  \& {Howard}}]{knutsonHD}
{Knutson}, H.~A., {Dragomir}, D., {Kreidberg}, L., {et~al.} 2014{\natexlab{b}},
  \href{http://dx.doi.org/10.1088/0004-637X/794/2/155}{\JournalTitle{\aj},
  794}, \href{http://arxiv.org/abs/1403.4602}{{\sffamily arXiv:1403.4602
  [astro-ph]}}

\bibitem[{{Kreidberg} {et~al.}(2014{\natexlab{a}}){Kreidberg}, {Bean},
  {D{\'e}sert}, {Line}, {Fortney}, {Madhusudhan}, {Stevenson}, {Showman},
  {Charbonneau}, {McCullough}, {Seager}, {Burrows}, {Henry}, {Williamson},
  {Kataria}, \& {Homeier}}]{Kreidberg_metallicity}
{Kreidberg}, L., {Bean}, J.~L., {D{\'e}sert}, J.-M., {et~al.}
  2014{\natexlab{a}},
  \href{http://dx.doi.org/10.1088/2041-8205/793/2/L27}{\JournalTitle{\apjl},
  793, L27}

\bibitem[{{Kreidberg} {et~al.}(2014{\natexlab{b}}){Kreidberg}, {Bean},
  {Désert}, {Benneke}, {Deming}, {Stevenson}, {Seager}, {Berta-Thompson},
  {Seifahrt}, \& {Homeier}}]{kreidbergGJ}
{Kreidberg}, L., {Bean}, J.~L., {Désert}, J.~M., {et~al.} 2014{\natexlab{b}},
  \href{http://dx.doi.org/10.1038/nature12888}{\JournalTitle{\nat}, 505, 69}

\bibitem[{{Kreidberg} {et~al.}(2014{\natexlab{c}}){Kreidberg}, {Bean},
  {D{\'e}sert}, {Benneke}, {Deming}, {Stevenson}, {Seager}, {Berta-Thompson},
  {Seifahrt}, \& {Homeier}}]{KreidbergB2014b}
{Kreidberg}, L., {Bean}, J.~L., {D{\'e}sert}, J.-M., {et~al.}
  2014{\natexlab{c}},
  \href{http://dx.doi.org/10.1038/nature12888}{\JournalTitle{\nat}, 505, 69}

\bibitem[{{Kreidberg} {et~al.}(2020){Kreidberg}, {Molli{\`e}re}, {Crossfield},
  {Thorngren}, {Kawashima}, {Morley}, {Benneke}, {Mikal-Evans}, {Berardo},
  {Kosiarek}, {Gorjian}, {Ciardi}, {Christiansen}, {Dragomir}, {Dressing},
  {Fortney}, {Fulton}, {Greene}, {Hardegree-Ullman}, {Howard}, {Howell},
  {Isaacson}, {Krick}, {Livingston}, {Lothringer}, {Morales}, {Petigura},
  {Rodriguez}, {Schlieder}, \& {Weiss}}]{Kreidberg_2020}
{Kreidberg}, L., {Molli{\`e}re}, P., {Crossfield}, I. J.~M., {et~al.} 2020,
  \JournalTitle{arXiv e-prints}, arXiv:2006.07444

\bibitem[{{Kundurthy} {et~al.}(2011){Kundurthy}, {Agol}, {Becker}, {Barnes},
  {Williams}, \& {Mukadam}}]{Kundurthy2011}
{Kundurthy}, P., {Agol}, E., {Becker}, A.~C., {et~al.} 2011,
  \href{http://dx.doi.org/10.1088/0004-637X/731/2/123}{\JournalTitle{\apj},
  731, 123}

\bibitem[{{Kurucz}(1970)}]{1970SAOSR.309.....K}
{Kurucz}, R.~L. 1970, \JournalTitle{SAO Special Report}

\bibitem[{{L{\'e}ger} {et~al.}(2004){L{\'e}ger}, {Selsis}, {Sotin}, {Guillot},
  {Despois}, {Mawet}, {Ollivier}, {Lab{\`e}que}, {Valette}, {Brachet},
  {Chazelas}, \& {Lammer}}]{Leger2004}
{L{\'e}ger}, A., {Selsis}, F., {Sotin}, C., {et~al.} 2004,
  \href{http://dx.doi.org/10.1016/j.icarus.2004.01.001}{\JournalTitle{\icarus},
  169, 499}

\bibitem[{{Lendl} {et~al.}(2017){Lendl}, {Ehrenreich}, {Turner}, {Bayliss},
  {Blanco-Cuaresma}, {Giles}, {Bouchy}, {Marmier}, \& {Udry}}]{Lendl2017}
{Lendl}, M., {Ehrenreich}, D., {Turner}, O.~D., {et~al.} 2017,
  \href{http://dx.doi.org/10.1051/0004-6361/201731278}{\JournalTitle{\aap},
  603, L5}

\bibitem[{Li {et~al.}(2015)Li, Gordon, Rothman, Tan, Hu, Kassi, Campargue, \&
  Medvedev}]{li_co_2015}
Li, G., Gordon, I.~E., Rothman, L.~S., {et~al.} 2015, \JournalTitle{The
  Astrophysical Journal Supplement Series}, 216, 15

\bibitem[{{Libby-Roberts} {et~al.}(2020){Libby-Roberts}, {Berta-Thompson},
  {Désert}, {Masuda}, {Morley}, {Lopez}, {Deck}, {Fabrycky}, {Fortney},
  {Line}, {Sanchis-Ojeda}, \& {Winn}}]{Roberts2020}
{Libby-Roberts}, J.~E., {Berta-Thompson}, Z.~K., {Désert}, J.~M., {et~al.}
  2020, \href{http://dx.doi.org/10.3847/1538-3881/ab5d36}{\JournalTitle{\aj},
  159, 29}

\bibitem[{{Lodders} \& {Fegley}(2006)}]{lodders}
{Lodders}, K., \& {Fegley}, B.~J. 2006,
  \href{http://dx.doi.org/10.1007/3-540-30313-8_1}{\JournalTitle{Springer
  praxis book}}

\bibitem[{{MacDonald} {et~al.}(2020){MacDonald}, {Goyal}, \&
  {Lewis}}]{MacDonald2020}
{MacDonald}, R.~J., {Goyal}, J.~M., \& {Lewis}, N.~K. 2020,
  \href{http://dx.doi.org/10.3847/2041-8213/ab8238}{\JournalTitle{\apjl}, 893,
  L43}

\bibitem[{{MacDonald} \& {Madhusudhan}(2019)}]{C_O}
{MacDonald}, R.~J., \& {Madhusudhan}, N. 2019,
  \href{http://dx.doi.org/10.1093/mnras/stz789}{\JournalTitle{\mnras}, 486,
  1292}

\bibitem[{{Maciejewski} {et~al.}(2014){Maciejewski}, {Niedzielski}, {Nowak},
  {Pall{\'e}}, {Tingley}, {Errmann}, \& {Neuh{\"a}user}}]{Maciejewski2014}
{Maciejewski}, G., {Niedzielski}, A., {Nowak}, G., {et~al.} 2014,
  \JournalTitle{\actaa}, 64, 323

\bibitem[{{Madhusudhan} \& {Seager}(2011)}]{GJ436b_2011}
{Madhusudhan}, N., \& {Seager}, S. 2011,
  \href{http://dx.doi.org/10.1088/0004-637X/729/1/41}{\JournalTitle{\apj}, 729,
  41}

\bibitem[{{Mansfield} {et~al.}(2018){Mansfield}, {Bean}, {Oklop{\v c}i{\'c}},
  {Kreidberg}, {D{\'e}sert}, {Kempton}, {Line}, {Fortney}, {Henry}, {Mallonn},
  {Stevenson}, {Dragomir}, {Allart}, \& {Bourrier}}]{Mansfield2018}
{Mansfield}, M., {Bean}, J.~L., {Oklop{\v c}i{\'c}}, A., {et~al.} 2018,
  \href{http://dx.doi.org/10.3847/2041-8213/aaf166}{\JournalTitle{\apjl}, 868,
  L34}

\bibitem[{{Markwardt}(2009)}]{Markwardt}
{Markwardt}, C.~B. 2009, \JournalTitle{ASP}, 411, 261

\bibitem[{Mikal-Evans \& et~al.({submitted})}]{Evans2020}
Mikal-Evans, T., \& et~al. {submitted}

\bibitem[{{Morley} {et~al.}(2012){Morley}, {Fortney}, {Marley}, {Visscher},
  {Saumon}, \& {Leggett}}]{morley}
{Morley}, C.~V., {Fortney}, J.~J., {Marley}, M.~S., {et~al.} 2012,
  \href{http://dx.doi.org/10.1088/0004-637X/756/2/172}{\JournalTitle{\aj},
  756}, \href{http://arxiv.org/abs/1206.4313}{{\sffamily arXiv:1206.4313
  [astro-ph]}}

\bibitem[{{Moses} {et~al.}({in press}){Moses}, {Cavalié}, \&
  {Cavalié}}]{Moses2020}
{Moses}, J., {Cavalié}, T., \& {Cavalié}, M.~T. {in press},
  \JournalTitle{{Phil. Trans. R. Soc. A}}

\bibitem[{Moses(2014)}]{Moses_2014}
Moses, J.~I. 2014,
  \href{http://dx.doi.org/10.1098/rsta.2013.0073}{\JournalTitle{Philosophical
  Transactions of the Royal Society A: Mathematical, Physical and Engineering
  Sciences}, 372, 20130073}

\bibitem[{Mousis {et~al.}(2020)Mousis, Deleuil, Aguichine, Marcq, Naar,
  Aguirre, Brugger, \& Gonçalves}]{Mousis_2020}
Mousis, O., Deleuil, M., Aguichine, A., {et~al.} 2020,
  \href{http://dx.doi.org/10.3847/2041-8213/ab9530}{\JournalTitle{\apjl}, 896}

\bibitem[{{Mugnai} {et~al.}(2020){Mugnai}, {Pascale}, {Edwards},
  {Papageorgiou}, \& {Sarkar}}]{Mugnai}
{Mugnai}, L.~V., {Pascale}, E., {Edwards}, B., {Papageorgiou}, A., \& {Sarkar},
  S. 2020,
  \href{http://dx.doi.org/10.1007/s10686-020-09676-7}{\JournalTitle{Experimental
  Astronomy}}, \href{http://arxiv.org/abs/2009.07824}{{\sffamily
  arXiv:2009.07824 [astro-ph.IM]}}

\bibitem[{{Newville} {et~al.}(2019){Newville}, {Otten}, {Nelson}, {Ingargiola},
  {Stensitzki}, {Allan}, {Fox}, {Carter}, {Micha{\l}}, {Pustakhod}, {Ram},
  {Glenn}, {Deil}, {Stuermer}, {Beelen}, {Frost}, {Zobrist}, {Pasquevich},
  {Hansen}, {Spillane}, {Caldwell}, {Polloreno}, {Andrewhannum}, {Zimmermann},
  {Borreguero}, {Fraine}, {Deep-42-Thought}, {Maier}, {Gamari}, \&
  {Almarza}}]{limfit}
{Newville}, M., {Otten}, R., {Nelson}, A., {et~al.} 2019, {lmfit/lmfit-py
  1.0.0}

\bibitem[{Oliphant(2006)}]{oliphant_numpy}
Oliphant, T.~E. 2006, A guide to NumPy, Vol.~1 (Trelgol Publishing USA)

\bibitem[{{Petigura} {et~al.}(2013){Petigura}, {Howard}, \&
  {Marcy}}]{Petigura2013}
{Petigura}, E.~A., {Howard}, A.~W., \& {Marcy}, G.~W. 2013,
  \href{http://dx.doi.org/10.1073/pnas.1319909110}{\JournalTitle{Proceedings of
  the National Academy of Science}, 110, 19273}

\bibitem[{{Petigura} {et~al.}(2017){Petigura}, {Howard}, {Marcy}, {Johnson},
  {Isaacson}, {Cargile}, {Hebb}, {Fulton}, {Weiss}, {Morton}, {Winn}, {Rogers},
  {Sinukoff}, {Hirsch}, \& {Crossfield}}]{Petigura2017}
{Petigura}, E.~A., {Howard}, A.~W., {Marcy}, G.~W., {et~al.} 2017,
  \href{http://dx.doi.org/10.3847/1538-3881/aa80de}{\JournalTitle{\aj}, 154,
  107}

\bibitem[{{Pluriel} {et~al.}(2020{\natexlab{a}}){Pluriel}, {Zingales},
  {Leconte}, \& {Parmentier}}]{Pluriel2020}
{Pluriel}, W., {Zingales}, T., {Leconte}, J., \& {Parmentier}, V.
  2020{\natexlab{a}},
  \href{http://dx.doi.org/10.1051/0004-6361/202037678}{\JournalTitle{\aap},
  636, A66}

\bibitem[{{Pluriel} {et~al.}(2020{\natexlab{b}}){Pluriel}, {Whiteford},
  {Edwards}, {Changeat}, {Yip}, {Baeyens}, {Al-Refaie}, {Fabienne Bieger},
  {Blain}, {Gressier}, {Guilluy}, {Yassin Jaziri}, {Kiefer},
  {Modirrousta-Galian}, {Morvan}, {Mugnai}, {Poveda}, {Skaf}, {Zingales},
  {Wright}, {Charnay}, {Drossart}, {Leconte}, {Tsiaras}, {Venot}, {Waldmann},
  \& {Beaulieu}}]{Pluriel2020_ares}
{Pluriel}, W., {Whiteford}, N., {Edwards}, B., {et~al.} 2020{\natexlab{b}},
  \href{http://dx.doi.org/10.3847/1538-3881/aba000}{\JournalTitle{\aj}, 160,
  112}

\bibitem[{Polyansky {et~al.}(2018)Polyansky, Kyuberis, Zobov, Tennyson,
  Yurchenko, \& Lodi}]{polyansky_h2o}
Polyansky, O.~L., Kyuberis, A.~A., Zobov, N.~F., {et~al.} 2018,
  \JournalTitle{Monthly Notices of the Royal Astronomical Society}, 480, 2597

\bibitem[{Rocchetto {et~al.}(2016)Rocchetto, Waldmann, Venot, Lagage, \&
  Tinetti}]{Rocchetto_2016}
Rocchetto, M., Waldmann, I.~P., Venot, O., Lagage, P.-O., \& Tinetti, G. 2016,
  \href{http://dx.doi.org/10.3847/1538-4357/833/1/120}{\JournalTitle{The
  Astrophysical Journal}, 833, 120}

\bibitem[{{Rodriguez} {et~al.}(2017){Rodriguez}, {Zhou}, {Vanderburg},
  {Eastman}, {Kreidberg}, {Cargile}, {Bieryla}, {Latham}, {Irwin}, {Mayo},
  {Calkins}, {Esquerdo}, \& {Mink}}]{Rodriguez2017}
{Rodriguez}, J.~E., {Zhou}, G., {Vanderburg}, A., {et~al.} 2017,
  \href{http://dx.doi.org/10.3847/1538-3881/aa6dfb}{\JournalTitle{\aj}, 153,
  256}

\bibitem[{{Rogers}(2015)}]{Rogers2015}
{Rogers}, L.~A. 2015,
  \href{http://dx.doi.org/10.1088/0004-637X/801/1/41}{\JournalTitle{\apj}, 801,
  41}

\bibitem[{{Rogers} {et~al.}(2011){Rogers}, {Bodenheimer}, {Lissauer}, \&
  {Seager}}]{Rogers2011}
{Rogers}, L.~A., {Bodenheimer}, P., {Lissauer}, J.~J., \& {Seager}, S. 2011,
  \href{http://dx.doi.org/10.1088/0004-637X/738/1/59}{\JournalTitle{\apj}, 738,
  59}

\bibitem[{{Rogers} \& {Seager}(2010{\natexlab{a}})}]{Rogers2010A}
{Rogers}, L.~A., \& {Seager}, S. 2010{\natexlab{a}},
  \href{http://dx.doi.org/10.1088/0004-637X/712/2/974}{\JournalTitle{\apj},
  712, 974}

\bibitem[{{Rogers} \& {Seager}(2010{\natexlab{b}})}]{Rogers2010B}
---. 2010{\natexlab{b}},
  \href{http://dx.doi.org/10.1088/0004-637X/716/2/1208}{\JournalTitle{\apj},
  716, 1208}

\bibitem[{{Rothman} {et~al.}(2010){Rothman}, {Gordon}, {Barber}, {Dothe},
  {Gamache}, {Goldman}, {Perevalov}, {Tashkun}, \& {Tennyson}}]{COandCO2}
{Rothman}, L.~S., {Gordon}, I.~E., {Barber}, R.~J., {et~al.} 2010,
  \href{http://dx.doi.org/10.1016/j.jqsrt.2010.05.001}{\JournalTitle{\jqsrt},
  111, 2139}

\bibitem[{{Sarkis} {et~al.}(2018){Sarkis}, {Henning}, {K{\"u}rster},
  {Trifonov}, {Zechmeister}, {Tal-Or}, {Anglada-Escud{\'e}}, {Hatzes},
  {Lafarga}, {Dreizler}, {Ribas}, {Caballero}, {Reiners}, {Mallonn}, {Morales},
  {Kaminski}, {Aceituno}, {Amado}, {B{\'e}jar}, {Hagen}, {Jeffers},
  {Quirrenbach}, {Launhardt}, {Marvin}, \& {Montes}}]{Sarkis2018}
{Sarkis}, P., {Henning}, T., {K{\"u}rster}, M., {et~al.} 2018,
  \href{http://dx.doi.org/10.3847/1538-3881/aac108}{\JournalTitle{\aj}, 155,
  257}

\bibitem[{{Skaf} {et~al.}(2020){Skaf}, {Bieger}, {Edwards}, {Changeat},
  {Morvan}, {Kiefer}, {Blain}, {Zingales}, {Poveda}, {Al-Refaie}, {Baeyens},
  {Gressier}, {Guilluy}, {Jaziri}, {Modirrousta-Galian}, {Mugnai}, {Pluriel},
  {Whiteford}, {Wright}, {Yip}, {Charnay}, {Leconte}, {Drossart}, {Tsiaras},
  {Venot}, {Waldmann}, \& {Beaulieu}}]{Skaf2020}
{Skaf}, N., {Bieger}, M.~F., {Edwards}, B., {et~al.} 2020,
  \href{http://dx.doi.org/10.3847/1538-3881/ab94a3}{\JournalTitle{\aj}, 160,
  109}

\bibitem[{{Skidmore} {et~al.}(2018){Skidmore}, {Anupama}, \&
  {Srianand}}]{Skidmore2018}
{Skidmore}, W., {Anupama}, G.~C., \& {Srianand}, R. 2018, \JournalTitle{arXiv
  e-prints}, arXiv:1806.02481

\bibitem[{{Stassun} {et~al.}(2017){Stassun}, {Collins}, \&
  {Gaudi}}]{Stassun2017}
{Stassun}, K.~G., {Collins}, K.~A., \& {Gaudi}, B.~S. 2017,
  \href{http://dx.doi.org/10.3847/1538-3881/aa5df3}{\JournalTitle{\aj}, 153,
  136}

\bibitem[{{Tinetti} {et~al.}(2018){Tinetti}, {Drossart}, {Eccleston},
  {Hartogh}, {Heske}, {Leconte}, {Micela}, {Ollivier}, {Pilbratt}, {Puig},
  {Turrini}, {Vandenbussche}, {Wolkenberg}, {Beaulieu}, {Buchave}, {Ferus},
  {Griffin}, {Guedel}, {Justtanont}, {Lagage}, {Machado}, {Malaguti}, {Min},
  {N{\o}rgaard-Nielsen}, {Rataj}, {Ray}, {Ribas}, {Swain}, {Szabo}, {Werner},
  {Barstow}, {Burleigh}, {Cho}, {du Foresto}, {Coustenis}, {Decin}, {Encrenaz},
  {Galand }, {Gillon}, {Helled}, {Morales}, {Mu{\~n}oz}, {Moneti}, {Pagano},
  {Pascale}, {Piccioni}, {Pinfield}, {Sarkar}, {Selsis}, {Tennyson}, {Triaud},
  {Venot}, {Waldmann}, {Waltham}, {Wright}, {Amiaux}, {Augu{\`e}res},
  {Berth{\'e}}, {Bezawada}, {Bishop}, {Bowles}, {Coffey}, {Colom{\'e}},
  {Crook}, {Crouzet}, {Da Peppo}, {Sanz}, {Focardi}, {Frericks}, {Hunt},
  {Kohley}, {Middleton}, {Morgante}, {Ottensamer}, {Pace}, {Pearson},
  {Stamper}, {Symonds}, {Rengel}, {Renotte}, {Ade}, {Affer}, {Alard}, {Allard},
  {Altieri}, {Andr{\'e}}, {Arena}, {Argyriou}, {Aylward}, {Baccani}, {Bakos},
  {Banaszkiewicz}, {Barlow}, {Batista}, {Bellucci}, {Benatti}, {Bernardi},
  {B{\'e}zard}, {Blecka}, {Bolmont}, {Bonfond}, {Bonito}, {Bonomo}, {Brucato},
  {Brun}, {Bryson}, {Bujwan}, {Casewell}, {Charnay}, {Pestellini}, {Chen},
  {Ciaravella}, {Claudi}, {Cl{\'e}dassou}, {Damasso}, {Damiano}, {Danielski},
  {Deroo}, {Di Giorgio}, {Dominik}, {Doublier}, {Doyle}, {Doyon}, {Drummond},
  {Duong}, {Eales}, {Edwards}, {Farina}, {Flaccomio}, {Fletcher}, {Forget},
  {Fossey}, {Fr{\"a}nz}, {Fujii}, {Garc{\'\i}a-Piquer}, {Gear}, {Geoffray},
  {G{\'e}rard}, {Gesa}, {Gomez}, {Graczyk}, {Griffith}, {Grodent}, {Guarcello},
  {Gustin}, {Hamano}, {Hargrave}, {Hello}, {Heng}, {Herrero}, {Hornstrup},
  {Hubert}, {Ida}, {Ikoma}, {Iro}, {Irwin}, {Jarchow}, {Jaubert}, {Jones},
  {Julien}, {Kameda}, {Kerschbaum}, {Kervella}, {Koskinen}, {Krijger}, {Krupp},
  {Lafarga}, {Landini}, {Lellouch}, {Leto}, {Luntzer}, {Rank-L{\"u}ftinger},
  {Maggio}, {Maldonado}, {Maillard}, {Mall}, {Marquette}, {Mathis}, {Maxted},
  {Matsuo}, {Medvedev}, {Miguel}, {Minier}, {Morello}, {Mura}, {Narita},
  {Nascimbeni}, {Nguyen Tong}, {Noce}, {Oliva}, {Palle}, {Palmer}, {Pancrazzi},
  {Papageorgiou}, {Parmentier}, {Perger}, {Petralia}, {Pezzuto},
  {Pierrehumbert}, {Pillitteri}, {Piotto}, {Pisano}, {Prisinzano}, {Radioti},
  {R{\'e}ess}, {Rezac}, {Rocchetto}, {Rosich}, {Sanna}, {Santerne}, {Savini},
  {Scandariato}, {Sicardy}, {Sierra}, {Sindoni}, {Skup}, {Snellen}, {Sobiecki},
  {Soret}, {Sozzetti}, {Stiepen}, {Strugarek}, {Taylor}, {Taylor}, {Terenzi},
  {Tessenyi}, {Tsiaras}, {Tucker}, {Valencia}, {Vasisht}, {Vazan}, {Vilardell},
  {Vinatier}, {Viti}, {Waters}, {Wawer}, {Wawrzaszek}, {Whitworth}, {Yung},
  {Yurchenko}, {Osorio}, {Zellem}, {Zingales}, \& {Zwart}}]{Tinetti2018}
{Tinetti}, G., {Drossart}, P., {Eccleston}, P., {et~al.} 2018,
  \href{http://dx.doi.org/10.1007/s10686-018-9598-x}{\JournalTitle{Experimental
  Astronomy}, 46, 135}

\bibitem[{{Tsiaras} {et~al.}(2016{\natexlab{a}}){Tsiaras}, {Waldmann},
  {Rocchetto}, {Varley}, {Morello}, {Damiano}, \& {Tinetti}}]{tsiaras_plc}
{Tsiaras}, A., {Waldmann}, I., {Rocchetto}, M., {et~al.} 2016{\natexlab{a}},
  ascl:1612.018

\bibitem[{{Tsiaras} {et~al.}(2016{\natexlab{b}}){Tsiaras}, {Waldmann},
  {Rocchetto}, {Varley}, {Morello}, {Damiano}, \& {Tinetti}}]{Tsiaras2016}
{Tsiaras}, A., {Waldmann}, I.~P., {Rocchetto}, M., {et~al.} 2016{\natexlab{b}},
  \href{http://dx.doi.org/10.3847/0004-637X/832/2/202}{\JournalTitle{\apj},
  832, 202}

\bibitem[{{Tsiaras} {et~al.}(2019){Tsiaras}, {Waldmann}, {Tinetti}, {Tennyson},
  \& {Yurchenko}}]{TsiarasK218b}
{Tsiaras}, A., {Waldmann}, I.~P., {Tinetti}, G., {Tennyson}, J., \&
  {Yurchenko}, S.~N. 2019,
  \href{http://dx.doi.org/10.1038/s41550-019-0878-9}{\JournalTitle{Nature
  Astronomy}, 451}

\bibitem[{{Tsiaras} {et~al.}(2016{\natexlab{c}}){Tsiaras}, {Rocchetto},
  {Waldmann}, {Venot}, {Varley}, {Morello}, {Damiano}, {Tinetti}, {Barton},
  {Yurchenko}, \& {Tennyson}}]{Tsiaras2016b}
{Tsiaras}, A., {Rocchetto}, M., {Waldmann}, I.~P., {et~al.} 2016{\natexlab{c}},
  \href{http://dx.doi.org/10.3847/0004-637X/820/2/99}{\JournalTitle{\apj}, 820,
  99}

\bibitem[{{Tsiaras} {et~al.}(2018{\natexlab{a}}){Tsiaras}, {Waldmann},
  {Zingales}, {Rocchetto}, {Morello}, {Damiano}, {Karpouzas}, {Tinetti},
  {McKemmish}, {Tennyson}, \& {Yurchenko}}]{Iraclis}
{Tsiaras}, A., {Waldmann}, I.~P., {Zingales}, T., {et~al.} 2018{\natexlab{a}},
  \href{http://dx.doi.org/10.3847/1538-3881/aaaf75}{\JournalTitle{\aj}, 155,
  156}

\bibitem[{{Tsiaras} {et~al.}(2018{\natexlab{b}}){Tsiaras}, {Waldmann},
  {Zingales}, {Rocchetto}, {Morello}, {Damiano}, {Karpouzas}, {Tinetti},
  {McKemmish}, {Tennyson}, \& {Yurchenko}}]{Tsiaras2018}
---. 2018{\natexlab{b}},
  \href{http://dx.doi.org/10.3847/1538-3881/aaaf75}{\JournalTitle{\aj}, 155,
  156}

\bibitem[{{Valencia} {et~al.}(2006){Valencia}, {O'Connell}, \&
  {Sasselov}}]{Valencia2006}
{Valencia}, D., {O'Connell}, R.~J., \& {Sasselov}, D. 2006,
  \href{http://dx.doi.org/10.1016/j.icarus.2005.11.021}{\JournalTitle{\icarus},
  181, 545}

\bibitem[{{Van Grootel} {et~al.}(2014){Van Grootel}, {Gillon}, {Valencia},
  {Madhusudhan}, {Dragomir}, {Howe}, {Burrows}, {Demory}, {Deming},
  {Ehrenreich}, {Lovis}, {Mayor}, {Pepe}, {Queloz}, {Scuflaire}, {Seager},
  {Segransan}, \& {Udry}}]{VanGrootel2014}
{Van Grootel}, V., {Gillon}, M., {Valencia}, D., {et~al.} 2014,
  \href{http://dx.doi.org/10.1088/0004-637X/786/1/2}{\JournalTitle{\apj}, 786,
  2}

\bibitem[{{Vanderburg} {et~al.}(2016){Vanderburg}, {Bieryla}, {Duev},
  {Jensen-Clem}, {Latham}, {Mayo}, {Baranec}, {Berlind}, {Kulkarni}, {Law},
  {Nieberding}, {Riddle}, \& {Salama}}]{Vanderburg2016}
{Vanderburg}, A., {Bieryla}, A., {Duev}, D.~A., {et~al.} 2016,
  \href{http://dx.doi.org/10.3847/2041-8205/829/1/L9}{\JournalTitle{\apjl},
  829, L9}

\bibitem[{{Venot} {et~al.}(2020){Venot}, {Cavali{\'e}}, {Bounaceur},
  {Tremblin}, {Brouillard}, \& {Lhoussaine Ben Brahim}}]{Venot2020}
{Venot}, O., {Cavali{\'e}}, T., {Bounaceur}, R., {et~al.} 2020,
  \href{http://dx.doi.org/10.1051/0004-6361/201936697}{\JournalTitle{\aap},
  634, A78}

\bibitem[{Waldmann {et~al.}(2015{\natexlab{a}})Waldmann, Rocchetto, Tinetti,
  Barton, Yurchenko, \& Tennyson}]{Waldmann_2015_2}
Waldmann, I.~P., Rocchetto, M., Tinetti, G., {et~al.} 2015{\natexlab{a}},
  \href{http://dx.doi.org/10.1088/0004-637x/813/1/13}{\JournalTitle{The
  Astrophysical Journal}, 813, 13}

\bibitem[{Waldmann {et~al.}(2015{\natexlab{b}})Waldmann, Tinetti, Rocchetto,
  Barton, Yurchenko, \& Tennyson}]{Waldmann_2015}
Waldmann, I.~P., Tinetti, G., Rocchetto, M., {et~al.} 2015{\natexlab{b}},
  \href{http://dx.doi.org/10.1088/0004-637x/802/2/107}{\JournalTitle{The
  Astrophysical Journal}, 802, 107}

\bibitem[{Yurchenko {et~al.}(2011)Yurchenko, Barber, \& Tennyson}]{ExoMol_NH3}
Yurchenko, S.~N., Barber, R.~J., \& Tennyson, J. 2011,
  \href{http://dx.doi.org/10.1111/j.1365-2966.2011.18261.x}{\JournalTitle{MNRAS},
  413, 1828}

\bibitem[{{Yurchenko} \& {Tennyson}(2014)}]{CH4}
{Yurchenko}, S.~N., \& {Tennyson}, J. 2014,
  \href{http://dx.doi.org/10.1093/mnras/stu326}{\JournalTitle{\mnras}, 440,
  1649}

\bibitem[{{Zeng} \& {Sasselov}(2013)}]{Zeng2013}
{Zeng}, L., \& {Sasselov}, D. 2013,
  \href{http://dx.doi.org/10.1086/669163}{\JournalTitle{\pasp}, 125, 227}

\bibitem[{{Zeng} {et~al.}(2016){Zeng}, {Sasselov}, \& {Jacobsen}}]{Zeng2016}
{Zeng}, L., {Sasselov}, D.~D., \& {Jacobsen}, S.~B. 2016,
  \href{http://dx.doi.org/10.3847/0004-637X/819/2/127}{\JournalTitle{\apj},
  819, 127}

\bibitem[{{Zeng} {et~al.}(2019{\natexlab{a}}){Zeng}, {Jacobsen}, {Sasselov},
  {Petaev}, {Vanderburg}, {Lopez-Morales}, {Perez-Mercader}, {Mattsson}, {Li},
  {Heising}, {Bonomo}, {Damasso}, {Berger}, {Cao}, {Levi}, \&
  {Wordsworth}}]{Zeng2019_waterworlds}
{Zeng}, L., {Jacobsen}, S.~B., {Sasselov}, D.~D., {et~al.} 2019{\natexlab{a}},
  \href{http://dx.doi.org/10.1073/pnas.1812905116}{\JournalTitle{Proceedings of
  the National Academy of Science}, 116, 9723}

\bibitem[{{Zeng} {et~al.}(2019{\natexlab{b}}){Zeng}, {Jacobsen}, {Sasselov},
  {Petaev}, {Vanderburg}, {Lopez-Morales}, {Perez-Mercader}, {Mattsson}, {Li},
  {Heising}, {Bonomo}, {Damasso}, {Berger}, {Cao}, {Levi}, \&
  {Wordsworth}}]{Zeng2019}
---. 2019{\natexlab{b}},
  \href{http://dx.doi.org/10.1073/pnas.1812905116}{\JournalTitle{Proceedings of
  the National Academy of Science}, 116, 9723}

\bibitem[{{Zhou} {et~al.}(2018){Zhou}, {Rodriguez}, {Vanderburg}, {Quinn},
  {Irwin}, {Huang}, {Latham}, {Bieryla}, {Esquerdo}, {Berlind}, \&
  {Calkins}}]{Zhou2018}
{Zhou}, G., {Rodriguez}, J.~E., {Vanderburg}, A., {et~al.} 2018,
  \href{http://dx.doi.org/10.3847/1538-3881/aad085}{\JournalTitle{\aj}, 156,
  93}

\end{thebibliography}
	
	\appendix
	
	\renewcommand{\thefigure}{\thesection.\arabic{figure}}
	\counterwithin{figure}{section}
	\renewcommand{\thetable}{A\arabic{table}}
	\setcounter{table}{0}
	%\counterwithin{figure}{section}
	%\appendix
	\section{Additional Figures}
\begin{figure*}[!ht]
		\raggedright
		\begin{subfigure}[]{}	
			\includegraphics[width=17.5cm]{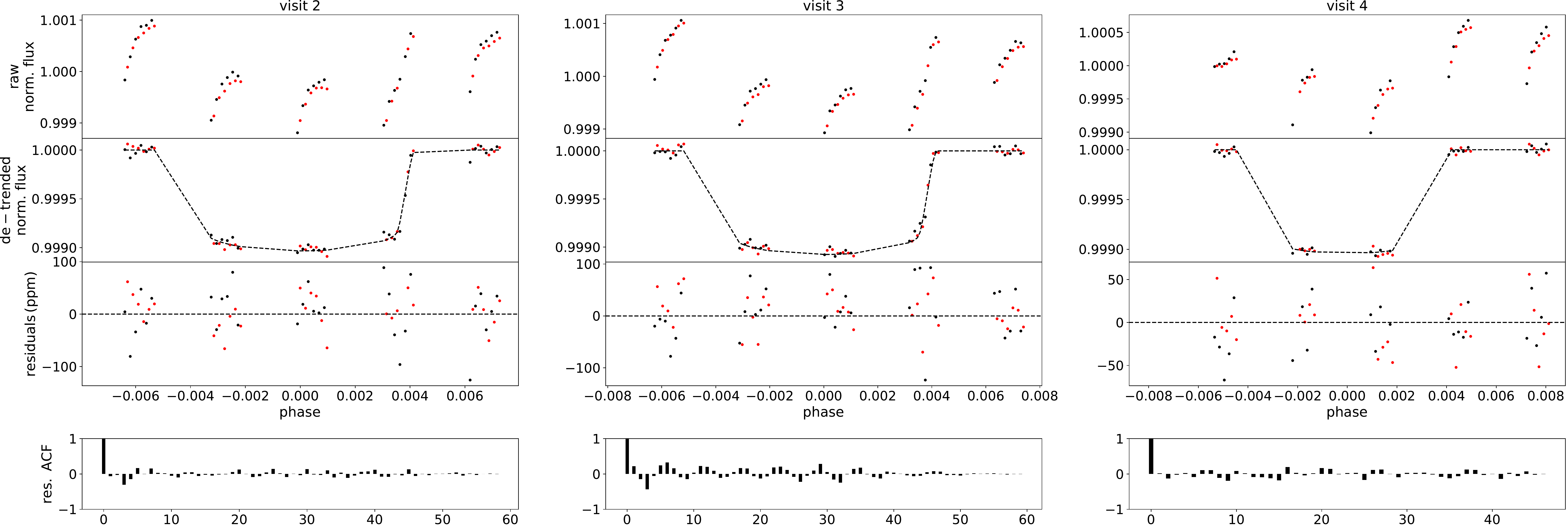}
		\end{subfigure} 
		\\
		\begin{subfigure}[]{}
			\raggedright
			\includegraphics[width=17.5cm]{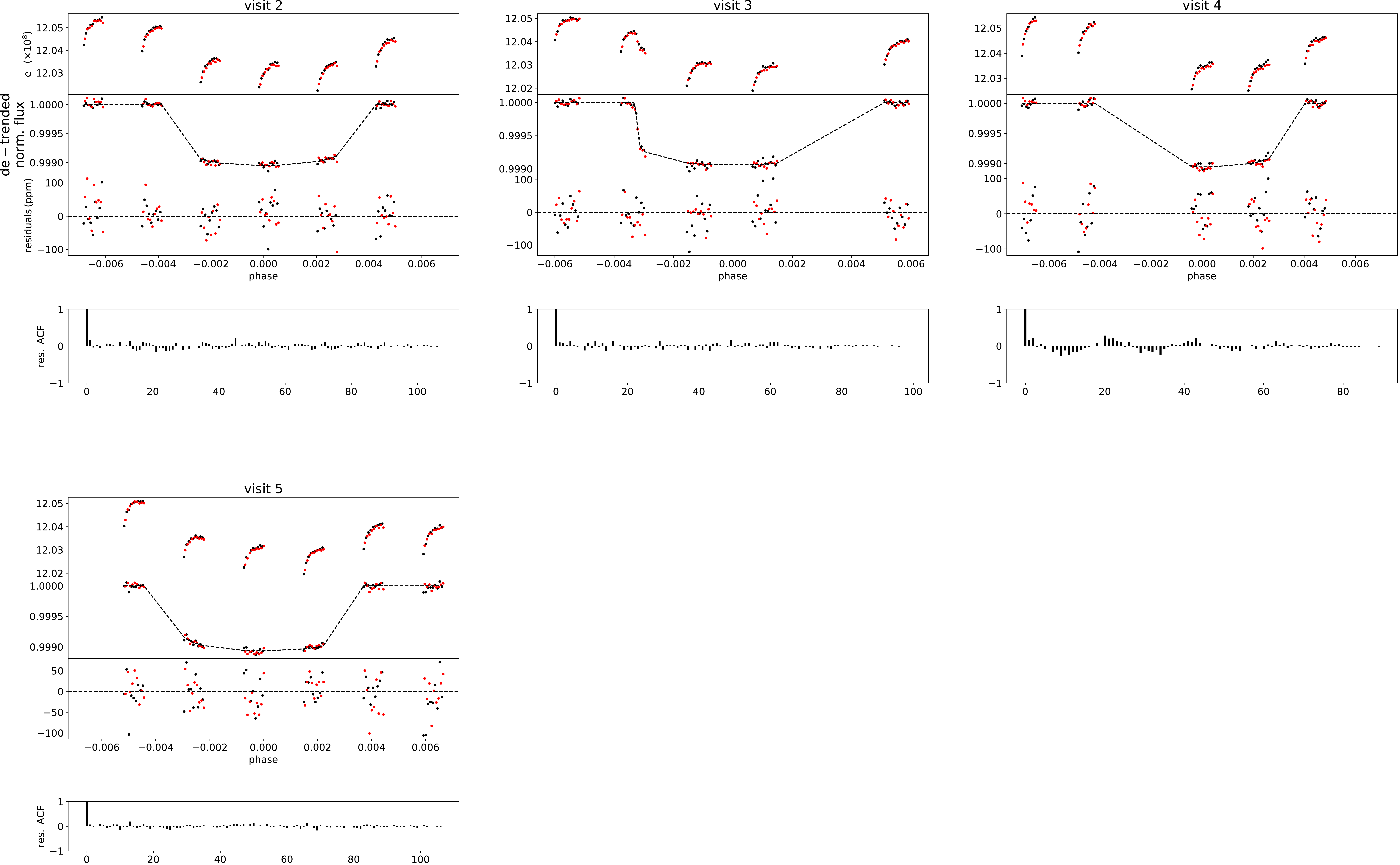}
		\end{subfigure}
		\caption{Same as Figure~\ref{fig2} for the other transits both for HD\,106315\,c (a panels) and for HD\,3167\,c (b panels).}
		\label{figA1}
	\end{figure*}

	In this appendix section, in Figure~\ref{figA1} we show the results of the white light-curve analysis for the transits not reported in Figure~\ref{fig2} for both the two exoplanets analysed in this work, whilst in Figure~\ref{figA2} we plot the HD\,3167\,c's orbits that showed contamination from HD\,3167\,b. The last Figure of the appendix (Figure~\ref{FigA3}) shows the posterior distribution we obtained by including also N$_2$ in the retrieval analysis of HD\,106315\,c.
	
	\begin{figure*}[!ht]
		\centering
		\begin{subfigure}[]{}	
			\includegraphics[width=11.5cm]{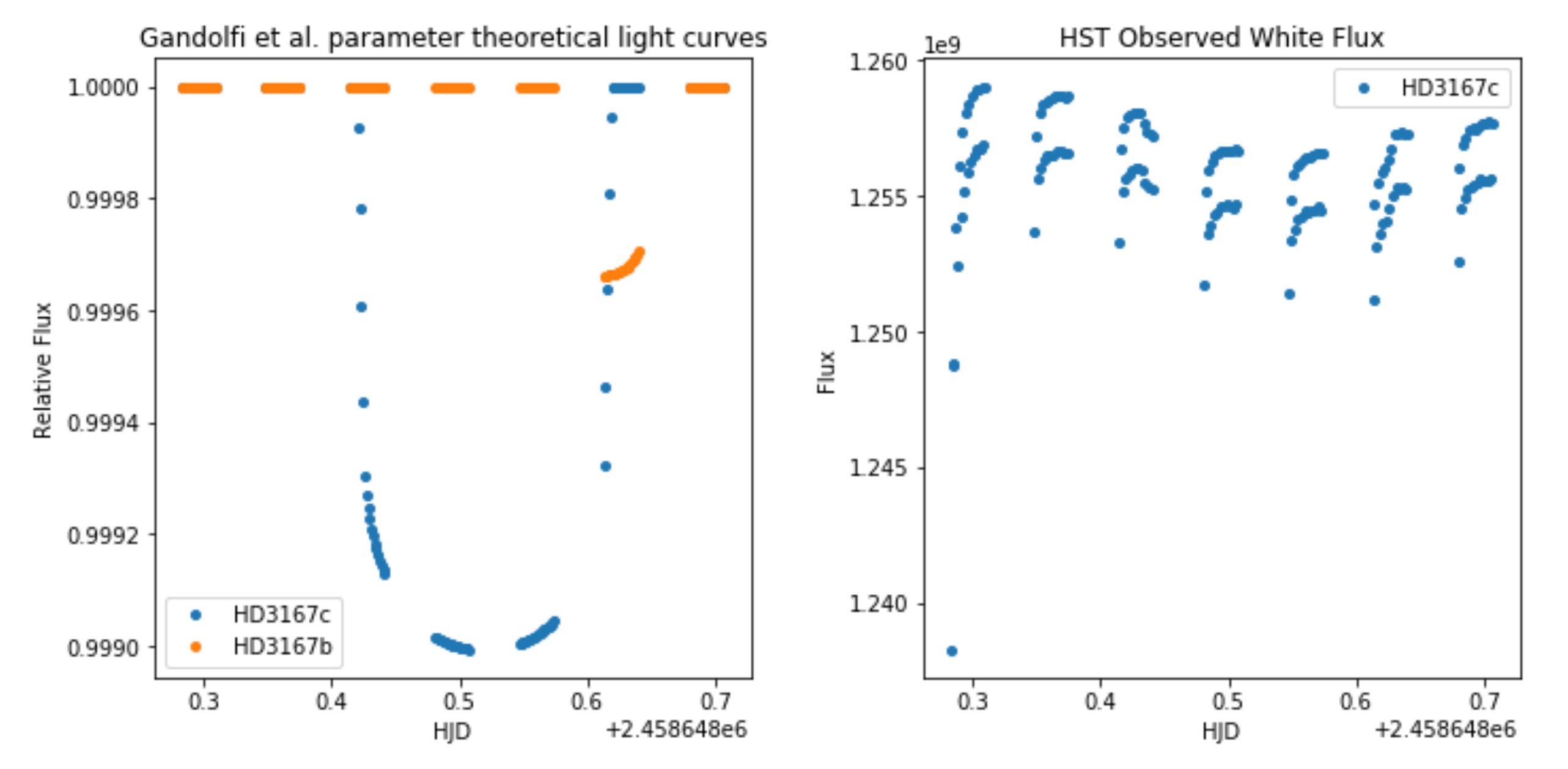}
		\end{subfigure} 
		\\
		\begin{subfigure}[]{}
			\centering
			\includegraphics[width=11.5cm]{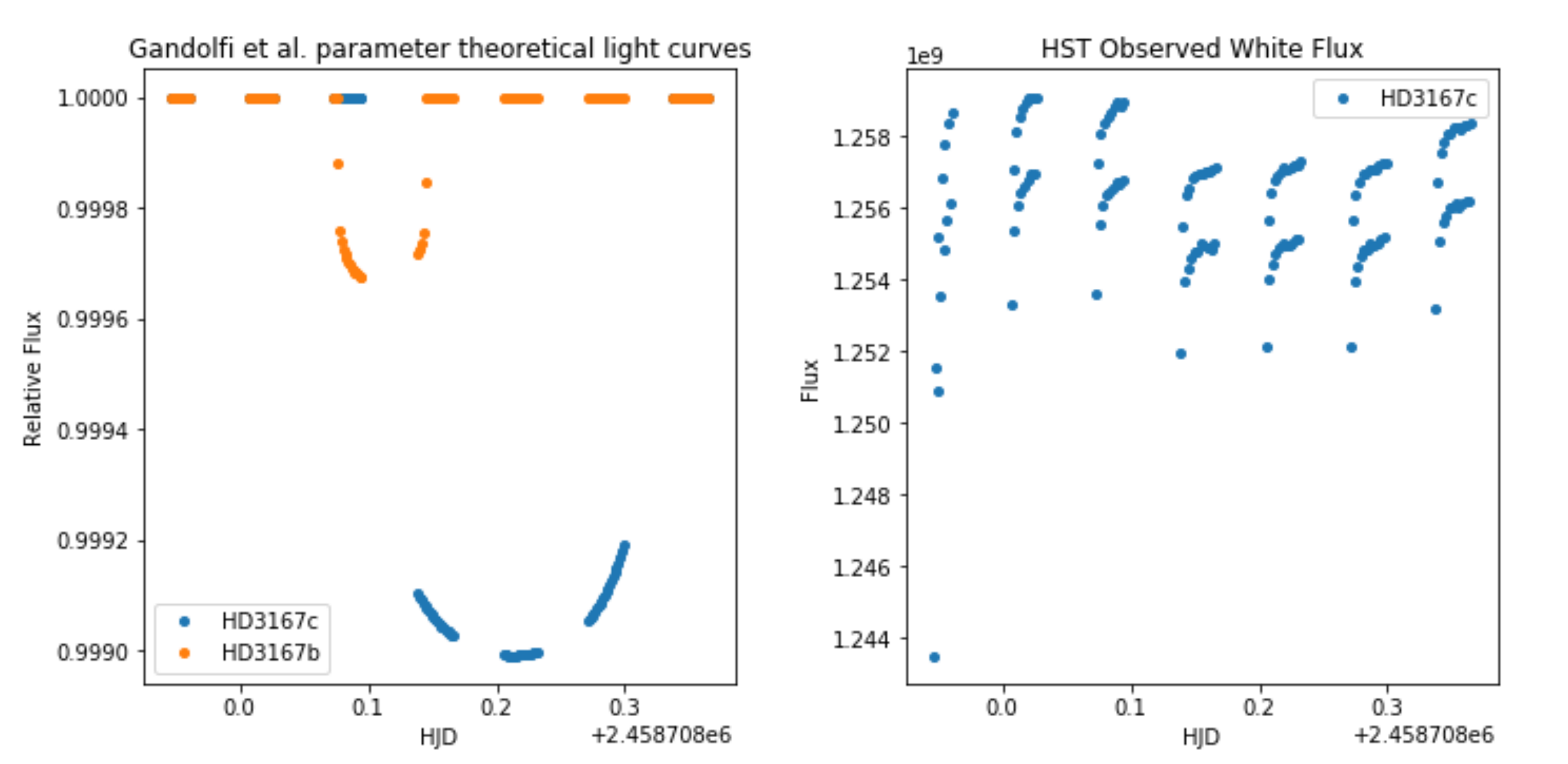}
		\end{subfigure}
		\caption{ HD\,3167\,c's orbits that showed contamination from HD\,3167\,b. Visit 3 and the impact on its sixth orbit is shown in a), while visit 4 and the contamination of its third orbit is plotted in b). }
		\label{figA2}
	\end{figure*}
	
	\begin{figure*}[!ht]
		\centering
		\includegraphics[width=\linewidth]{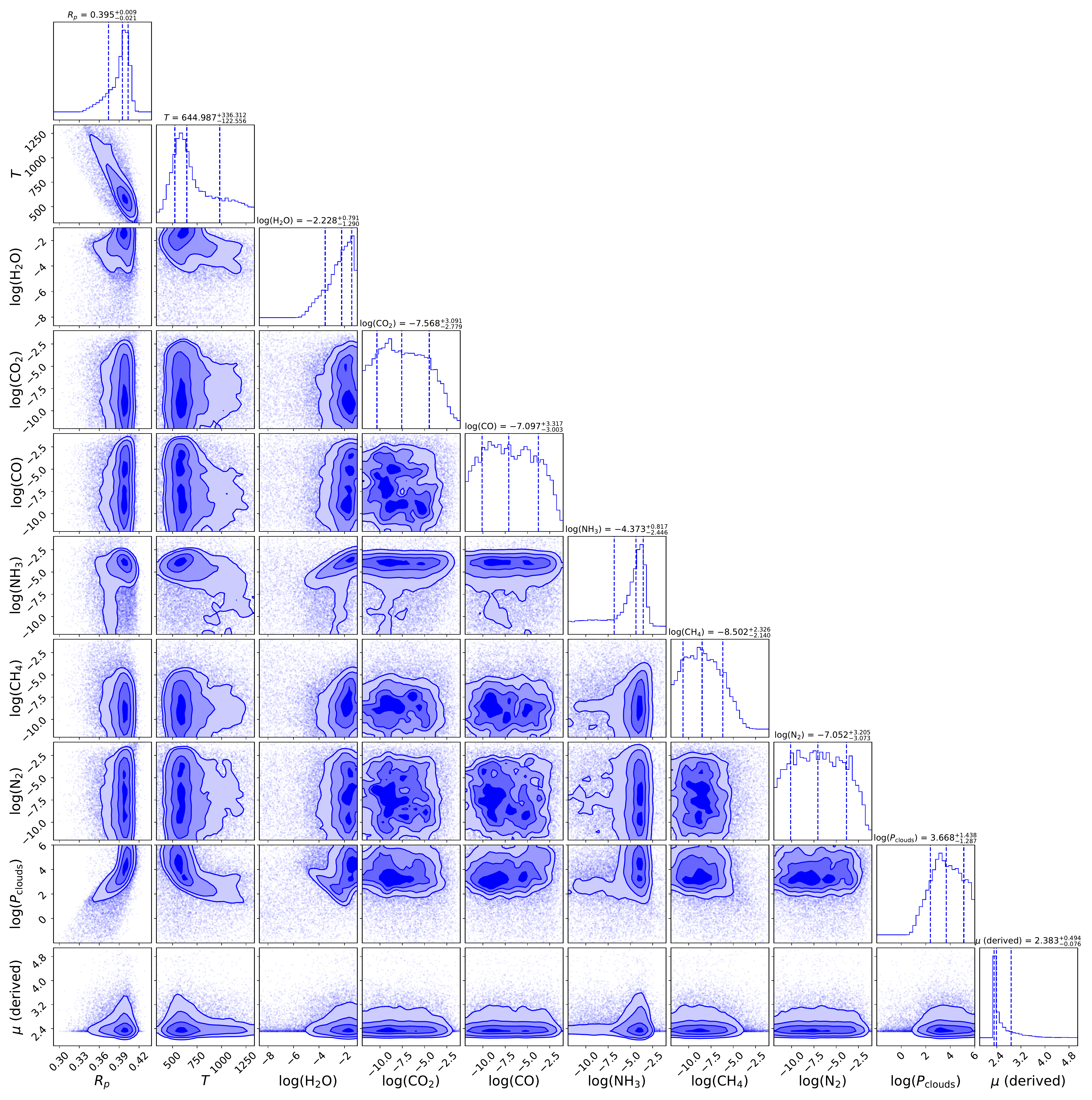}
		\caption{ HD\,106315\,c's posterior distribution including also N$_2$ to the full chemical scenario. As this Figure shows, the inclusion of nitrogen does not affect the mean molecular weight. Moreover, the detection of NH$_3$ remains around 10$^{-4}$.}
		\label{FigA3}
	\end{figure*}

\end{document}